\documentclass[3p]{elsarticle}
\usepackage{float}
\usepackage{color}
\usepackage{graphicx}
\usepackage{amsmath,amssymb,verbatim,bm}
\usepackage{ctable} 
\usepackage{slashed}
\usepackage{longtable}
\usepackage{url}
\usepackage{ragged2e}
\usepackage[ruled]{algorithm}
\usepackage{algorithmic}
\newcommand{\R}{\mathbb{R}}
\newcommand{\bone}{\bm{1}}
\newcommand{\bzero}{\bm{0}}
\newcommand{\bfi}{\bm{\phi}}
\newcommand{\bFi}{\bm{\Phi}}
\newcommand{\bPi}{\bm{\Pi}}
\newcommand{\bb}{\bm{b}}
\newcommand{\bd}{\bm{d}}
\newcommand{\be}{\bm{e}}

\newcommand{\bu}{\bm{u}}
\newcommand{\bv}{\bm{v}}
\newcommand{\bx}{\bm{x}}
\newcommand{\bA}{\bm{A}}
\newcommand{\dA}{\bm{{d\mspace{-1mu}A}}}
\newcommand{\bB}{\bm{B}}

\newcommand{\bD}{\bm{D}}

\newcommand{\bI}{\bm{I}}
\newcommand{\bJ}{\bm{J}}
\newcommand{\bL}{\bm{L}}
\newcommand{\bM}{\bm{M}}
\newcommand{\bQ}{\bm{Q}}
\newcommand{\bR}{\bm{R}}
\newcommand{\tR}{\widetilde{\bm{R}}}
\newcommand{\bS}{\bm{S}}
\newcommand{\bU}{\bm{U}}
\newcommand{\bV}{\bm{V}}

\newcommand{\bw}{\bm{w}}
\newcommand{\bT}{\bm{T}}
\newcommand{\bX}{\bm{X}}
\newcommand{\bY}{\bm{Y}}
\newcommand{\by}{\bm{y}}

\newcommand{\tz}{\widetilde{\bm{z}}}
\newcommand{\tZ}{\widetilde{\bm{Z}}}
\newcommand{\tZo}{\tZ^{(1)}}
\newcommand{\tZt}{\tZ^{(2)}}
\newcommand{\tZoT}{[\tZ^{(1)}]^T}
\newcommand{\tZtT}{[\tZ^{(2)}]^T}

\newcommand{\bSm}{\bm{\Sigma}}
\newcommand{\bnabla}{\bm{\nabla}}

\newcommand{\bck}{\mspace{-4mu}}
\newtheorem{definition}{Definition}
\newtheorem{theorem}{Theorem}
\newtheorem{proof}{Proof of Theorem}
\newtheorem{corollary}{Corollary}
\newtheorem{cproof}{Proof of Corollary}
\newtheorem{axiom}{Axiom}
\newtheorem{principle}{Principle}

\newtheorem{lemma}{Lemma}
\newtheorem{lproof}{Proof of Lemma}
\newtheorem{remark}{Remark}

\DeclareMathOperator{\argmin}{argmin}
\DeclareMathOperator{\diag}{diag}
\DeclareMathOperator{\tr}{tr}
\DeclareMathOperator{\E}{\mathbb{E}}
\DeclareMathOperator{\proba}{Prob}

\DeclareMathOperator{\rp}{rp}
\DeclareMathOperator{\DC}{DC}
\DeclareMathOperator{\ki}{{KI}}

\DeclareMathOperator{\drpo}{d_{\rp 1}}
\DeclareMathOperator{\drpt}{d_{\rp 2}}
\DeclareMathOperator{\drpp}{d_{\rp(p)}}
\DeclareMathOperator{\drh}{\widehat{d}_{\rp(p)}}
\newcommand{\hR}{\widehat{R}}
\DeclareRobustCommand{\er}[1]{{\hR_{#1}}}

\DeclareRobustCommand{\dbar}{{\overline{d_n}}}
\DeclareRobustCommand{\droot}[2]{{d_\text{rootED}\left({#1},{#2}\right)}}
\DeclareRobustCommand{\dcon}[2]{{d_{\DC_0}\left({#1},{#2}\right)}}
\DeclareRobustCommand{\dw}[1]{{\Delta w_{#1}}}

\DeclareRobustCommand{\O}[1]{{\cal O} \left(#1\right)}
\DeclareRobustCommand{\T}[1]{{\Theta} \left(#1\right)}

\journal{Discrete Applied Mathematics}
\usepackage{hyperref}
\hypersetup{
  bookmarks=true,
  bookmarksopen=true,
  bookmarksnumbered=tue,
  pdfpagemode = {UseOutlines},
  pdfstartview = {FitH},
  citecolor={green},
  linkcolor={red}
  colorlinks = true,
  linkbordercolor = {white},
  colorlinks=true
  }
\begin{document}
\begin{frontmatter}

  \title{The Resistance Perturbation Distance:\\  A Metric for the Analysis of Dynamic Networks}
  \author{Nathan D. Monnig\fnref{fnt1}}
  \address{Numerica Corporation, Fort Collins, CO 80528}
  \fntext[fnt1]{nathan.monnig@colorado.edu}
  \author{Fran\c{c}ois G. Meyer\fnref{fnt2}}
  \address{University of Colorado at Boulder, Boulder CO 80305}
  \fntext[fnt2]{Corresponding author: fmeyer@colorado.edu}

  \begin{abstract}
    To quantify the fundamental evolution of time-varying networks, and detect abnormal behavior, one
    needs a notion of temporal difference that captures significant organizational changes between two
    successive instants.  In this work, we propose a family of distances that can be tuned to
    quantify structural changes occurring on a graph at different scales: from the local scale
    formed by the neighbors of each vertex, to the largest scale that quantifies the connections
    between clusters, or communities. Our approach results in the definition of a true distance, and
    not merely a notion of similarity. We propose fast (linear in the number of edges) randomized
    algorithms that can quickly compute an approximation to the graph metric. The third contribution
    involves a fast algorithm to increase the robustness of a network by optimally decreasing the
    Kirchhoff index. Finally, we conduct several experiments on synthetic graphs and real networks,
    and we demonstrate that we can detect configurational changes that are directly related to the
    hidden variables governing the evolution of dynamic networks.
  \end{abstract}

\end{frontmatter}
\section{Introduction}
Many complex systems are well represented as graphs or networks, with the agents represented as
vertices and edges symbolizing relationships or similarities between them. In many instances, the
relationships between vertices evolve as a function of time: edges may appear and disappear, the
weights along the edges may change. The study of such {\em dynamic graphs} often involves the
identification of patterns that couple changes in the network topology with the latent dynamical
processes that drive the evolution of the connectivity of the network \cite{akoglu14, karsai14,
  kovanen13, lafond14,ranshous15}.

To quantify the temporal and structural evolution of time-varying networks, and detect abnormal
behavior, one needs a notion of temporal difference that captures significant configurational changes
between two successive instants. The design of similarity measures for the pairwise comparison of
graphs \cite{soundarajan14} is therefore of fundamental importance. 

Because we are interested in detecting changes between two successive instants, we focus on the
problem of measuring the distance between two graphs on the same vertex set, with known vertex
correspondence (see Fig. \ref{dynagraph}). We note that determining whether two graphs are
isomorphic under a permutation of the vertex labels is a combinatorially hard problem (e.g.,
\cite{mckay2014}, and references therein, but see the recent results \cite{babai15}). Several notion
of similarities (e.g., \cite{berlingerio13,papadimitriou2010,koutra16}, and references therein)
have been proposed.  Unlike a true metric, a similarity merely provides a notion of
resemblance. Most approaches rely on the construction of a feature vector that provides a signature
of the graph characteristics; the respective feature vectors of the two graphs are then compared
using some norm, or distance. A similarity function is typically not injective (two graphs can be
perfectly similar without being the same), and rarely satisfies the triangular inequality.

Instead of comparing two feature vectors, several researchers (e.g.,
\cite{ahmed15,borgwardt07,bunke11,bai13,foggia14,shervashidze11,vishwanathan10} and references
therein) have proposed to use a kernel function. This approach offers the same advantage as the
computation of a similarity:%
\begin{figure}[H]
  \centerline{
    \includegraphics[width=0.4\textwidth]{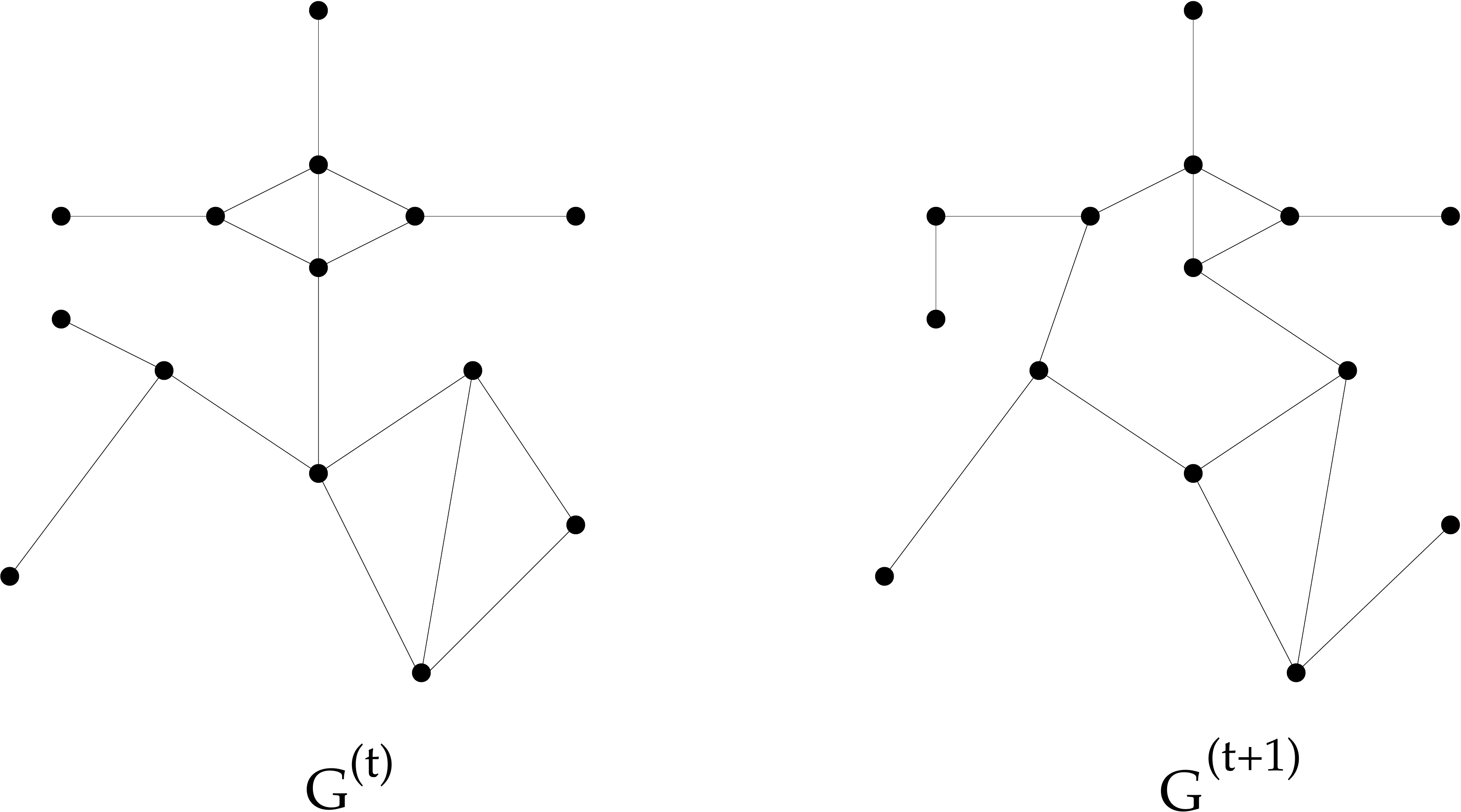}
  }
  \caption{Dynamic graph $G^{(t)}$ at time $t$ (left) and $t+1$ (right)
    \label{dynagraph}}
\end{figure}
\noindent the isomorphism problem need not be solved. Unfortunately, the kernels
do not define proper metrics, and we are left with a weaker notion of similarity.

Several distances between two graphs with the same size have been proposed
(e.g., \cite{baur05,bunke11}, and references therein).  As detailed in section
\ref{existing_metrics_intro}, we argue that existing distances either fail to
capture a notion of structural similarity, or lead to algorithms that have a
high computational complexity. 
\subsection{Contribution and Organization of the Paper}
The contributions of this work are threefold.  First, we propose a family of distances that can be
tuned to quantify configurational changes that occur on a graph at different scales: from the local scale
formed by the local neighbors of each vertex, to the largest scale that quantifies the connections
between clusters, or communities. Our approach results in the definition of a true distance, and not
merely a notion of similarity. The second contribution encompasses fast computational algorithms to
evaluate the metrics developed in the first part. We developed fast (linear in the number of edges)
randomized algorithms that can quickly compute an approximation to the graph metric. The third
contribution involves fast algorithms to increase the robustness of a network by optimally
decreasing the Kirchhoff index. Finally, we conduct several experiments on synthetic and real
dynamic networks, and we demonstrate that the resistance perturbation distance can detect the
significant changes in the hidden latent variables that control the network dynamics.

The remainder of this paper is organized as follows.  In the next section we introduce the main
mathematical concepts and corresponding nomenclature. In section \ref{definition} we formally define
the problem and review the existing literature. In section \ref{general_distance_section} we propose
a novel framework for constructing graph distances; we focus the rest of the paper on the {\bf
  resistance perturbation distance}, which is defined in section \ref{rp_dist_def_sec}. In section
\ref{analytic_results_section}, we study simple perturbations of several prototypical graphs for
which the resistance perturbation distance can be computed analytically. Fast randomized algorithms
are described in section \ref{rp2_computation_section}. The optimization of the robustness of a
network, based on optimally decreasing the Kirchhoff index, is described in section
\ref{optimization}. In section \ref{experiments}, we use the resistance perturbation metric to
detect significant changes in synthetic and real dynamic networks.  We conclude in Section
\ref{discussion} with a discussion on future work. Some technical details and proofs are left aside
in the Appendix. A list of the main notations used in the paper is provided in section \ref{notation}.
\section{Preliminaries and Notation}
We introduce in this section the main concepts and associated nomenclature.

\noindent We denote by $\be_i$ the $i^\text{th}$ vector of the canonical basis in $\R^n$.
The space of matrices of size $n \times m$ with entries in $\R$ is denoted by
$\bM_{n\times m}$; to alleviate notations we write $\bM_n$ to denote
$\bM_{n\times n}$.\\

\noindent We denote by $G=(V,E,w)$ an undirected weighted graph with a vertex set
$V=\{1,\ldots,n\}$, an edge set $E$, and a symmetric weight function $w$ that quantifies the
similarity between any two vertices $i$ and $j$. In this work, we use the terms graph and network
exchangeably. 

The weighted adjacency matrix, $\bA \in \bM_n$, is given by
\begin{equation}
  A_{ij} = A_{ji} = 
  \begin{cases}
    w_e & \text{if the edge} \; e = [i,j] \in E, \\
    0 & \text{otherwise.} 
  \end{cases}
  \label{adjacency_def_eqn}
\end{equation}
For simplicity, we will always assume $G$ is connected and does not contain any self-loops.  We
further define the combinatorial Laplacian matrix,
\begin{equation}
  \bL = \bD - \bA,
  \label{combinatorial_lap_def_eqn}
\end{equation}
where the degree matrix $\bD$ is the diagonal matrix of vertex degrees,
\begin{equation*}
  D_{ii}=\sum_{j=1}^n A_{ij}.
\end{equation*}
The matrix $\bL$ is symmetric and positive semi-definite. We denote by $\bfi_k$
the $k^\text{th}$ eigenvector of $\bL$ corresponding to $\lambda_k$, with
$0=\lambda_1 < \lambda_2 \leq \ldots \leq \lambda_n$. We can write $\bL$ in
terms of its spectral decomposition, 
\begin{equation}
  \bL= \sum_{k=2}^n \lambda_k \bfi_k \bfi_k^T.
  \label{spectral}
\end{equation}
$\bL^\dagger$ denotes the Moore-Penrose pseudoinverse of $\bL$.  Because $\bL$ is symmetric,
$\bL^\dagger$ is also symmetric. The pseudoinverse is easily formulated from the spectral
decomposition of $\bL$,
\begin{equation}
  \bL^\dagger = \sum_{k=2}^n \frac{1}{\lambda_k} \bfi_k \bfi_k^T.
  \label{Ldagger_eqn}
\end{equation}
We  can also express $\bL^\dagger$ in terms of the inverse of $\bL +
\frac{1}{n}\bJ$, which is full-rank,
\begin{equation}
  \bL^\dagger = \left(\bL+ \frac{1}{n}\bJ\right)^{-1} - \frac{1}{n}\bJ,
  \label{Lpseudo}
\end{equation}
where 
\begin{equation}
  \bJ = \bone\bone^T,\;\text{with}\quad \bone^T = \begin{bmatrix} 1 & 1 & \cdots & \end{bmatrix}.
  \label{J}
\end{equation}
For the purpose of defining a concept of gradient on the graph, we
assign an (arbitrary) orientation to each edge $e$. With this orientation, we define a notion of
gradient, captured by the signed edge incidence matrix, $\bB \in \bM_{m \times n}$,
\begin{equation}
  B_{ei}=
  \begin{cases}
    1 & \text{if vertex $i$ is at the head of $e$,} \\
    -1 & \text{if vertex $i$ is at the tail of $e$,} \\
    0 & \text{otherwise.}
  \end{cases}
  \label{edgeincidence}
\end{equation}
We define the diagonal edge weight matrix $\dA \in \bM_{m\times n}$ with diagonal entries
$\dA_{ee}=w_e$.  The (gradient) edge incidence matrix can be used to express the combinatorial
Laplacian matrix as
\begin{equation}
  \bL= \bB^T \dA \bB.
\end{equation}
In this paper we are concerned with two undirected weighted graphs $G^{(1)}$ and $G^{(2)}$ with a
common vertex set $V=\{1,\ldots,n\}$, two edge sets $E^{(1)}$ 
and $E^{(2)}$, and two symmetric weight functions $w^{(1)}$ and $w^{(2)}$. We denote by $\bA^{(1)}$
and $\bA^{(2)}$ the corresponding weighted adjacency matrices.
\section{Statement of the Problem and Related Work}
\label{definition}
Inspired by the work of Koutra {\em et al.} \cite{koutra16}, we propose to characterize distances
between graphs using a set of axioms and principles. After defining these axioms and principles, we
use these to review the existing literature on graph distance and similarity.
\subsection{Metrics Between Graphs: an Axiomatic Definition}
\label{axioms_section}
\begin{axiom}[Definition of a Distance]
  A distance on a space of graphs should meet all the conditions of a distance: non-negativity,
  identity, symmetry, and subadditivity.
\end{axiom}
The set of axioms in Koutra {\em et al.} \cite{koutra16} are somewhat similar to our single axiom, given
the translation of a distance into a similarity measure.  Our axiom is stronger in that it also
implies the triangle inequality, in addition to symmetry and identity (the first two axioms in
\cite{koutra16}).  

We note that Axiom 3 from Koutra {\em et al.} \cite{koutra16} is the {\em Zero property}:\\
$\text{sim}(G^{(1)},G^{(2)}) \rightarrow 0$ as $n \rightarrow \infty$, if $G^{(1)}$ is the complete
graph and $G^{(2)}$ is the empty graph. As explained in Remark \ref{target_drpo}, the zero property
holds in our case: if $G^{(2)}$ is obtained by disconnecting $G^{(1)}$, then our distance goes to
infinity; in other words the similarity goes to zero.\\

\noindent In addition to the above axiom, we argue that a useful distance should obey the following
the four principles proposed by Koutra {\em et al.} \cite{koutra16}.
\begin{principle}[Edge Importance] 
  Changes that create disconnected components should be penalized more than changes
  that maintain the connectivity properties of the graphs. 
  \label{edge_importance_principle}
\end{principle}
\begin{principle}[Weight Awareness] 
  In weighted graphs, the larger the weight of the removed edge is, the greater the
  impact on the distance should be.
  \label{weight_awareness_principle}
\end{principle}
\begin{principle}[Edge-``Submodularity''] 
  A specific change is more important in a graph with few edges than in a much
  denser, but equally sized graph.
  \label{edge_submodularity_principle}
\end{principle}
\begin{principle}[Focus Awareness] 
  Random changes in graphs are less important than targeted changes of the same extent.
  \label{focus_awareness_principle}
\end{principle}
The first three principles are intuitive and self-explanatory.  The principle of focus awareness
requires some interpretation.  Koutra {\em et al.} \cite{koutra16} test for focus awareness by
either removing all edges connected to a vertex (a targeted change) or randomly removing the same
number of edges from the whole graph (a random change of the same extent).  In most applications,
the targeted removal of all edges connecting a single vertex would be viewed as a more significant
change to the network topology compared with most realizations of random edge removal.  An ideal
distance should account for the relative importance of these types of changes.

We propose an alternative interpretation of the ``focus awareness'' principle.  We first observe
that edges can be partitioned in terms of their ``functionality'' in the network. In this work, the
notion of functionality is measured in terms of connectivity, and is quantified with the concept of
{\bf effective resistance}. Now, if we consider the distribution of effective resistances across all
edges in $E$, some edges will contribute to rare events because they have very large effective
resistance. We argue that such edges are unlikely to be removed by a random selection of edges in
the network.  In other words, a {\em targeted change} would correspond to a perturbation of these
rare edges with very high effective resistance. Our definition of focus awareness recovers the
intuitive notion introduced by Koutra {\em et al.} \cite{koutra16}. This point is further discussed in
section \ref{discussion}.

\noindent Adherence to the axioms and principles does not imply that a distance will be useful in
practice.  {\em A distance must also be computable}.  Modern applications require algorithms to
compute or approximate the distance in nearly linear time in the number of edges.
\subsection{Existing Notions of Similarities Between Graphs
\label{existing_metrics_intro}}
Unlike a true metric that satisfies the three axioms of a metric, similarities
are merely providing a notion of resemblance. This approach relies on the
construction of a feature vector that provides a signature of the graph
characteristics; the respective feature vectors of the two graphs are then
compared using some norm, or distance (e.g.,
\cite{berlingerio13,koutra16,papadimitriou2010}, and references therein).  The
similarity function is typically not injective (two graphs can be perfectly
similar without being the same), and rarely satisfies the triangular
inequality. The authors in \cite{koutra16} offer a list of properties that a
``good'' similarity should obey.  They define the DeltaCon$_0$ similarity as
follows,
\begin{equation}
  \text{sim}_{\text{DC}_0}\left( G^{(1)},G^{(2)} \right)
  =\frac{1}{1+\droot{G^{(1)}}{G^{(2)}}},  
\label{delta_con_def}
\end{equation}
where the root Euclidean distance is defined as
\begin{equation}
  \droot{G^{(1)}}{G^{(2)}} =
  \left\{
    \sum_{i,j=1}^n  \left(\sqrt{{\bS}^{(1)}_{ij}}-\sqrt{{\bS}^{(2)}_{ij}} \right)^2 
  \right\}^{1/2}\mspace{-24mu},
  \label{root}
\end{equation}
and where ${\bS}^{(i)} $ is the fast belief propagation matrix defined by
\begin{equation}
  {\bS}^{(i)} = \left[ \bI + \varepsilon^2 \bD^{(i)} - \varepsilon {\bA}^{(i)} \right]^{-1}\mspace{-24mu},
  \label{belief}
\end{equation}
and $\varepsilon = 1/(1+\max_i D_{ii})$.  To gain some intuition about the role of the fast belief
propagation matrix $\bS$ , we assume $\varepsilon \ll 1$, and drop the term $\varepsilon^2 \bD$ in
$\bS$ to arrive at
\begin{equation}
  \bS \approx \left( \bI - \varepsilon \bA \right)^{-1} = \bI + \varepsilon \bA+
  \varepsilon^2 {\bA}^2 + \varepsilon^3 {\bA}^3 + \ldots . 
\end{equation}
In the unweighted case, $A^k_{ij}$ is the count of paths of length $k$ between
vertices $i$ and $j$. In the weighted case, $A^k_{ij}$ is the sum, over all
paths of length $k$ between vertices $i$ and $j$, of the product of the weights
along the corresponding paths. We conclude that $\bS$ encapsulates information
about the connectivity between vertices at all scales (with longer paths having
a reduced impact).

We provide in \ref{DeltaCon_appendix} a detailed analysis of the DeltaCon$_0$
distance that uncovers unexpected behavior.  Our analysis is based on a
single-edge%
  \begin{figure}[H]
    \centerline{
      \includegraphics[width=0.5\textwidth]{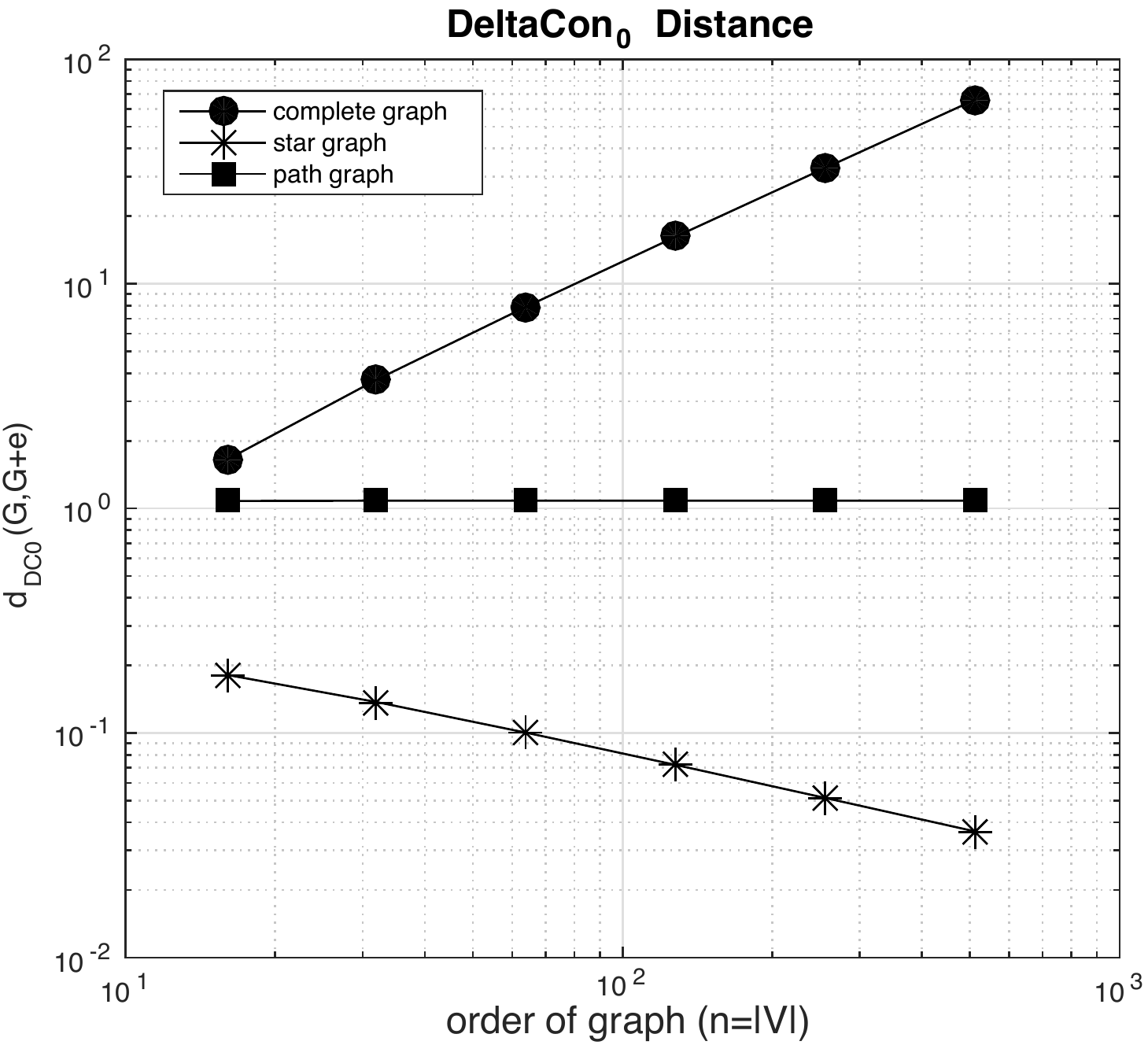}
      \includegraphics[width=0.5\textwidth]{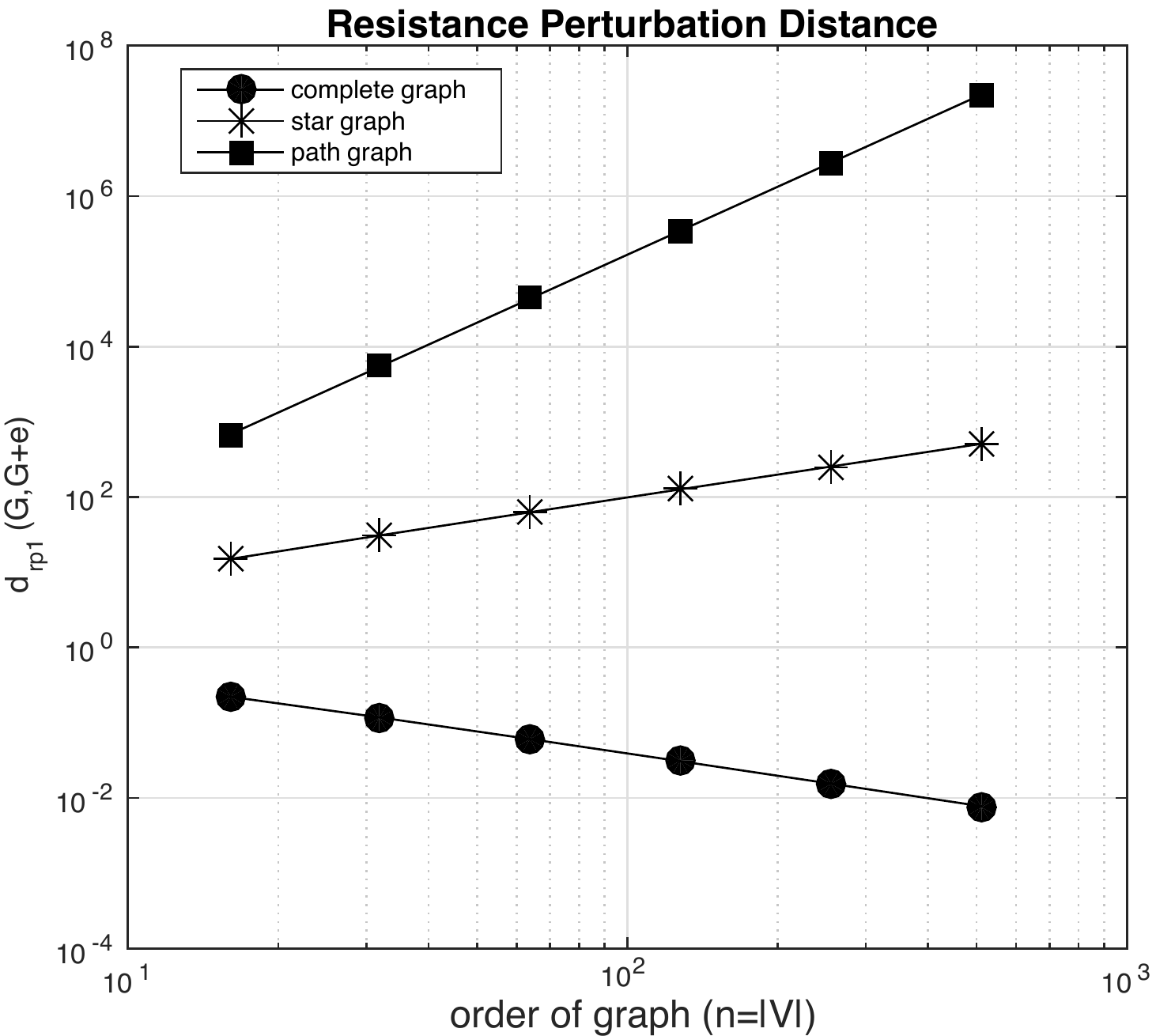}
    }
    \caption{Experimental scaling of the DeltaCon$_0$ similarity (left) and the resistance
      perturbation distance (right) for single edge perturbation of simple graphs.
      \label{RP_vs_DC}}
  \end{figure}
  \noindent  perturbation of several simple graphs, where the DeltaCon$_0$
distance between the original and perturbed graphs can be computed analytically.\\

\noindent In the case of the complete graph, $K_n$, the root Euclidean distance created by the
perturbation grows with $n=|V|$,
\begin{equation}
  \droot{K_n}{K_n+\dw{kl})} = \left\vert\frac{1}{\sqrt{2}} - \frac{1}{\sqrt{2 + \dw{kl}}} \right\rvert n + \O{1}.
\end{equation}
However, in the case of the much sparser star graph, $S_n$, the root Euclidean distance decays with
$n=|V|$, 
\begin{equation}
  \droot{S_n}{S_n+\dw{kl}} = 
 \frac{\sqrt{2\dw{kl}}}{\sqrt{n}}- \frac{\sqrt{2}}{n} + \O{1/n^{3/2}}.
\end{equation}
These leading-order analyses are confirmed experimentally in Fig. \ref{RP_vs_DC}, where we compare
the DeltaCon$_0$ similarity with the resistance perturbation presented in this paper.

  We can interpret these results in terms of the graph density.  The density of the complete
  graph $K_n$, as measured by the average degree $\dbar$, is $n-1$, whereas the densities of the
  star $S_n$ and path $P_n$ graphs are $2(1 - 1/n)$ and $1$ respectively,
  \begin{equation}
    \dbar(P_n) = 1 < \dbar(S_n) = 2\left(1 -1/n\right)  < \dbar(K_n) = n-1, \quad n \ge 3.
  \end{equation}
  The DeltaCon distances for a single edge perturbation are ordered as follows, 
  \begin{equation}
    \begin{split}
      \droot{S_n}{S_n +\dw{kl}} =  \T{\displaystyle \frac{1}{\sqrt{n}}} & < \droot{P_n}{P_n +\dw{kl}}  = \T{1}\\
      & < \droot{K_n}{K_n +\dw{kl}} = \T{n},
    \end{split}
  \end{equation}
  while the  RP distances for the respective graphs are ordered as follows,
  \begin{equation}
    \begin{split}
      \drpo\left(K_n, K_n +\dw{kl}\right) = \T{\displaystyle \frac{1}{n}} 
      & < \drpo\left(S_n, S_n +\dw{kl}\right) = \T{n}\\
      & < \drpo\left(P_n, P_n +\dw{kl}\right) = \T{n^3}.
    \end{split}
  \end{equation}
  We conclude that, on these three graphs, the RP distance for a single edge perturbation decreases
  as a function of the graph density, which is consistent with Principle 3 from Koutra et
  al. \cite{koutra16}, which asserts that {\em ``A specific change is more important in a graph with
    few edges than in a much denser, but equally sized graph.''}

  The ordering of the DeltaCon distances is not exactly the reverse of the ordering of the
  RP-distances. Nevertheless, when comparing the complete graph to either the star, or the path
  graphs, we conclude that the DeltaCon distance for a single edge perturbation increases as a
  function of the graph density, which is inconsistent with Principle 3.

  Indeed, a principled distance should ascribe greater significance to changing an edge weight in
  the star graph (a sparser graph in which each edge is more important) relative to the complete
  graph (a dense graph in which no single edge is crucial to the overall connectivity).

  When comparing the star to the path, we note that DeltaCon respects Principle 3. Because
  both the star and the path graphs have a constant density, we find this comparison to be less of a
  concern.  We complement our theoretical analysis of DeltaCon$_0$  with an experimental
  evaluation conducted in Section~\ref{experiments}.\\

  In the context of the analysis of dynamic graphs, the authors in \cite{sricharan14} describe an
  algorithm to localize edges that most significantly contribute to dynamical structural changes.
  To tackle this question, the authors define the following distance to quantify structural changes 
  as the graph $G^{(n)}$ evolves to $G^{(n+1)}$,
  \begin{equation}
    d_{\text{CAD}}(G^{(n)}, G^{(n+1)}) = \mspace{-32mu}\sum_{(u,v) \in F\subseteq E^{(n)}}\mspace{-32mu}
    \left|A^{(n+1)}(u,v) - A^{(n)}(u,v)\right|\left|\kappa^{(n+1)}(u,v) - \kappa^{(n)}(u,v)\right|,
    \label{CAD}
  \end{equation}
  where $F$ is a subset of the edge set $E^{(n)}$ of the graph $G^{(n)}$, and $\kappa^{(n)}_{u,v}$ is
  the {\em commute time} between vertices $u$ and $v$ in the graph $G^{(n)}$ (see Definition \ref{def_commute}).

  The authors in \cite{sricharan14} propose to minimize this distance to identify the maximal ``core''
  subset of edges $F$ that contribute to the least structural changes between time $n$ and $n+1$. The
  complement of the core set $F$ consists of edges that trigger large structural changes.

  While the goal of our work is quite different from that of \cite{sricharan14}, our notion of
  effective resistance, defined in (\ref{rp_dist_def_eqn}), is indeed similar to the distance
  (\ref{CAD}). As explained in section \ref{resistances_intro_section}, the commute time is -- up to a
  renormalization by the volume of the graph $m=|E^{(n)}|$ -- the same as the effective resistance.

  Because of the presence of the term $\left|A^{(n+1)}(u,v) - A^{(n)}(u,v)\right|$, the distance
  $d_{\text{CAD}}$ does not satisfy the triangle inequality. We suspect that $d_{\text{CAD}}$ is not
  injective. An increase (decrease) in the commute time $\kappa_{uv}$ throughout the graph could in
  principle be cancelled by a corresponding increase (decrease) in the volume $m$, to keep the
  effective resistance the same (see (\ref{kappa_eq_R})). This argument is not in
  contradiction with the Rayleigh's Monotonicity Principle that only applies to effective resistance,
  and not the commute time.

  Because of the similarity between the distance $d_{\text{CAD}}$ and the resistance perturbation
  distance, we evaluated $d_{\text{CAD}}$ in all experiments conducted in section~\ref{experiments}.

Another similarity that captures the geometry of the graph at all scale is
provided by the {\em spectral similarity} which quantifies the distance between the respective
spectra $\{ \lambda^{(1)}_i \}_{i=1}^n$ and $\{ \lambda^{(2)}_i \}_{i=1}^n$ of $G^{(1)}$ and
$G^{(2)}$ The spectra can be computed from the adjacency, Laplacian, or normalized Laplacian
matrices \cite{bunke2007,peabody2002,wilson2008}. The spectral similarity is defined by
\begin{equation*}
  d_\lambda\left( G^{(1)},G^{(2)} \right)=\sqrt{ \displaystyle \sum_{i=1}^n \left( \lambda^{(1)}_i
      - \lambda^{(2)}_i \right)^2  }. 
\end{equation*}
The existence of iso-spectral graphs prevents $d_\lambda$ to be a distance,
since\\
$d_\lambda\left( G^{(1)},G^{(2)} \right) = 0$ does not necessarily imply
that $G^{(1)}=G^{(2)}$.  In addition, the spectral methods are costly since they
require computation of the full graph spectrum.

Signature similarity is another method considered in Koutra {\em et al.} \cite{koutra16}.  The signature
similarity compares two graphs by first computing a large number of features from the two graphs.
These features are then projected onto a random lower-dimensional feature space within which the
similarity between the two graphs is computed.  This method was found to be the best performing
method in Papadimitriou {\em et al.} \cite{papadimitriou2010}.  Unfortunately, Koutra et
al. \cite{koutra16} proved that the signature similarity, along with the graph edit distance, and
all variants of the $\lambda$-distance fail to conform to Principles \ref{edge_importance_principle}
and \ref{edge_submodularity_principle}.

Other notions of similarity, which do not necessarily define a proper distance, can be defined.  For
example, Spielman and Teng \cite{spielman2011} (see also \cite{batson}) introduced another notion of {\em spectral
  similarity}. Two graphs $G^{(1)}$ and $G^{(2)}$, with Laplacians $\bL^{(1)}$ and $\bL^{(2)}$, on
the same vertex set $V$ are said to be {\em $\sigma$-spectrally similar} if \cite{spielman2011},
\begin{equation}
  \frac{1}{\sigma}\;
  \bx^T \bL^{(2)}\mspace{-2mu}\bx 
  \leq 
  \bx^T \bL^{(1)} \mspace{-2mu}\bx 
  \leq 
  \sigma \; \bx^T \bL^{(2)}\mspace{-2mu}\bx, 
  \quad 
  \forall \bx \in \R^n. 
\end{equation}
\subsection{Graph Kernels}
Instead of comparing the feature vectors, which represent the graphs $G^{(1)}$ and $G^{(2)}$
respectively, several researchers (e.g.,
\cite{ahmed15,borgwardt07,bunke11,bai13,foggia14,shervashidze11,vishwanathan10} and references
therein) have proposed to use a kernel function. This approach offers the same advantage as the
computation of a similarity: the isomorphism problem need not be solved. Unfortunately, the kernels
do not define proper metrics, and we are left with weaker notions of resemblance.
\subsection{Existing True Metrics on the Space of Connected Graphs of a Fixed Size.}
\noindent Finally, we review the distances between two graphs with the same size
$n$ that lead to true metrics
\cite{baur05,bunke11}.  \\

\noindent The {\bf edit distance} between  $G^{(1)}$ and $G^{(2)}$ is defined by 
\begin{equation*}
  d_1(G^{(1)},G^{(2)}) = 
  \left\lVert {\bA}^{(1)} - {\bA}^{(2)} \right\rVert_1 = 
  \sum_{i,j} \left\lvert A^{(1)}_{ij} - A^{(2)}_{ij} \right\rvert. 
\end{equation*}
The edit distance does not reflect structural differences: all edges are treated
equally.  A more useful notion of distance is provided by the {\bf cut distance}
defined by
\begin{equation*}
  d_C(G^{(1)},G^{(2)}) = 
  \underset{S,T\subseteq V}{\text{max}} 
  \left\lvert E_{G^{(1)}}(S,T)-E_{G^{(2)}}(S,T) \right\rvert,
\end{equation*}
where $E_G(S,T)$ denotes the sum of the weights along the edges  connecting the vertices in
$S\subseteq V$ to the vertices in $T\subseteq V$.  The computation of the cut norm
requires optimizing over ${\cal O}(2^{2n})$ pairs of subsets of $V$, and is therefore
prohibitively expensive even for moderately sized graphs.\\

\noindent The {\bf difference in path lengths} \cite{chartrand1998} is based on the pairwise
difference between the shortest distances in the two graphs,
\begin{equation*}
  \min_{\bPi} \sum_{u,v \in V} | d_{G^{(1)}} (u,v) - d_{G^{(2)}} (\bPi(u),\bPi(v))|
\end{equation*}
where the minimum is computed over any permutation $\bPi$ of the vertices, and $d_{G^{(i)}} (u,v)$
is the shortest distance from $u$ to $v$ in the graph $G^{(i)}$.  Although this method defines a
metric between unweighted graphs, it only defines a pseudo-metric on the space of weighted graphs
(it is not injective).\\

\noindent Finally, a set-theoretical notion of distance can be derived from computing the size (number
of vertices) of the largest edge-, or vertex-induced subgraph that is common to $G^{(1)}$ and
$G^{(2)}$. It can be shown that this concept yields a metric on graphs with the same size
\cite{baur05}. Unfortunately, the detection of a {\bf maximum common subgraph}
is an NP-complete problem.\\

\noindent We conclude this section with the observation that many existing distances fail to conform
to the set of axioms and principles presented in the previous section, which were inspired by the
work of \cite{koutra16}.  Furthermore, many true distances suffer from a prohibitive
computational cost (e.g., the cut distance).  The limitations of existing distances and similarity
measures demonstrate the need for novel distances between graphs. In the next section, we introduce a
very general framework for constructing distances between two graphs. This novel approach allows the
user to customize the distance to specific needs. We study one specific instance of this framework,
and introduce the {\bf resistance perturbation distance}, as a metric that obeys all the axioms and
principles. In addition, we develop fast algorithms to compute this metric.
\section{A Unified Framework for Graph Distances}
\label{general_distance_section}
We first make the following simple observation: if we consider a distance $d$ on $\bM_n$, then we
can induce a family of distances between any two graphs $G^{(1)}$ and $G^{(2)}$ on the same set of
vertices by measuring the distance, $d(\bA^{(1)},\bA^{(2)})$ between the corresponding adjacency
matrices $\bA^{(1)}$ and $\bA^{(2)}$ . More generally, one can compute the distance between any
matrix-to-matrix function $\varphi$ of $\bA^{(1)}$ and $\bA^{(2)}$, as explained in the following definition.

\begin{definition}[General graph distance]
  \label{pseudo_dist_def}
  Given a matrix-to-matrix function, (or more simply a matrix function), $\varphi $,
  \begin{equation*}
    \varphi: \bM_n \rightarrow \bM_n,
  \end{equation*}
  and a distance $d$ on $\bM_n$,  we define the pseudo-distance $d_\varphi$
  between two graphs  $G^{(1)}$ and $G^{(2)}$ as follows,
  \begin{equation}
    d_\varphi(G^{(1)},G^{(2)})=d(\varphi({\bA}^{(1)}),\varphi({\bA}^{(2)})),
  \end{equation}
  where $\bA^{(1)}$ and $\bA^{(2)}$ are the adjacency matrices representing $G^{(1)}$ and $G^{(2)}$,
  respectively.  If $\varphi$ is injective, then $d_\varphi$ defines a distance.
\end{definition}
Definition \ref{pseudo_dist_def} is significant because it provides a natural mechanism to construct
new distances by decoupling two aspects of the distance $d_\varphi$. First, the matrix function
$\varphi$ extracts from each graph a property of interest. The function $\varphi$ extracts
configurational or geometric properties about each graph. The distance $d$ can then be used to
emphasize large or small variations in the matrix function $\varphi$.  In addition, the choice of
$d$ can also be guided by the existence of fast algorithms to compute $d_\varphi$ (as is the case in
our work).

We note that the structure introduced in Definition \ref{pseudo_dist_def} is quite general since
many existing (pseudo-) distances can be recast using this formalism.  For example, if $\varphi$ is
the identity map, and $d$ is the entrywise 1-norm of the difference, then $d_\varphi$ is the edit
distance.  Alternatively, if $d$ is the cut norm of the difference between the adjacency matrices
\cite{Frieze99}, we arrive at the cut distance.  If $\varphi$ returns the diagonal matrix of sorted
eigenvalues of either the adjacency, Laplacian, or normalized Laplacian matrices, and $d$ is chosen
as the Frobenius norm of the difference, then $d_\varphi$ is the spectral pseudo-distance. If
$\varphi(\bA)=[\bI+\varepsilon^2 \bD-\varepsilon \bA]^{-1}$ is the fast belief propagation matrix,
and $d$ is the root Euclidean distance (\ref{root}), then $d_\varphi$ is the DeltaCon$_0$
similarity.  Finally, if $\varphi$ computes the matrix of pairwise shortest distance between two
nodes, and $d$ is the $l_1$ norm, then $d_\varphi$ is the difference in path lengths.

In this paper, we propose to use the matrix function $\varphi$ that maps the adjacency matrix $\bA$
to the corresponding matrix, $\bR$, of pairwise effective resistances. We study various norms for the
distance $d$.  As we will see, the matrix function $\varphi$ is injective, and therefore $d_\varphi$
is a proper distance. As illustrated in several examples, the choice of $\varphi$ yields a distance
that adheres to the axioms and principles defined in section \ref{axioms_section}. Because the
effective resistance can be understood in terms of the commute time, our new distance shares some
similarity with the difference in path lengths \cite{chartrand1998}, albeit with a richer choice of
distances $d$. The effective resistance can also be expressed using the eigenvalues and
corresponding eigenvectors of the graph Laplacian, and thus this new distance can resolve changes in
the graphs occurring at multiple spectral scale in a manner similar to the spectral distance.
\section{The Resistance Perturbation Distance}
\label{rp_dist_def_sec}
For the sake of completeness, we review the concept of effective resistance.  Our discussion
focuses on those aspects that are relevant for the definition of the new distance. Excellent
references on the topic include, for instance, \cite{klein1993,doyle1984,ghosh2008,ellens2011}. The
reader familiar with these concepts can jump to section \ref{rp_dist_def_subsec}.
\subsection{The Effective Resistance}
\label{resistances_intro_section}
There are many different ways to present the concept of effective resistance. We use the electrical
analogy, which is very standard (e.g., \cite{doyle1984}). Given a graph $G=(V,E)$, we transform $G$
into a resistor network by replacing each edge $e$ by a resistor with conductance $w_e$ (i.e., with
resistance $1/w_e$).
\begin{definition}[Effective resistance \cite{klein1993}]
  The {\bf effective resistance} between two vertices $u$ and $v$ in $V$ is defined as the voltage
  applied between $u$ and $v$ that is required to maintain a unit current through the terminals
  formed by $u$ and $v$.
\end{definition}
A simple derivation (see e.g., \cite{bapat10}, chapter 9) yields the following expression of the
effective resistance,
\begin{equation}
  R_{ij} =  L^\dagger_{ii} + L^\dagger_{jj} - 2 L^\dagger_{ij},
  \label{effective_resistances_formula}
\end{equation}
or equivalently in matrix form
\begin{equation}
  \bR = \diag(\bL^\dagger)\bone^T + \bone \diag(\bL^\dagger)^T - 2 \bL^\dagger,
  \label{effective_resistance_matrix_eqn}
\end{equation}
where $\diag(\bL^\dagger)$ is the column vector formed by the diagonal entries of $\bL^\dagger$,
\begin{equation}
  \diag(\bL^\dagger) = \begin{bmatrix}
    L^\dagger_{11} \\ 
    \vdots \\
    L^\dagger_{nn} 
  \end{bmatrix}
\end{equation}
In this paper, we will often compute the Kirchhoff index to quantify the robustness of a network
(e.g., \cite{wang2014}).
\begin{definition}[Kirchhoff Index \cite{ellens2011}]
  The total resistance, or Kirchhoff index,  $\ki(G)$ of a graph $G$ is defined as the sum of the
  effective resistances between all pairs of vertices in a graph,
  \begin{equation}
    \ki(G) = \sum_{i,j\in V} R_{ij}.
    \label{kirchhoff_index_eqn}
  \end{equation}
\end{definition}
The relevance of the effective resistance in graph theory stems from the fact that it provides a
distance on a graph \cite{klein1993} that quantifies the connectivity between any two vertices, not
simply the length of the shortest path. In problems related to diffusion on a graph, or propagation
of infections or gossips \cite{draief10,keeling05,pastor15}, the redundancy of paths affects the
dynamics of the corresponding processes. Formally, the effective resistance provides the correct
notion of distance for a random walk on a graph, also known as the {\bf commute time}.
\begin{definition}[Commute Time \cite{chandra1996}]
  Consider a random walk $\{X_t\}_{t=1}^\infty$ on the set of vertices $V$, with the probability
  transition matrix $P_{ij}=P[X_{t+1}=j|X_t=i]=A_{ij}/D_{ii}$, then the {\em commute time} between
  vertices $i$ and $j$, $\kappa_{ij}$, is defined as the expected time for the random walk to travel
  from $i$ to $j$, and back to $i$,
  \begin{equation}
    \kappa_{ij}= \E[T_{ij}]+\E[T_{ji}],
    \label{commute_time_eqn}
  \end{equation}
  where $\E[T_{ij}]$ is the expected number of steps needed for the random walk, initialized
  at $i$, to reach $j$, 
  \begin{equation}
    \E[T_{ij}]=\E[\argmin_{t \geq 1} \{ X_t = j | X_0 = i \}].
  \end{equation}
  \label{def_commute}
\end{definition}
Chandra {\em et al.} \cite{chandra1996} showed that the commute time and the effective resistance are
equivalent up to a rescaling by the volume of the graph, $m = |E|$,
\begin{equation}
  \kappa_{ij} = 2 m R_{ij}, \quad \forall i,j \in V.
  \label{kappa_eq_R}
\end{equation}
\subsection{The Resistance Perturbation Distance}
\label{rp_dist_def_subsec}
We are now in a position to introduce the {\bf resistance perturbation distance} between two graphs
with known node correspondence.  This distance, which is a particular instance of the general
construction proposed in Definition~\ref{pseudo_dist_def}, obeys all the axioms and principles laid
out in section \ref{axioms_section}. In addition, we propose fast algorithms to compute the
distance.
\begin{definition}[Resistance Perturbation Distance]
  Let $G^{(1)} = (V,E^{(1)},w^{(1)})$ and $G^{(2)} = (V,E^{(2)},w^{(2)})$ be two connected,
  weighted, undirected graphs on the same vertex set, with respective effective resistance
  matrices, $\bR^{(1)}$ and $\bR^{(2)}$, respectively.  The {\em RP-p distance}, $\drpp$, between
  $G^{(1)}$ and $G^{(2)}$ is defined as the element-wise p-norm of the difference between their
  effective resistance matrices.  For $1 \leq p < \infty$,
  \begin{equation}
    \drpp(G^{(1)},G^{(2)}) = \left\lVert \bR^{(1)} - \bR^{(2)} \right\rVert_{p} =  \left[
      \sum_{i,j \in V} \left\lvert R^{(1)}_{ij} - R^{(2)}_{ij} \right\rvert^p  \right]^{1/p}\mspace{-24mu}, 
    \label{rp_dist_def_eqn}
  \end{equation}
  and for $p=\infty$,
  \begin{equation}
    d_{rp(\infty)}\left(G^{(1)},G^{(2)}\right) = \left\lVert \bR^{(1)} - \bR^{(2)} \right\rVert_{\infty} =
    \max_{i,j \in V} \left \lvert R^{(1)}_{ij} - R^{(2)}_{ij} \right \rvert. 
  \end{equation}
\end{definition}
\begin{theorem}[Resistance perturbation distance]
  \label{distance_thm}
  For $1 \leq p \leq \infty$, the RP-p distance defines a distance on the space of connected,
  weighted, undirected graphs with the same vertex set. 
\end{theorem}
\begin{proof}
  According to (\ref{effective_resistance_matrix_eqn}), the Laplacian $\bL$ uniquely
  identifies its effective resistance matrix $\bR$.  Additionally, for $1 \leq p \leq \infty$, the
  element-wise p-norm $\| \cdot \|_p$ is a norm on $\bM_n$.  As a result, non-negativity, symmetry,
  and the triangle inequality are satisfied.  Additionally, we observe that if $G^{(1)}=G^{(2)}$,
  then $\drpp(G^{(1)},G^{(2)})=0$, since $\bR^{(1)}=\bR^{(2)}$.  It remains to show that if
  $\drpp(G^{(1)},G^{(2)})=0$, or equivalently $\bR^{(1)}=\bR^{(2)}$, then $G^{(1)}=G^{(2)}$.  The
  following lemma completes the proof of the theorem, by showing that a resistance matrix uniquely
  identifies a weighted graph. \hfill$\square$
\end{proof}
\begin{lemma}[Injective property]
  If $G^{(1)}$ and $G^{(2)}$ are two graphs with the same effective resistance
  matrix, $\bR^{(1)}=\bR^{(2)}$, then $G^{(1)}=G^{(2)}$. 
  \label{injective_lemma}
\end{lemma}
\begin{lproof}  
  We proceed as follows: since $G^{(1)}$ and $G^{(2)}$ do not contain self-loops, the equality of
  their respective Laplacian matrices implies the equality of their adjacency matrices. We will
  therefore prove that if $\bR^{(1)}=\bR^{(2)}$ then $\bL^{(1)}=\bL^{(2)}$. In fact, we show that in
  general $\bL$ is uniquely determined from $\bR$. The first observation is that since
  $\bL^\dagger \bone = \bzero$ we have
  \begin{equation}
    \sum_{j=1}^nL^\dagger_{ij} = 0.
  \end{equation}
  We also have $\bone^T\bL^\dagger = \bzero^T$, since $\bL^\dagger$ is symmetric. Thus
  \begin{equation}
    \sum_{i=1}^nL^\dagger_{ij} = 0.
  \end{equation}
  Starting from the expression of $R_{ij}$ given by (\ref{effective_resistances_formula}), one 
  should be able to express $L^\dagger_{ij}$ in terms of $R_{ij}$ by using the cancellations above. In
  fact, a simple calculation shows that
  \begin{equation}
    L^\dagger_{ij} = -\frac{1}{2} \left[ R_{ij} - \frac{1}{n} ([R J]_{ij} + [J
      R]_{ij}) + \frac{1}{n^2} [J R J]_{ij} \right], 
    \label{LtoR}
  \end{equation}
  where $\bJ = \bone\bone^T$. We conclude the proof by injecting in (\ref{LtoR}) the expression of $\bL^\dagger$ given
  by (\ref{Lpseudo}) to recover $\bL$ as a function of $\bR$,
  \begin{equation}
    \bL= \left( -\frac{1}{2} \left[ \bR - \frac{1}{n} (\bR \bJ + \bJ \bR) + \frac{1}{n^2} \bJ \bR \bJ \right] +
      \frac{1}{n} \bJ \right)^{-1} - \frac{1}{n}\bJ.
  \end{equation}
\hfill $\square$
\end{lproof}
We note that the resistance perturbation distance is related to changes in the Kirchhoff index,
as described in the following result.
\begin{corollary}[Monotonicity]
  If $G^{(2)}$ is obtained from $G^{(1)}$ by monotone changes in edge weights,
  $w^{(2)}_{ij} \geq (\leq) w^{(1)}_{ij}$ for all $i,j$, then
  \begin{equation}
    \drpo(G^{(1)},G^{(2)}) = \left\lvert \ki(G^{(1)})-\ki(G^{(2)}) \right\rvert.
  \end{equation}
  \label{monotonicity_corollary}
\end{corollary}
\begin{cproof} 
  If $G^{(2)}$ is obtained from $G^{(1)}$ by monotone changes in edge weights,
  $w^{(2)}_{ij} \geq (\leq) w^{(1)}_{ij}$ for all $i,j$, then $R^{(1)}_{ij} \leq (\geq)
  R^{(2)}_{ij}$ for all $i,j \in V$, due to Rayleigh's Monotonicity Principle.  Thus,
  \begin{equation*}
    \begin{split}
      \drpo(G^{(1)},G^{(2)}) & = \sum_{i,j \in V} \left\lvert R^{(1)}_{ij} - R^{(2)}_{ij} \right\rvert 
      = \left\lvert \sum_{i,j \in V}  \left( R^{(1)}_{ij} - R^{(2)}_{ij}\right) \right\rvert \\
      &= \left\lvert \sum_{i,j \in V}   R^{(1)}_{ij}  - \sum_{i,j \in V} R^{(2)}_{ij} \right\rvert 
      = \left\lvert \ki(G^{(1)})-\ki (G^{(2)})\right\rvert. 
    \end{split}
  \end{equation*}
\hfill  $\square$
\end{cproof}
In the remainder of the paper we will restrict our attention to the RP-1 and RP-2 distances.  We
dedicate our attention to these two instances of the RP-p distance for the following reasons: in
some contexts, the RP-1 distance is directly analogous to the Kirchhoff index, and the RP-2 distance
can be computed with a fast randomized algorithm.
\begin{remark}
  The resistance metric is not properly defined when the vertices are not within the same connected
  component. To remedy this, we use a standard approach, and use the conductance instead of the
  resistance. Let $u$ and $v$ be two vertices. If $u$ and $v$ are connected, with effective
  resistance $R_{uv}$, then $C_{u\, v} = R_{uv}^{-1}$ is the connectivity between these vertices. If
  $u$ and $v$ belong to different connected components, then we set $C_{u\, v} = 0$.

  We proceed to define the following similarity measure
\begin{equation}
\er{uv} = \frac{1}{1+C_{u\, v}} = \frac{R_{u\, v}}{1 + R_{u\, v}},
\end{equation}
which we refer to as the \emph{renormalized effective resistance}. The renormalized resistance
perturbation distance is defined as follows.
\begin{definition}
  Let $G^{(1)}=(V^{(1)},E^{(1)})$ and $G^{(2)}=(V^{(2)},E^{(2)})$ be two graphs (with possibly
  different vertex sets).  We consider $V = V^{(1)}\cup V^{(2)}$, and relabel the union of vertices
  using $[n]$, where $n=|V|$. Let $\hR^{(1)}$ and $\hR^{(2)}$ denote the renormalized effective
  resistances in $\widetilde{G^{(1)}} = (V,E^{(1)})$ and $\widetilde{G^{(2)}} = (V,E^{(2)})$
  respectively.

  We define the renormalized resistance distance to be 
  \begin{equation}
    \drh(G^{(1)},G^{(2)}) = \left[\sum_{u,v=1,\ldots,n} \left\lvert \er{uv}^{(1)} - \er{uv}^{(2)}\right\rvert^p\right]^{1/p}.
    \label{eq:metric-defn}  
  \end{equation}
  \label{defn:metric}
\end{definition}
\noindent The following lemma confirms that the distance defined by \eqref{eq:metric-defn} remains a
metric when we compare graphs with the same vertex set.
\begin{lemma}[\cite{wills17a}]
  Let $V$ be a vertex set. The distance  $\drh$ defined by \eqref{eq:metric-defn} is a metric on the space of
  unweighted undirected graphs defined on the same vertex set $V$.
\end{lemma}
  The metric given in Definition \ref{defn:metric} can be used to compare two graphs of different
  sizes, by adding isolated vertices to both graphs until they have the same vertex set (this is why
  we must form the union $V = V^{(1)} \cup V^{(2)}$ and compare the graphs over this vertex set). This method
  will give reasonable results when the overlap between $V^{(1)}$ and $V^{(2)}$ is large.
  
When the graphs $G^{(1)}$ and $G^{(2)}$ have different sizes, the distance $\drh$ still satisfies the
triangle inequality, and is symmetric. However, $\drh$ is no longer injective: it is a
pseudo-metric. Indeed, as explained in the following lemmas, if $\drh(G^{(1)},G^{(2)}) = 0$, then the
connected components of $G^{(1)}$ and $G^{(2)}$ are the same, but the respective vertex sets may
differ by an arbitrary number of isolated vertices.
\begin{lemma}[\cite{wills17a}]
  Let $G=(E,V)$ be an unweighted undirected graph, and let $V^{(i)}$ be a set of isolated vertices, to wit
  $V^{(i)} \cap V = \emptyset$ and $\forall e \in E, \text{endpoints} \,(e) \notin V^{(i)}$. Define $G^\prime
  =(V \cup V^{(i)}, E)$, then we have $\drh(G,G^\prime) = 0$.
\end{lemma}
\noindent The following lemma shows that the converse is also true.
\begin{lemma}[\cite{wills17a}]
  Let $G^{(1)} = (V,E^{(1)})$ and $G^{(2)} = (V,E^{(2)})$ be two unweighted, undirected
  graphs, where $|V^{(1)}| > |V^{(2)}$. 

  \noindent If $\drh(G^{(1)}, G^{(2)}) = 0$, then $E^{(1)} = E^{(2)}$. Furthermore, there exists a set
  $V^{(i)}$ of isolated vertices, such that $V^{(1)} = V^{(2)} \cup V^{(i)}$.
\end{lemma}
\noindent In summary, one can easily extend the $\drpp$ distance to unconnected graphs using the 
$\drh$ distance. To simplify the exposition,  we focus on the distance $\drpp$ in the remainder of the paper,
and we only consider graphs that are connected with high probability. 
\end{remark}

\subsection{RP-1 Distance  After a Single Edge Perturbation
  \label{edge_mod_sec}}
We consider the case where a single edge is modified. This case is useful because it provides a
baseline scenario to compare various graph perturbations in the context of dynamic graphs. Our
analysis is based on the following two ideas. First, one can compute analytically changes in the
effective resistance that result from the modification of a single edge. Indeed, we can apply the
Sherman--Morrison--Woodburry theorem \cite{golan12} to compute the low-rank perturbation of the
pseudo-inverse $\bL^\dagger$. The second idea is to express $\bL^\dagger$ in terms of its spectral
decomposition (\ref{Ldagger_eqn}).  We use this result to derive a closed-form expression of the
RP-1 distance between a graph and a rank-one perturbation of that graph.
\begin{theorem}[RP-1 edge modification]
  If $G+\dw{i_0j_0}$ is the graph obtained from $G$ by a perturbation $\dw{i_0j_0}$ to
  the edge $[i_0,j_0]$, then 
  \begin{equation}
    \begin{split}
      \drpo(G,G+\dw{i_0j_0}) & = 
      \frac{2 n  \left\lvert \dw{i_0j_0} \right\rvert}    {1 + \dw{i_0j_0} R_{i_0j_0}}\;
      \sum_{k=2}^n \frac{1}{\lambda_k^2} \left[\bfi_k(i_0) - \bfi_k(j_0) \right]^2\\
      & =  2 n  \left\lvert \dw{i_0j_0} \right\rvert
      \frac{
        \displaystyle  \sum_{k=2}^n \frac{1}{\lambda_k^2} \left[\bfi_k(i_0) - \bfi_k(j_0)
        \right]^2
      }
      {
        1 + \dw{i_0j_0} \displaystyle  \sum_{k=2}^n   \frac{1}{\lambda_k}
        \left[\bfi_k(i_0)-\bfi_k(j_0)\right]^2
      }  
    \end{split}
    \label{edge_mod_eq}
  \end{equation}
  \label{edge_mod_thm}
\end{theorem}
\begin{proof}  
  The proof is given  in \ref{edge_mod_thm_proof}.
\end{proof}
\begin{remark}
  \label{edge_mod_scaling}
  It is important to understand the behavior of the term
  \begin{equation}
    \frac{\dw{i_0j_0} }{1 + \dw{i_0j_0} R_{i_0j_0}},
    \label{ratio_drpo}
  \end{equation}
  that controls the size of $\drpo(G,G+\dw{i_0j_0})$. A quick computation shows that the derivative
  of the ratio (\ref{ratio_drpo}) with respect to $\dw{i_0j_0}$ is equal to
  $1/(1 + \dw{i_0j_0} R_{i_0j_0})^2$, and thus (\ref{ratio_drpo}) is an increasing function of
  $\dw{i_0j_0}$. We also note that the smallest value that $\dw{i_0j_0}$ can take without
  disconnecting the edge $[i_0,j_0]$ is $-w_{i_0j_0}$. Because we always have
  $R_{i_0j_0} \leq 1/w_{i_0j_0}$, we confirm that the denominator of
  (\ref{ratio_drpo}) is always non negative, $1 + \dw{i_0j_0} R_{i_0j_0} \ge 0$. \\

  \noindent In general, $R_{i_0j_0} < 1/w_{i_0j_0}$, to wit $i_0$ and $j_0$ are connected by at
  least another path other than the direct edge $[i_0,j_0]$. In this case, we can disconnect the
  edge $[i_0,j_0]$ with the perturbation $\dw{i_0j_0} = -w_{i_0j_0}$, and the ratio
  (\ref{ratio_drpo}) becomes
  \begin{equation}
    - \frac{w_{i_0j_0} }{1 - w_{i_0j_0} R_{i_0j_0}}.
  \end{equation}
  This is the smallest value of (\ref{ratio_drpo}), which really corresponds to an increase in the
  effective resistance of $G+\dw{i_0j_0}$ (because of the absolute value around
  $\dw{i_0j_0}$ in (\ref{edge_mod_eq})). \\

  \noindent We conclude that $\drpo(G,G+\dw{i_0j_0})$ in (\ref{edge_mod_eq}) decreases for
  increasing $\dw{i_0j_0}$ in the interval $[ -w_{i_0j_0},0]$, reaches a minimum at
  $\dw{i_0j_0}=0$, and increases for $\dw{i_0j_0}$ in the interval $[0,\infty)$. As
  $\dw{i_0j_0}\rightarrow \infty$, the resistance perturbation distance no longer depends on
  $\dw{i_0j_0}$.
\end{remark}
\begin{remark}
  \label{target_drpo}
  \noindent We further note that the case $1 + \dw{i_0j_0} R_{i_0j_0} = 0$ corresponds to a
  targeted change $\dw{i_0j_0} = -w_{i_0j_0}$ along an edge $[i_0,j_0]$ where
  $1/R_{i_0j_0} = w_{i_0j_0}$. Such a change will disconnect the graph, since the condition
  $R_{i_0j_0} = 1/w_{i_0j_0}$ indicates that the edge $[i_0,j_0]$ is the only path between $i_0$ and
  $j_0$, and setting its weight to zero cuts the
  graphs into two parts. In this case, $\drpo(G,G+\dw{i_0j_0}) = \infty$.
\end{remark}
\begin{remark}
  \label{block_drpo}
  \noindent The size of the sum $\sum_{k=2}^n \left[\bfi_k(i_0) - \bfi_k(j_0) \right]^2 /\lambda_k^2$
  in (\ref{edge_mod_eq}) can be analyzed as follows. For large $k$, eigenvectors $\bfi_k$
  ``oscillate'' very quickly on the graph, making it difficult to estimate the contribution of
  $\left[\bfi_k(i_0) - \bfi_k(j_0) \right]^2$. This issue is mitigated by the fact that the weights
  $1/\lambda_k^2$ are relatively small, since the eigenvalues $\lambda_k$ are large.

  For small $k$, the eigenvalues $\lambda_k$ are small, and the corresponding eigenvectors $\bfi_k$
  ``oscillate'' very slowly on the graph, i.e. $\bfi_k(i_0) - \bfi_k(j_0) \approx 0$ unless $i_0$
  and $j_0$ belong to different nodal regions. In this latter case, the effect of the edge perturbation
  $\dw{i_0j_0}$ will be maximal. An example of this phenomenon corresponds to a network
  formed by densely connected communities, which are weakly connected to one another. For the same
  $\dw{i_0j_0}$, $\drpo(G,G+\dw{i_0j_0})$  will be maximal if $i_0$ and
  $j_0$ are in different communities.
\end{remark}
\section{The RP-1 Metric Created by Small Perturbations of Simple Graphs}
\label{analytic_results_section}
To understand the manner in which the RP distance quantifies changes in graph connectivity, we study
this distance on several graphs that epitomize limiting cases of general graph
topology. Specifically, we compute analytically (either by spectral decomposition of the graph
Laplacian, or by simplification of the corresponding resistor networks) the distance between a graph
and a slightly perturbed version of it.

Our goal is to demonstrate that the RP distance can detect edge perturbations that have a profound
effect on the functionality of the network, while remaining unaffected by edge changes that have
harmless consequences for the graph.

To simplify the analysis we perturb a single edge, and we denote by $G + \dw{i_0j_0}$ the graph 
formed by altering the edge weight between vertices $i_0$ and $j_0$ according to
$w_{i_0j_0} \rightarrow w_{i_0j_0} +\dw{i_0j_0}$.  In this section we will not discuss the
edit distance, but simply note that the edit distance is trivially constant for all the following
examples: $d_1(G,G+\dw{i_0j_0})=\left\lvert\dw{i_0j_0}\right\rvert$.

Because the RP-1 distance $\drpo (G,G+\dw{i_0j_0})$ can either decrease or increase with $n$,
as $n$ goes to infinity, we also compute a normalized RP-1 distance by dividing by the $l_1$ norm of
the matrix $\bR$ (Kirchhoff index). As is shown in this section, this normalized distance is able to
quantify the importance of the perturbation on the geometry of the graph.
\subsection{Complete graph}
\noindent We consider a complete graph, $K_n$, with $n$ vertices.
\begin{theorem}
  \label{complete_graph_thm}
  If we perturb the weight of the edge $[i_0,j_0]$ by $\dw{i_0j_0}$, then the RP-1 distance
  between the original and the perturbed graph is 
  \begin{equation}
    \drpo(K_n,K_n+\dw{i_0j_0}) = \frac{ 4 | \dw{i_0j_0} |}{n + 2 \dw{i_0j_0}}.
    \label{complete_graph_eqn}
  \end{equation}
\end{theorem}
\begin{proof}
  See  \ref{complete_graph_proof}.
\end{proof}
The Kirchhoff index for the complete graph is
\begin{equation}
  \ki(K_n) = 2(n-1),
\end{equation}
and therefore the normalized $\drpo$ distance created by modifying the edge
$w_{i_0j_0}$ is given by
\begin{equation}
  \frac{\drpo(K_n,K_n+\dw{i_0j_0})}{\ki(K_n)} = {\cal O} \left( \frac{1}{n^2} \right).
\end{equation}
The scaling of $\drpo(K_n,K_n+\dw{i_0j_0})/\ki(K_n)$ suggests that individual edges in the
complete graph rapidly lose significance with increasing $n$.  This matches our intuition about the
complete graph, which is the most robust to the removal of edges, due to the maximal redundancy in
paths between all pairs of vertices.
\begin{remark}
  It is interesting to compare the complete graph to a dense Erd\H{o}s-R\'enyi
  graph, $G(n,p)$, when $p > \log (n)/n$. As shown in \cite{sood04,lowe14},
  \begin{equation}
    n (1 + o(1)) \leq \E[\kappa_{i,j}] \leq n (2 + o(1)).
  \end{equation}
  Since the expected number of edges, $\E[m] \sim pn^2/2$, we obtain the following estimate of
  the effective resistance,
  \begin{equation}
    \E[R_{ij}] \sim \frac{2}{n p}.
  \end{equation}
  We can compute the RP-1 distance between one random graph $G$ in $G(n,p)$,
  and a perturbed version of $G$, obtained by randomly adding or removing one edge,
  \begin{equation}
    \E\left[\drpo(G,G+\Delta 1_{i_0j_0})\right] \sim \frac{ 2 }{n p}.
    \label{gnp_eqn}
  \end{equation}
  We conclude that this RP-1 distance has the same behavior as that of the complete graph, given by
  (\ref{complete_graph_eqn}). 
\end{remark}
\subsection{Star graph}
\noindent We consider the star graph $S_n$, which is a tree where every leaf
node $2,\ldots, n$ is connected to the root node (hub) $1$.
\begin{theorem}
  \label{star_graph_thm}
  If we perturb the edge $[1,i_0]$, which connects the hub $1$ to the leaf $i_0\neq 1$, by
  $\dw{i_0j_0}$, then the RP-1 distance between the original and the perturbed graph is
  \begin{equation}
    \drpo(S_n,S_n+\dw{1i_0}) = \frac{ 2 (n-1) | \dw{1i_0} |}{1 +
      \dw{1i_0}}.
    \label{star_graph_eqn}
  \end{equation}
  If we add an edge with weight $\dw{i_0j_0} \ge 0$ between two leaves $i_0$ and $j_0$,
  $i_0,j_0\neq 1$, then the RP-1 distance between the original and the perturbed graph is
  \begin{equation}
    \drpo(S_n,S_n+\dw{i_0j_0}) = \frac{ 4 n  \dw{i_0j_0} }{1 + 2
      \dw{i_0j_0}}. 
  \end{equation}
\end{theorem}
\begin{proof}
  See  \ref{star_graph_proof}.
\end{proof}
\noindent The Kirchhoff index for the star graph is
\begin{equation}
  \ki(S_n) = 2(n-1)^2,
\end{equation}
and therefore the normalized $\drpo$ distance created by modifying the edge
$w_{i_0j_0}$ is given by
\begin{equation}
  \frac{\drpo(S_n,S_n+\dw{i_0j_0})}{\ki(S_n)} = {\cal O} \left( \frac{1}{n} \right).
\end{equation}
For the star graph, $\drpo(S_n,S_n+\dw{i_0j_0})/\ki(S_n)$ decays more slowly with $n$ than 
with the complete graph.  This matches our intuition, since the star graph is a tree (i.e. it has no
redundant paths).
\subsection{Path graph}
\noindent We consider the path graph, $P_n$, on $n$ vertices. 
\begin{theorem}
  \label{path_graph_thm}
  If we add an edge with weight $\dw{i_0j_0} \ge 0$ between the vertices $i_0$ and
  $j_0 > i_0$, then the RP-1 distance between the original and the perturbed graph is
  \begin{equation}
    \begin{split}
      \drpo(P_n,P_n+\dw{i_0j_0}) &= \\
      |\dw{i_0j_0}|  (j_0 - i_0) 
      & \frac{ 
        2n \left[1 + (j_0-i_0)(2j_0 + 4i_0 -3)\right] -3 (j_0 - i_0)(i_0 + j_0 -1)^2 
      }
      {6\left(\dw{i_0j_0}(j_0 - i_0) + 1\right)}.
    \end{split}
    \label{rp1path}
  \end{equation}
\end{theorem}
\begin{proof}
  See \ref{path_graph_proof}.
\end{proof}
\noindent The Kirchhoff index for the path graph is
\begin{equation}
  \ki(P_n) = \frac{1}{3}(n-1)n(n+1).
\end{equation}
If we assume that $i_0={\cal O}(1)$ and ${\cal O}(1) \leq j_0 \leq {\cal O}(n)$, then the normalized
$\drpo$ distance created by modifying the edge weight $w_{i_0j_0}$ is given by
\begin{equation}
  \frac{\drpo(P_n,P_n+\dw{i_0j_0})}{\ki(P_n)} 
  = {\cal O} \left( \left[\frac{j_0}{n}\right]^2 \right).
\end{equation}
If $j_0={\cal O}(1)$, then $i_0$ and $j_0$ remain close, and the new edge has little impact on the
graph.  However, if $j_0={\cal O}(n)$, then the new edge acts as a short circuit that joins the
beginning and the end of the path. In this case, $\drpo(P_n,P_n+\dw{i_0j_0})$ grows at the
same rate as $\ki(P_n)$.  In other words, the addition of the edge has a profound effect that
remains constant, as the graph grows.

We note that this behavior is very different from that of the star graph, even though both graphs
are trees. Indeed, in the star graph, all the nodes are well connected: a distance of 1 between a
leaf and the hub, and a distance of 2 between two leaves. On the contrary, in the path graph the
head and the tail of the graph are at a distance $n$, and the addition of a short circuit has a
significant effect. Clearly, the RP-1 distance provides a very useful tool for the analysis of
perturbations of both graph models.

It is interesting to note, that although the distance in (\ref{rp1path}) is correlated with
$|i_0-j_0|$, the values of $i_0$ and $j_0$ also play a role.  In particular, the maximum of $\drpo$
does not occur when we add an edge between the endpoints of the path.  If the shortcut were at the
extreme, it would create a cycle of perimeter $n$.  However, if the shortcut connects nodes $n/8$
and $7n/8$, then the path becomes a cycle of perimeter $3n/4$, with two small tails of length $n/8$.
On average, the diffusion will move faster across this geometry than around the larger cycle.
\subsection{Cycle graph}
\noindent Finally, we consider the cycle on $n$ vertices, $C_n$.
\begin{theorem}
  \label{ring_graph_thm}
  If we add an edge with weight $\dw{i_0j_0} \ge 0$ between the vertices $i_0$ and
  $j_0 < i_0$, then the RP-1 distance between the original and the perturbed graph is
  \begin{equation}
    \begin{split}
      \drpo(C_n,C_n+\dw{i_0j_0})  & = \\
      \frac{1}{6}
      \dw{i_0j_0}
      \left[i_0 \ominus j_0\right] n  & \;
      \frac{
        \left[i_0 \ominus j_0 \right]^3 
        -2n \left[(i_0 \ominus  j_0 )^2 -1\right] 
        + \left[i_0 \ominus j_0 \right](n^2 -2)
      }
      {n^2 + n \dw{i_0j_0} \left[i_0 \ominus j_0 \right]
        \left[n - (i_0 \ominus j_0)\right]
      },
    \end{split}
    \label{ring_graph_eqn}
  \end{equation}
  with $i_0 \ominus j_0 = i_0 -j_0 \pmod{n}$.
\end{theorem}
\begin{proof}
  See  \ref{ring_graph_proof}. 
\end{proof}
\noindent The Kirchhoff index for the cycle graph is
\begin{equation}
  \ki(C_n) = \frac{1}{6}(n-1)n(n+1).
\end{equation}
If we assume that $O(1) \leq i_0 \ominus j_0 \leq O(n)$, we observe the
following scaling, 
\begin{equation}
  \frac{
    \drpo(C_n,C_n+\dw{i_0j_0})
  }{
    \ki(C_n)
  } = {\cal O} \left( \frac{i_0\ominus j_0}{n} \right).
\end{equation}
The interpretation of the scaling for the cycle graph is very similar to that of the path graph.
One can show that the largest change in the RP-1 distance in (\ref{ring_graph_eqn}) is achieved with
$i_0-j_0 = n/2$. This edge creates a short circuit in the middle of the cycle, and leads to a ``small
world'' model.
\section{Fast Computation of the RP-2 Distance}
\label{rp2_computation_section}
Our discussion so far has focused on the relevance of the RP-p distance to detect structural changes
between graphs. We now consider the second fundamental question: can this new distance be computed
efficiently? \\

\noindent A naive evaluation of $\drpp(G^{(1)},G^{(2)})$ suggests that one first needs to compute
the pseudo-inverse of $\bL$, in order evaluate the distance as follows
\begin{equation}
  \begin{split}
    \drpp(G^{(1)},G^{(2)}) &=\\
    \left\lVert \diag(\bL^{(1)\dagger}-\bL^{(2)\dagger}) \right. & \left. \bone^T + \bone
      \diag(\bL^{(1)\dagger}-\bL^{(2)\dagger})^T - 2 (\bL^{(1)\dagger}-\bL^{(2)\dagger}) \right\rVert_p. 
  \end{split}
\end{equation}
Equivalently, one could compute the eigenvectors and eigenvalues of $\bL^{(1)}$ and $\bL^{(2)}$, and
estimate 
\begin{equation}
  \begin{split}
    \drpp(G,\widetilde{G})  &=\\
    \left\{ \sum_{i=1}^n \sum_{j=1}^n \left\lvert \sum_{k=2}^n \frac{1}{\lambda^{(1)}_k}  \right.\right.&
    \left.\left. \mspace{-12mu}[\phi^{(1)}_k(i) -\phi^{(1)}_k(j)]^2 - \sum_{k=2}^n
        \frac{1}{\lambda^{(2)}_k}[\phi^{(2)}_k(i) -\phi^{(2)}_k(j)]^2 \right\lvert^p
    \right\}^{1/p} \mspace{-24mu}. 
  \end{split}
\end{equation}
This direct computation involves a full spectral decomposition of two Laplacian matrices of
potentially very large size, followed by the computation of the element-wise $p$-norm of the
difference of two (dense) resistance matrices, at a total cost of at least ${\cal O}(n^2)$.
Clearly, a direct computation is prohibitively expensive for large networks, which motivates the
development of a scalable randomized approximation algorithm.

We consider two general scenarios.  The first one is the general problem of computing the
resistance perturbation distance between two graphs, which we address in this section.  In section
\ref{optimization}, we explore the restricted problem of computing the resistance perturbation
distance between a graph and a slightly perturbed version of that graph (for example, a second graph
obtained by adding one or several edges, or perturbing the weight of an edge).  The second problem
has applications in a variety of settings including anomaly detection in streaming graphs, and edge
addition or protection for purposes of improving or maintaining network robustness.
\subsection{Fast Approximation of Pairwise Resistances}
\label{fast_resistances_section}
A key ingredient of our linear-time algorithm for approximation of the RP-2 distance is the
linear-time algorithm of Spielman and Srivastava \cite{spielman2008,srivastava10} for approximating
pairwise effective resistances.  The algorithm relies on a bi-Lipschitz embedding of the vertices
in $\R^{{\cal O}(\log n)}$ that preserves the pairwise effective resistances.  Specifically, given
$\varepsilon >0$, there exists an $\widetilde{{\cal O}}(m \log \overline{w} /\varepsilon^2)$ time
algorithm \cite{spielman2008}, where $\overline{w}=w_\text{max}/w_\text{min}$, that computes a
$(24 \log n /\varepsilon^2) \times n$ matrix $\tZ$ such that with probability at least $1-1/n$,
\begin{equation}
  (1-\varepsilon) R_{ij} \leq \left\lVert \tZ (\be_i - \be_j)
  \right\rVert_2^2 \leq (1 + \varepsilon) R_{ij}, \quad
  \forall i,j \in V,
  \label{SS08_embedding_eqn}
\end{equation}
where we recall that $\be_i$ is the $i^\text{th}$ vector of the canonical basis in $\R^n$;
$w_\text{min}$ and $w_\text{max}$ are the minimum and maximum edge weights, respectively.  The
algorithm \cite{spielman2008} combines some crucial ideas, which we recall succinctly in the
following. The reader can consult \cite{spielman2008,srivastava10} for further details about the
algorithm.

The first observation is that the vertices can be embedded in an $m$-dimensional space where the
pairwise squared Euclidean distance is equal to the effective resistance between the corresponding vertices in the graph,
\begin{equation}
  \begin{split}
    R_{ij} 
    &= (\be_i - \be_j)^T \bL^\dagger (\be_i - \be_j) 
    = (\be_i - \be_j)^T \bL^\dagger \bL \bL^\dagger (\be_i - \be_j) \\
    &= \left( (\be_i - \be_j)^T \bL^\dagger \bB^T \dA^{1/2} \right) 
    \left(\dA^{1/2} \bB \bL^\dagger (\be_i - \be_j) \right) \\
    &= \left\lVert \dA^{1/2} \bB \bL^\dagger (\be_i - \be_j) \right\rVert_2^2.
  \end{split}
\end{equation}
The second idea is to replace $\dA^{1/2} \bB $ with a randomized version
$\bY \bck = \bQ \dA^{1/2} \bB$ of size $s \times n$, where $s = 24 \log n/\varepsilon^2$. The matrix
$\bQ \in \R^{s \times m}$ is populated with random entries $\pm 1/\sqrt{s}$. The matrix $\tZ$ in
(\ref{SS08_embedding_eqn}) is then defined as $\tZ^T = \bL^\dagger \bY^T$.  Instead of computing
directly the pseudo-inverse $\bL^\dagger$, one approximates the $i^{th}$ column of $\tZ^T$ by
solving the linear system $\bL \tz_i = \by_i$, for $i=1,\ldots,s$, where $\by_i$ is the $i^{th}$
column of $\bY^T$. In summary, the matrix $\tZ$ in (\ref{SS08_embedding_eqn}) is constructed using
the algorithm \cite{spielman2008,srivastava10} described in Algorithm \ref{algo1}.  The algorithm
runs in expected time $\widetilde{{\cal O}}(m \log(1/\delta))$, where $m$ is the number of edges in
$G$.  The algorithm returns the matrix
$\tZ = \begin{bmatrix} \widetilde{z}_1 & \cdots & \widetilde{z}_s \end{bmatrix}^T \in \R^{s \times
  n}$, which meets the bi-Lipschitz condition of (\ref{SS08_embedding_eqn}).
%
\begin{algorithm}[H]
  \sffamily
  \caption{Compute the matrix $\tZ$ in  (\ref{SS08_embedding_eqn}) \cite{spielman2008,srivastava10}}
  \label{algo1}
  \begin{algorithmic}[1]
    \STATE Generate a realization $\bQ  \in  \bM_{s \times m}$, with random entries $\pm 1/\sqrt{s}$, and
    $s= 24 \log n/\varepsilon^2$. 
    \STATE  $\bY \leftarrow \bQ \dA^{1/2} \bB$ \\
    \COMMENT{$\delta$ controls the relative error, $\left\lVert x - \bL^\dagger y \right\rVert_L
      \leq \delta \left\lVert \bL^\dagger y \right\rVert_L$, where $\| y \|_L = \sqrt{y^T \bL y}$}
    \STATE  $\displaystyle \delta \leftarrow \frac{\varepsilon}{3} \sqrt{ \frac{2}{n^3}
      \left(\frac{1-\varepsilon}{1+\varepsilon }\right) \frac{w_\text{min}}{w_\text{max}} }$.\\
    \COMMENT{Use Laplacian solver {\tt STSolve} of Spielman and Teng \cite{spielman2006,spielman2004}} 
    \STATE Compute: $\tz_i \leftarrow \text{\sf STSolve}(\bL, \by_i, \delta),\quad  \forall i=1,\ldots, s.$\\
  \end{algorithmic}
\end{algorithm}
\subsection{Fast Computation of the $\drpt$ distance}
\noindent Based on $\tZ$, we can approximate the effective resistance matrix as
follows,
\begin{equation}
  \bR \approx \tR = \diag(\tZ^T \tZ) \bone^T +
  \bone \diag(\tZ^T \tZ)^T - 2 \tZ^T
  \tZ. 
  \label{R_approx}
\end{equation}
If we approximate the RP-2 distance using (\ref{SS08_embedding_eqn}), then we obtain the following
error bound.
\begin{theorem}
  If $\tZo$ and $\tZt$ are matrices satisfying (\ref{SS08_embedding_eqn}) for the graphs $G^{(1)}$
  and $G^{(2)}$ respectively, then we have the following inequalities
  \begin{equation}
    \begin{split}
      \left\lVert\bR^{(1)}-\bR^{(2)}\right\rVert_F 
      - \varepsilon \left\lVert\bR^{(1)}+\bR^{(2)}\right\rVert_F & \leq 
      \left\lVert\tR^{(1)}-\tR^{(2)}\right\rVert_F \\
      \leq \left\lVert\bR^{(1)}-\bR^{(2)}\right\rVert_F &+ \varepsilon
      \left\lVert\bR^{(1)}+\bR^{(2)}\right\rVert_F, 
    \end{split}
  \end{equation}
  where $\tR^{(1)}$ and $\tR^{(2)}$ are defined in (\ref{R_approx}).
  \label{frobenius_bounds_thm}
\end{theorem}
\begin{proof}
  Several applications of the triangle inequality prove the result; see \ref{frobenius_bounds_thm_proof}.
\end{proof}
\subsection{Fast Frobenius norm
  \label{fast_frobenius_section}}
Using the results of section \ref{fast_resistances_section} we can approximate
the RP-2 distance as follows, 
\begin{equation}
  \begin{split}
    \drpt(G^{(1)},G^{(2)})  \approx \left\lVert \tR^{(1)} - \tR^{(2)} \right\rVert_F 
    =  \left\lVert \diag\left(\tZoT \tZo-\tZtT \tZt\right) \bone^T \right.\\
    \left. + \bone \diag\left(\tZoT\tZo-\tZtT \tZt\right)^T 
      - 2 \left( \tZ^{(1)T} \tZo -\tZtT \tZt\right) \right\rVert_F.
  \end{split}
  \label{frobenius_approx}
\end{equation}
Direct computation of the Frobenius norm is quadratic in the number of vertices, $n$.  However, the
structure of our problem permits us to compute (\ref{frobenius_approx}) in near linear time in $n$.
\begin{theorem}[Fast Frobenius]
  $\left\lVert \tR^{(1)} - \tR^{(2)} \right\rVert_F$ can be computed in $\widetilde{{\cal O}}(n)={\cal O}(n
  \log^2 n)$ time. 
  \label{fast_frobenius_thm}
\end{theorem}
\begin{proof}
  Let 
  \begin{equation}
    \bd =\diag\left(\tZoT \bck\tZo - \tZtT \bck\tZt\right)  \in \R^n.
  \end{equation}
  Using the invariance of the trace under cyclic permutations, we show in
  \ref{fast_frobenius_thm_proof} that
  \begin{equation}
    \begin{split}
      \left\lVert \tR^{(1)} - \tR^{(2)} \right\rVert^2_F  = 
      2 & \left\lgroup
        \left[\bone^T \bck \bd \right]^2 
        + n\left\lVert \bd \right\rVert_2^2 
        + 4 \left(\bone^T \tZtT \tZt \bd  - \bone^T \tZoT \tZo \bd\right) 
      \right.\\ 
      + 2 & \left. \left(
          \left\lVert \tZo \tZoT \right\rVert^2_F 
          + \left\lVert \tZt \tZtT \right\rVert_F^2 
          - 2 \left\lVert \tZt \tZoT \right\rVert_F^2
          \right) \right\rgroup, 
    \end{split}
  \end{equation}
  which can be computed in $\widetilde{{\cal O}}(n)={\cal O}(n \log^2 n)$ time.
\hfill $\square$
\end{proof}
\subsection{A Nearly Linear-time Algorithm for the RP-2 Distance}
Combining the results of sections \ref{fast_resistances_section} and \ref{fast_frobenius_section},
we now build an algorithm to approximate the RP-2 distance between two graphs in nearly linear time.
In the following theorem, let $G^{(1)}=(V,E^{(1)},w^{(1)})$ and $G^{(2)}=(V,E^{(2)},w^{(2)})$ be two
graphs with the same vertex set, and let $m^{(1)}=| E^{(1)} |$, $m^{(2)}=| E^{(2)} |$,
$\overline{w}^{(1)}=w^{(1)}_{max}/w^{(1)}_{min}$, and $\overline{w}^{(2)}=w^{(2)}_{max}/w^{(2)}_{min}$.
Further, let $m \log \overline{w}=\text{max}(m^{(1)}\log \overline{w}^{(1)}$, $m^{(2)}\log \overline{w}^{(2)})$.
\begin{theorem}[Fast RP-2 algorithm]
  \label{full_algorithm_thm}
  There is an $\displaystyle \widetilde{{\cal O}}\left( n + \frac{m \log
      \overline{w}}{\varepsilon^2}  \right)$ algorithm that computes
  $\widetilde{d}_{rp2}(G^{(1)},G^{(2)})$, an approximation of the RP-2 distance, such that with
  probability at least $1-2/n$, 
  \begin{equation}
    \begin{split}
      \drpt(G^{(1)},G^{(2)}) - \varepsilon  \left\lVert\bR^{(1)}+\bR^{(2)}\right\rVert_F \leq
      & \widetilde{d}_{rp2}(G^{(1)},G^{(2)}) \\
      \leq & \drpt(G^{(1)},G^{(2)}) + \varepsilon \left\lVert\bR^{(1)}+\bR^{(2)}\right\rVert_F. 
    \end{split}
  \end{equation}
\end{theorem}
\begin{proof}
  Direct consequence of (\ref{SS08_embedding_eqn}) and theorems \ref{frobenius_bounds_thm}, and
  \ref{fast_frobenius_thm}. \hfill $\square$
\end{proof}
The algorithm for the fast computation of the $\drpt$ distance is described in Algorithm
\ref{algo2}.  A MATLAB implementation of Spielman and Srivastava's algorithm written by Richard
Garcia-Lebron was used \cite{fastResistances} to compute Algorithm \ref{algo1}.  The code utilizes
an implementation of the combinatorial multigrid solver \cite{koutis2011} written by Ioannis Koutis,
and Gary Miller.%
\begin{algorithm}[H]
  \sffamily
  \caption{Compute $\widetilde{d}_{rp2}(G^{(1)},G^{(2)})$}
  \label{algo2}
  \begin{algorithmic}[1]
    \STATE Input: $E^{(1)}$, $w^{(1)}$, $E^{(2)}$, $w^{(2)}$, tolerance $\varepsilon>0$.
    \STATE Compute $\tZo, \tZt \in \bM_{s \times n}$ using Algorithm~\ref{algo1}.  
    \STATE  $\bd \leftarrow \diag(\tZoT \tZo - \tZtT \tZt)$.
    \STATE \RaggedRight $\widetilde{d}_{rp2} \leftarrow \sqrt{2} 
    \left\lgroup \left[\bone^T \bck \bd \right]^2 + n\left\lVert \bd \right\rVert_2^2 
      + 4 \left(\bone^T \tZtT \tZt \bd  - \bone^T \tZoT \tZo \bd \right) \right.$
    \hspace*{4em}$\left. + 2 \left( \left\lVert \tZo \tZoT   \right\rVert^2_F 
        + \left\lVert \tZt \tZtT \right\rVert_F^2 
        - 2 \left\lVert \tZt \tZoT \right\rVert_F^2
      \right)
    \right\rgroup^{1/2}\mspace{-24mu}.$ 
    \RETURN $\widetilde{d}_{rp2}$.
  \end{algorithmic}
\end{algorithm}
\noindent The scalability of the algorithm was verified experimentally on a set of sparse random
graphs with $m={\cal O}(n)$ edges.  The graphs generated for this experiment were latent space
  random path graphs with a power law kernel edge probability; the probability of connecting nodes
  $i$ and $j$ is given by $P (i \sim j) =100/|i-j|$.
\begin{figure}[H]
  \begin{center}
    \includegraphics[width=0.87\textwidth]{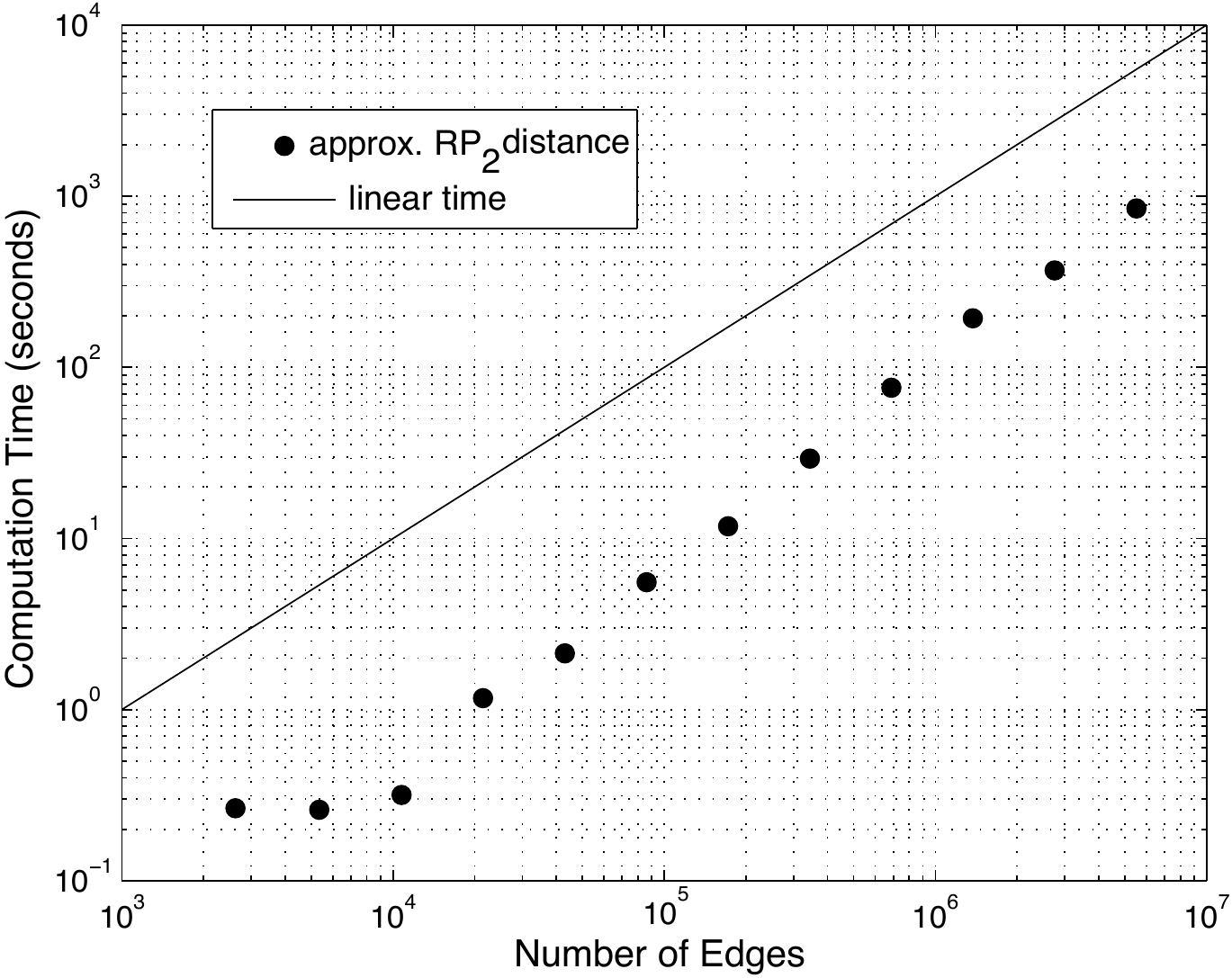}
  \end{center}
  \caption{Computation time for $\widetilde{d}_{rp2}$ as a function of the number of edges $m={\cal
      O}(n)$.} 
  \label{benchmark}
\end{figure}
This generates sparse random graphs with edge counts approximately proportional to the vertex
counts.

In Fig. \ref{benchmark} we see that the
computation time scales nearly linearly in the number of edges.
\section{Fast Optimal Design of Networks Using the RP-1 Distance
  \label{optimization}}
Improving network robustness via targeted edge addition is a problem with considerable applications.
The Kirchhoff index is often used as a measure of network robustness (see e.g., Wang et
al. \cite{wang2014} and references therein).  A lower Kirchhoff index is indicative of a more robust
network, since lower effective resistances between pairs of vertices is indicative of short and/or
redundant paths between vertices.  The greedy approach, which consists in connecting the pair of
vertices with the highest effective resistance, is known to be suboptimal (e.g., Ellens et
al. \cite{ellens2011}).  Wang {\em et al.} \cite{wang2014} demonstrate however, that choosing the maximum
effective resistance is often close to optimal, and can be accomplished in $O(n^3)$ time rather than
$O(n^5)$ as required for an exhaustive search.

We make the following significant contribution to this question: we propose an algorithm with
complexity ${\cal O}(n)$ to approximate the optimal edge addition.  Specifically, this novel
algorithm combines a low-rank approximation of the exact $\drpo$ distance given by theorem
\ref{partial_sum_thm} with a fast heuristic. We describe these two components in the next sections.
\subsection{Low-rank Approximation of the RP-1 Distance}
Theorem \ref{edge_mod_thm} provides an exact formula for computing the perturbation of the Kirchhoff
index due to changes (addition, or removal) in a single edge.  The optimal edge addition can thus be
computed with a complexity $O(n^3)$ time. Indeed, $O(n^3)$ operations are needed to compute the
spectral decomposition of $\bL$; another $O(n) \times O(n^2)$ operations are then required to
exhaustively compute the exact $\drpo(G,G+\dw{i_0j_0})$ distance (in $O(n)$ operations
using (\ref{edge_mod_eq})), for every pair of vertices $i_0$ and $j_0$.

The $O(n^3)$ complexity is a significant improvement over the $O(n^5)$ algorithm described in Wang
{\em et al.} \cite{wang2014}.  However, $O(n^3)$ is still prohibitively expensive for large networks,
which motivates us to consider a low-rank approximation strategy to reduce the cost of solving the
optimal edge modification problem.

Many graphs exhibit a concentration of the bulk of the eigenvalues of the graph Laplacian
\cite{chung06}. In this case, the bulk is well separated from the smallest eigenvalues, and because
it is well confined, it can be replaced by a single ``representative'' eigenvalue. This idea leads
to the following approximations, which prove to be very accurate in practice, for the summations in
(\ref{edge_mod_eq}).
\begin{theorem}[Low-rank approximation]
  The sums in the numerator and denominator in (\ref{edge_mod_eq}) can be approximated using the
  following lower and upper bounds,
  \begin{equation}
    \begin{split}
      \frac{2}{\lambda^2_n}+\sum_{k=2}^p
      \left\{\bck \frac{1}{\lambda^2_k}-\frac{1}{\lambda^2_n\bck}\right\}
      & \bck \left[\bfi_k(i)-\bfi_k(j)\right]^2   
      \leq \sum_{k=2}^n \frac{1}{\lambda^2_k}\bck  \left[\bfi_k(i)-\bfi_k(j)\right]^2\\
      & \leq  \frac{2}{\lambda^2_p}+\sum_{k=2}^p
      \bck \left\{\bck \frac{1}{\lambda^2_k}-\frac{1}{\lambda^2_p}\right\}
      \bck \left[\bfi_k(i)-\bfi_k(j)\right]^2\bck, 
    \end{split}
    \label{low_rank_sq_approx_bound_eqn}
  \end{equation}
  and
  \begin{equation}
    \begin{split}
      \frac{2}{\lambda_n}+\sum_{k=2}^p
      \left\{\bck \frac{1}{\lambda_k}-\frac{1}{\lambda_n}\bck \right\}
      & \bck \left[\bfi_k(i)-\bfi_k(j)\right]^2  
      \leq  \sum_{k=2}^n \frac{1}{\lambda_k} \bck  \left[\bfi_k(i)-\bfi_k(j)\right]^2 \\
      & \leq   \frac{2}{\lambda_p}+  \sum_{k=2}^p
      \bck \left\{\bck \frac{1}{\lambda_k}-\frac{1}{\lambda_p\bck}\right\}
      \bck  \left[\bfi_k(i)-\bfi_k(j)\right]^2\bck. 
    \end{split}
    \label{low_rank_approx_bound_eqn}
  \end{equation}
  \label{partial_sum_thm}
\end{theorem}
\begin{proof}
  The proof relies on the orthonormality of the eigenvectors to bound the
  contribution of the bulk of the spectrum ($\lambda_p,\ldots,\lambda_n$) from
  above and below. See details in \ref{partial_sum_thm_proof}.
\end{proof}
Using the above result, we can approximate (\ref{edge_mod_eq}) using a partial set of eigenpairs.
Corollary \ref{approx_opt_cor} in \ref{corollary2} provides the corresponding bounds.  In the next
section we evaluate numerically the quality of the low-rank approximations provided by
theorem \ref{partial_sum_thm}.  Our experiments indicate that close-to-optimal results (as
measured by the reduction in the Kirchhoff index) can be achieved with $p \ll n$ eigenpairs.

\begin{figure}[H]
  \centerline{\small \sffamily Erd\H{o}s-R\'{e}nyi random graph}
  \centerline{\includegraphics[width=0.875\textwidth]{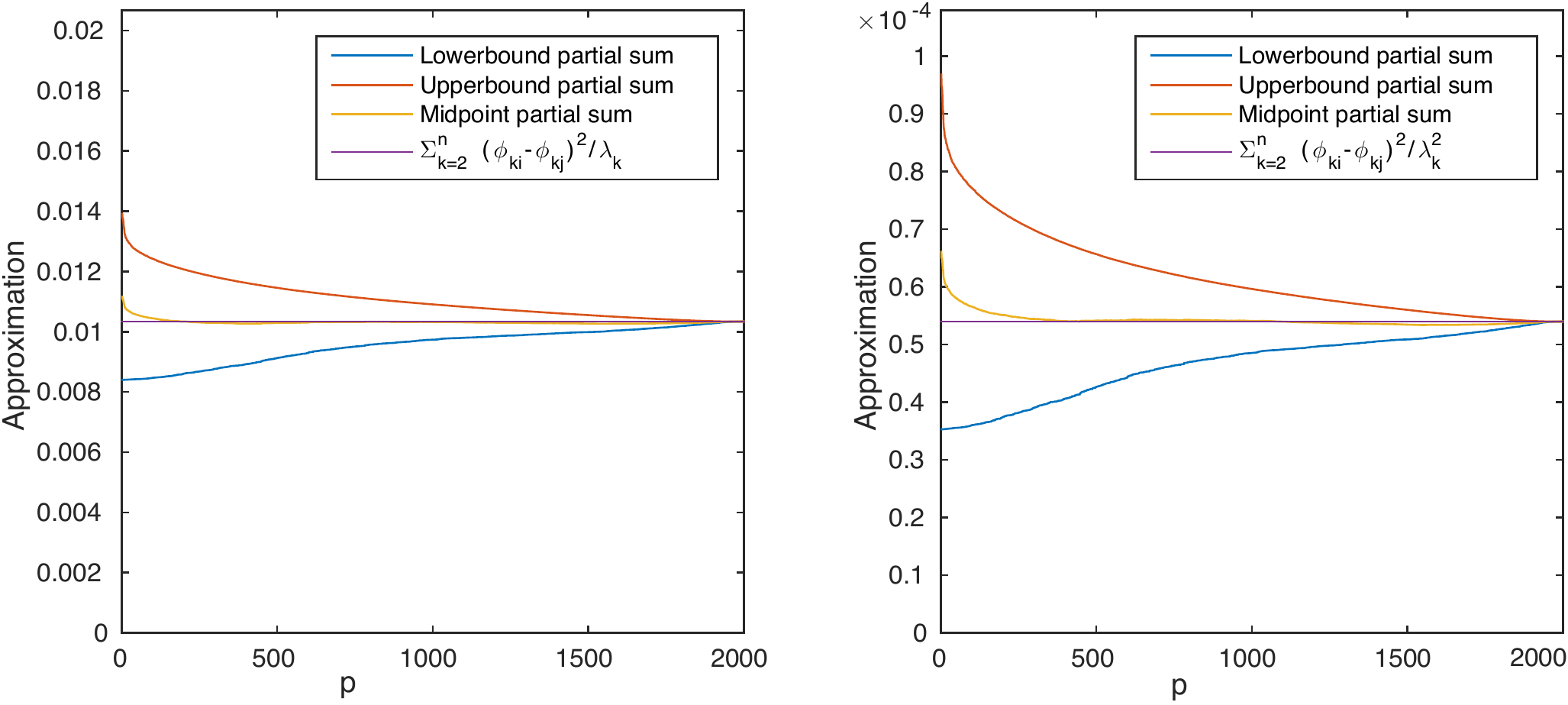}}
  \centerline{\small \sffamily Two communities planted partition model}
  \centerline{\includegraphics[width=0.875\textwidth]{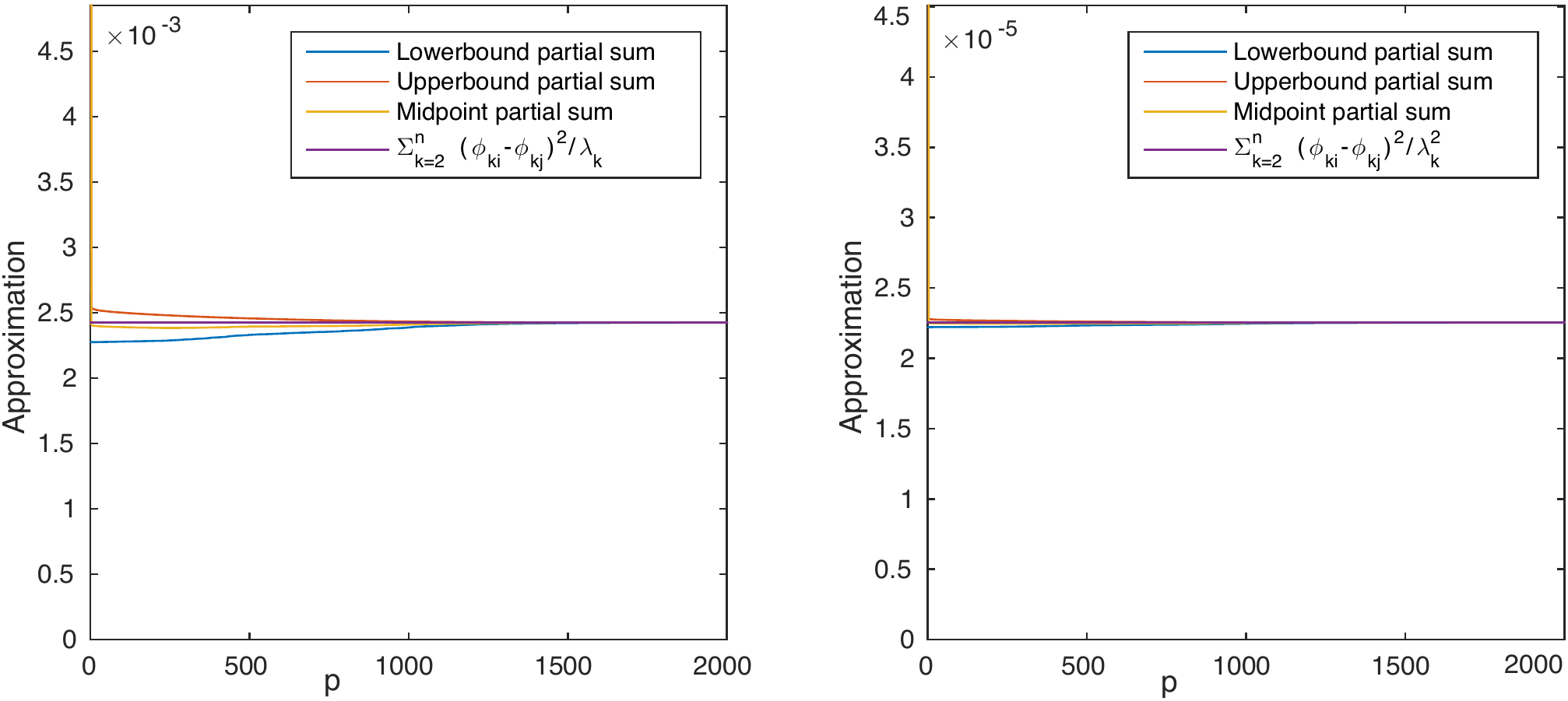}}
  \centerline{\small \sffamily Small world (Watts and Strogatz) model}
  \centerline{\includegraphics[width=0.875\textwidth]{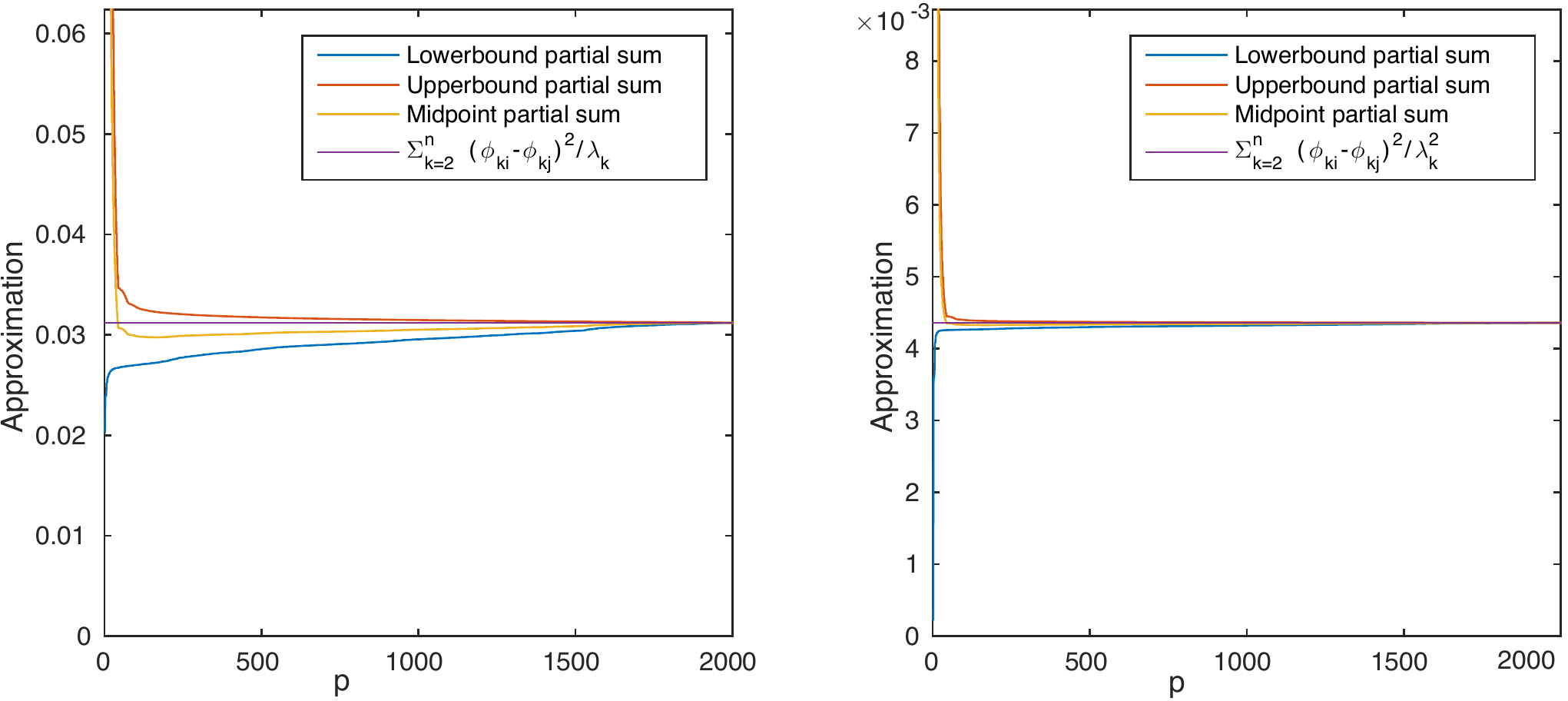}}
  \centerline{
    \hfill \hspace*{2em}
    {\sffamily $\sum_{k=2}^n \bck  \left[\bfi_k(i)-\bfi_k(j)\right]^2/\lambda_k$}
    \hfill
    {\sffamily $\sum_{k=2}^n  \bck  \left[\bfi_k(i)-\bfi_k(j)\right]^2/\lambda_k^2$}
    \hfill
  }
  \caption{The lower, upper, and average bounds given by theorem \ref{partial_sum_thm}, as well as the 
    exact sum for $\sum_{k=2}^n \bck  \left[\bfi_k(i)-\bfi_k(j)\right]^2/\lambda_k$ (left) and
    $\sum_{k=2}^n  \bck  \left[\bfi_k(i)-\bfi_k(j)\right]^2/\lambda_k^2$ (right). All the quantities
    are displayed as a function of $p$, the number of eigenpairs used in the partial sums,
    (\ref{low_rank_approx_bound_eqn}) and (\ref{low_rank_sq_approx_bound_eqn}).  All graphs have
    $n=2,000$ vertices. See main  text for details.  
    \label{partial_sum_fig}}
\end{figure}

We generated several graphs from ensembles of random graphs, and computed the upper and lower bounds
for both sums (\ref{low_rank_sq_approx_bound_eqn}), and (\ref{low_rank_approx_bound_eqn}).  To
further improve the approximation, we noticed that the average of the lower and upper bounds in
(\ref{low_rank_sq_approx_bound_eqn}) and (\ref{low_rank_approx_bound_eqn}) produced a very accurate
estimates of the corresponding sum.  Indeed, the idea is that the bulk is approximated by the
average of the largest and (one of) the smallest eigenvalue in the bulk. Fig. \ref{partial_sum_fig}
displays the various approximations. The left column shows the approximation of
$\sum_{k=2}^n \bck \left[\bfi_k(i)-\bfi_k(j)\right]^2/\lambda_k$, while the left column displays the
approximation of $\sum_{k=2}^n \bck \left[\bfi_k(i)-\bfi_k(j)\right]^2/\lambda_k^2$.

Each row corresponds to a different graph. All graphs have $2,000$ vertices. The top row is a
realization of an Erd\H{o}s-R\'{e}nyi random graph with edge probability equal to $0.1$.  The middle
row corresponds to a block stochastic model composed of two communities of equal sizes (also know as
a planted partition model), where the within-community edge probability is $p_\text{in} = 0.9$, and
the between-community edge probability is $p_\text{out}= 0.005$. Finally, the bottom row  corresponds to a small world (Watts and Strogatz) model constructed by randomly re-wiring
a regular ring lattice of constant degree 80, where each edge is rewired with a probability
$\beta=0.01$.  We conclude that for all three graphs, the average of the lower and upper bounds in
(\ref{low_rank_sq_approx_bound_eqn}) and (\ref{low_rank_approx_bound_eqn}) provided an accurate
estimate of the numerator and the denominator of $\drpo(G,G+\dw{i_0j_0})$.
\subsection{Fast greedy Optimization of the Kirchhoff Index
  \label{greedy}}
To avoid the exhaustive search of the optimal edge over all pairs of vertices, we designed the
following fast greedy search method.  The algorithm iteratively constructs a sequence of edges that
converges toward a local optimum of (\ref{edge_mod_eq}). The initial edge is constructed by choosing
randomly a vertex $i_0$. The algorithm then visits the other $n-1$ vertices, and select that vertex
$j_0$ that maximizes the decrease in the Kirchhoff index, as measured by (\ref{edge_mod_eq}).  The
vertex $j_0$ is then kept fixed, and the algorithm visits the remaining $n-2$ vertices to replace
$i_0$ by $i_1$ in order to further decrease (\ref{edge_mod_eq}) using the edge $[i_1,j_0]$.  The
process is repeated until (\ref{edge_mod_eq}) can no longer be improved. This algorithm runs in
$\O{n}$ time, a significant improvement over the $O(n^2)$ exhaustive search.
\subsection{Experimental Validation of the Optimization of the Kirchhoff Index}
\begin{figure}[H]
  \begin{center}
    \includegraphics[width=0.49\textwidth]{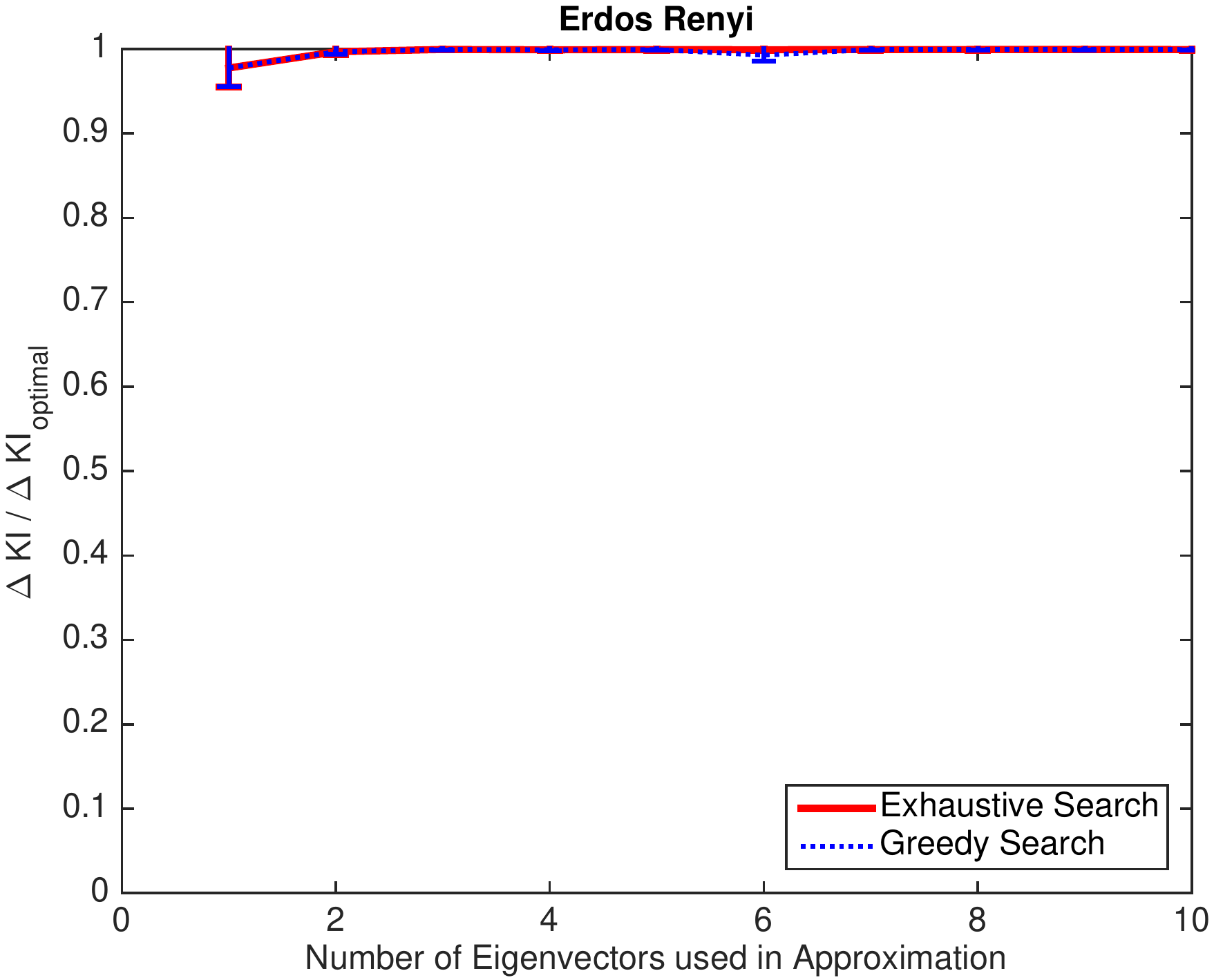}
    \includegraphics[width=0.49\textwidth]{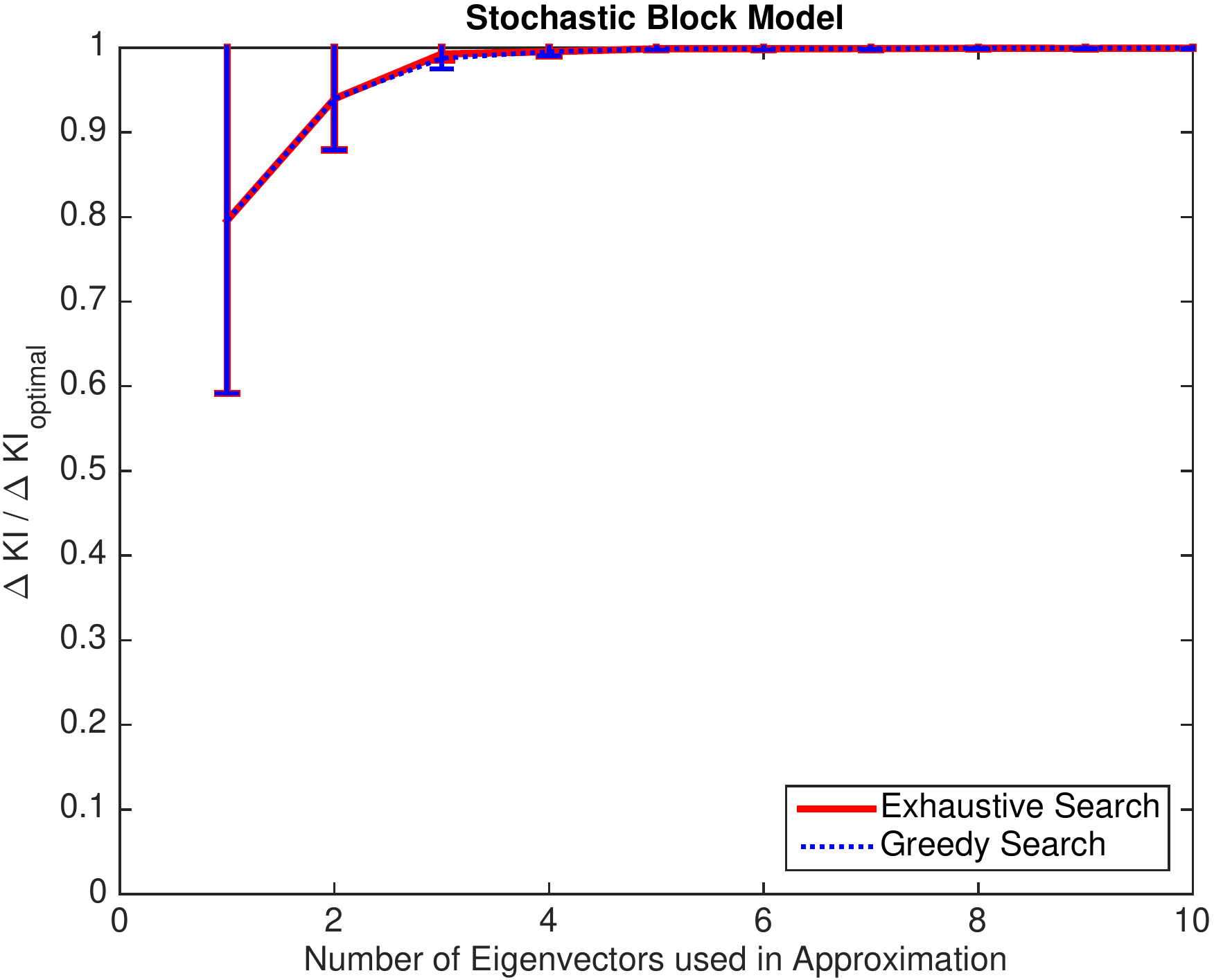}
    \includegraphics[width=0.49\textwidth]{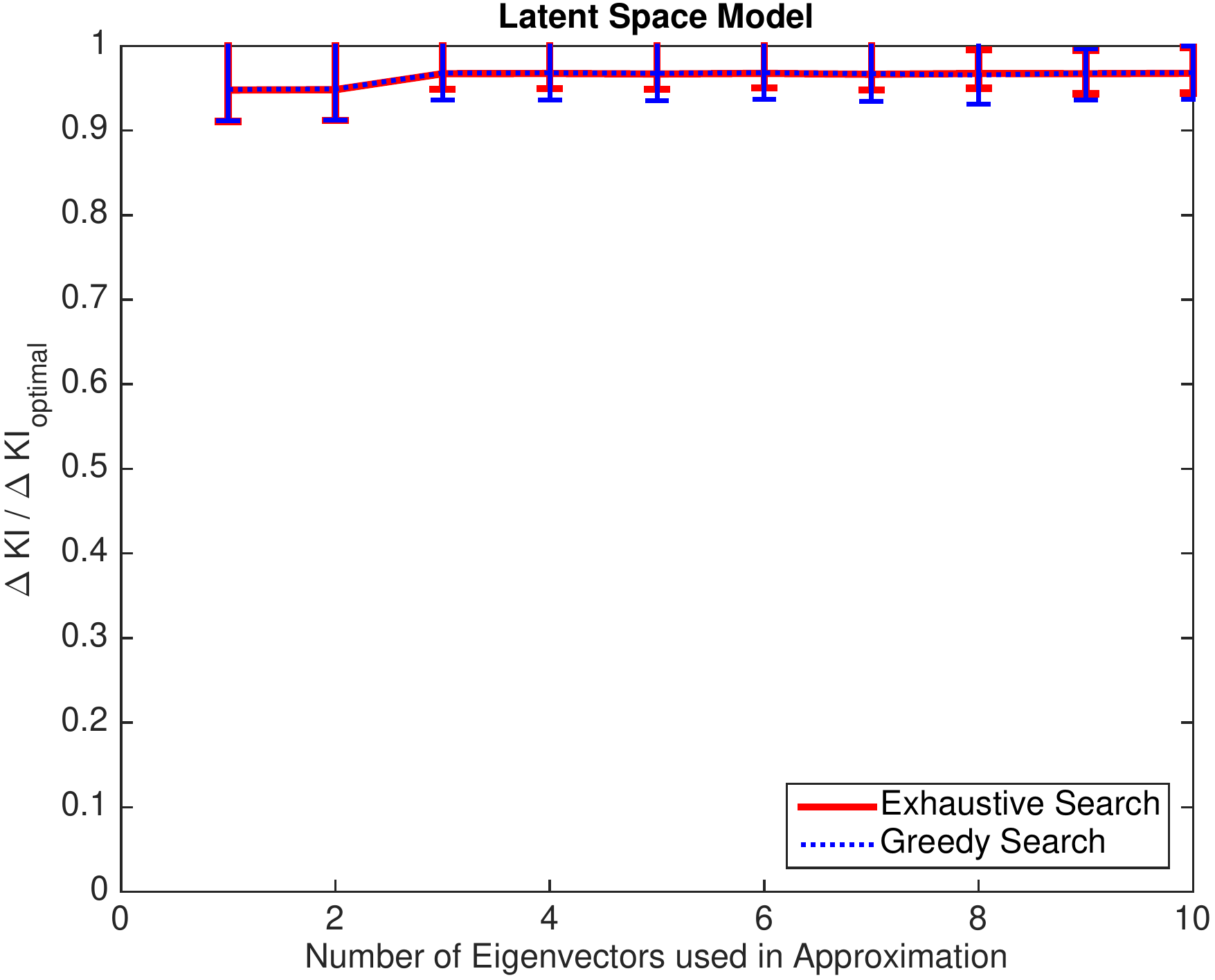}
    \includegraphics[width=0.49\textwidth]{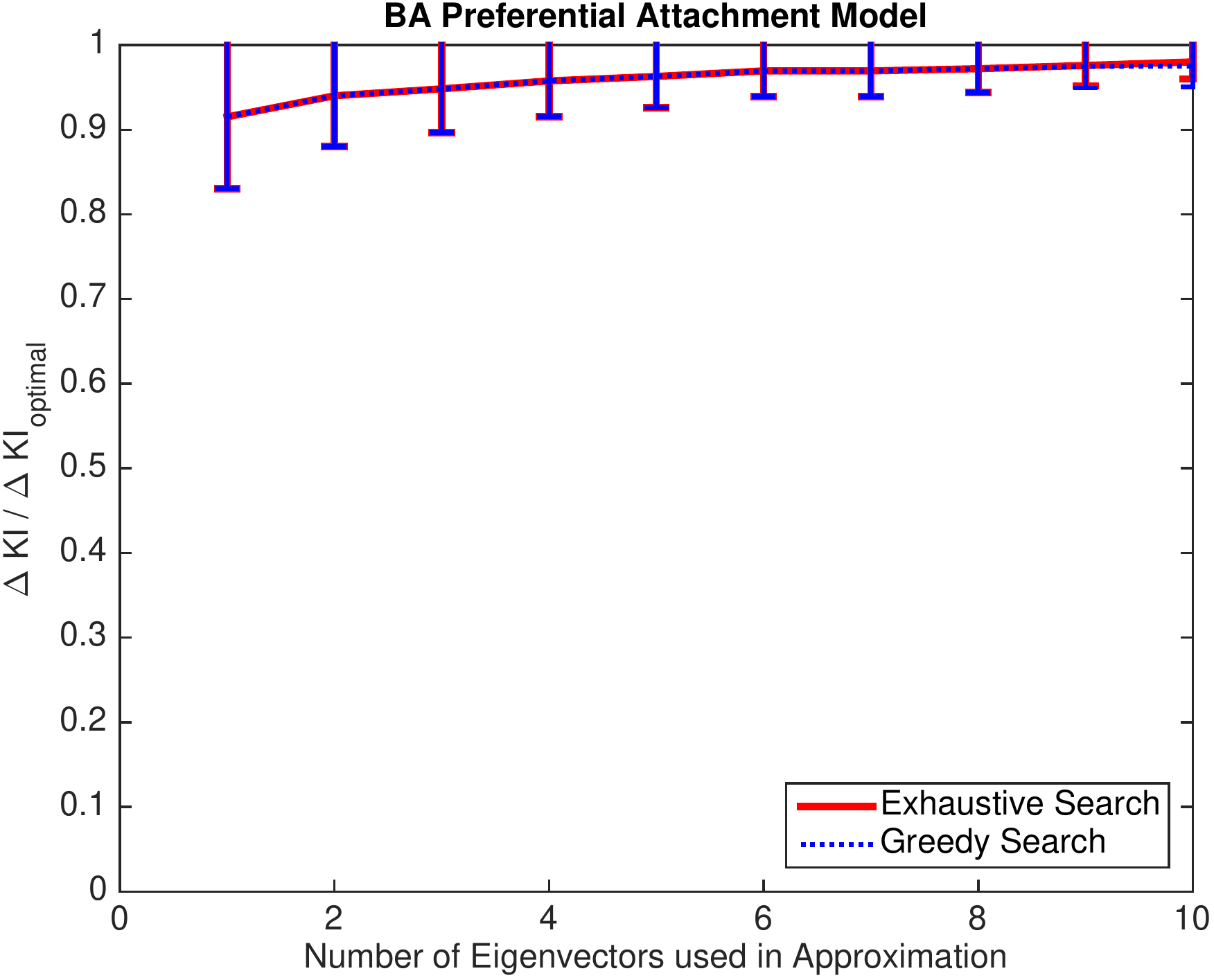}
    \includegraphics[width=0.49\textwidth]{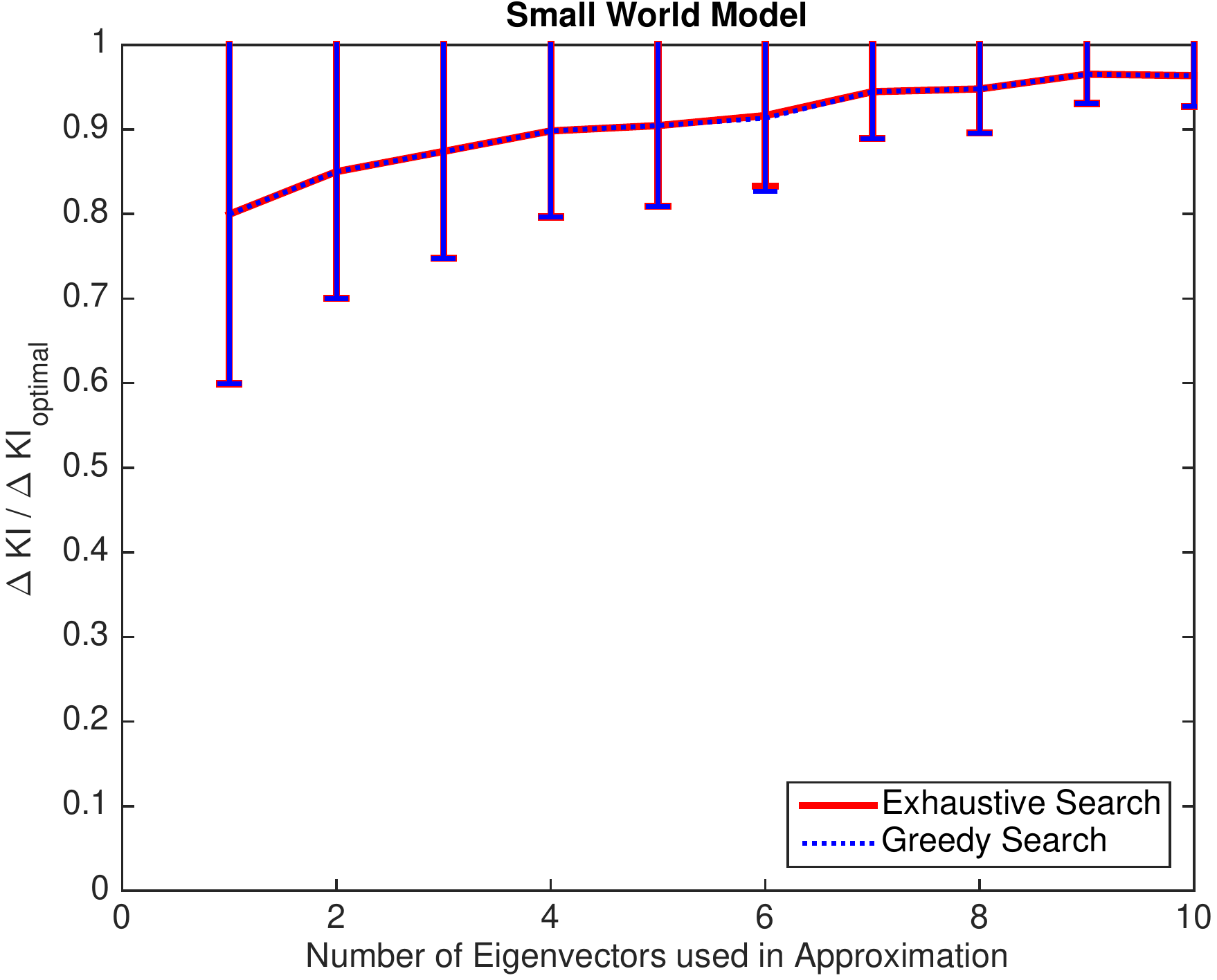}
  \end{center}
  \caption{Relative reduction in the Kirchhoff index $\Delta \ki/\Delta \ki_{\text{optimal}}$ as a
    function of $p$, the number of eigenvectors used to approximate (\ref{edge_mod_eq}). We compare
    the exhaustive (solid red line) with the fast greedy search (dotted purple line). From left to
    right and top to bottom: Erd\H{o}s-R\'{e}nyi, stochastic block model, latent space model,
    Barab\'{a}si-Albert preferential attachment, and Watts and Strogatz model. The mean relative
    reduction, as well as the range (minimum to maximum values, over 50 random realizations) are shown.}
  \label{all_KI_reduction_figs}
\end{figure}

\noindent To validate the fast optimization of the Kirchhoff index, we designed a second set of
experiments, using graphs generated from archetypal ensembles of random graphs. In this set of
experiments, all graphs have 500 vertices.  For all experiments we approximated the $\drpo$ distance
using the average of the lower and upper bounds (\ref{low_rank_sq_approx_bound_eqn}) and
(\ref{low_rank_approx_bound_eqn}) for the numerator and denominator of (\ref{edge_mod_eq}),
respectively. This led to an estimate of the decrease of the Kirchhoff index, $\Delta \ki$, that was
computed using $p$ eigenvectors.  As $p$ increases and approaches $n$, we recover the exact
expression given by (\ref{edge_mod_eq}). The gold standard,
$\Delta \ki_{\text{optimal}} = \drpo(G,G+\dw{\text{optimal}})$, is the optimal decrease of the
Kirchhoff index that would result from the optimal edge addition if we were to use an exhaustive
search. Each plot in Fig. \ref{all_KI_reduction_figs} displays the relative error,
$\Delta \ki/\Delta \ki_{\text{optimal}}$ as a function of $p$. For each random graph model, the
experiment was repeated 50 times. 

The mean and the range (minimum to maximum, shown as an error-bar) of the relative reduction in the
Kirchhoff index is plotted in Fig. \ref{all_KI_reduction_figs}.  We note that this error compounds
two approximations: the low-rank approximation in (\ref{partial_sum_thm}), and the greedy algorithm
described in section \ref{greedy}.

We now describe the five graph models.

{\noindent \bf Unit Circle Latent Space Model.}  We sampled 500 points using a uniform distribution
on the unit circle in $\R^2$,
\begin{equation*}
  \bx_i = 
  \begin{bmatrix}
    \cos(\theta_i)\\
    \sin(\theta_i) 
  \end{bmatrix},
  \quad \text{where} \; \theta_i \sim \text{U}[0,2\pi],\quad i=1,\ldots,500. 
\end{equation*}
\noindent An unweighted graph $G=(V,E)$ was then generated by randomly connecting each pair of
vertices $\{i,j\}$ with an edge $[i,j]$ according to a probability prescribed by a Gaussian kernel
in the latent space,
\begin{equation}
  P([i,j] \in E)=\frac{10}{\sqrt{\pi}}\exp \left( -100 \| \bx_i - \bx_j \|^2 \right), \quad \text{for} \; i \neq j.
  \label{latent}
\end{equation}
{\noindent \bf Erd\H{o}s-R\'{e}nyi random graph.} We constructed a random graph with edge
probability equal to 0.05. 

{\noindent \bf Two communities stochastic block model.} We generated a stochastic block
model formed by two communities of equal sizes, where the within-community edge probability was
$p_\text{in}=0.1$, and the between-community edge probability was $p_\text{out} = 0.01$. 

{\noindent \bf Barab\'{a}si-Albert preferential attachment model.} The graph was constructed by
sequentially adding two edges from each new vertex, attaching to other vertices with probability
proportional to their current degrees.

{\noindent \bf Watts and Strogatz model.}  The small world model was designed by randomly re-wiring
a regular ring lattice of constant degree 40 and a rewiring probability $\beta=0.1$.\\

\noindent We first notice in Fig. \ref{all_KI_reduction_figs} that, for all graphs, the greedy
search performed as well, or nearly as well, as the exhaustive search. With regard to the quality of
the low-rank approximation, using only the Fiedler vector ($\bfi_2$), we were able to capture 95\%
of the optimal increase in the Kirchhoff index. The Erd\H{o}s R\'enyi graph only required $\bfi_2$ to
estimate the optimal $\Delta \ki_{\text{optimal}}$. As expected, the two-communities stochastic
block model required two eigenvectors $\bfi_2$ and $\bfi_3$ to achieve near-optimal
approximation. The latent space model required more eigenvectors to completely recover the optimal
$\Delta \ki_{\text{optimal}}$. Nevertheless, a very good estimate was obtained with $\bfi_2$ only,
which was able to capture the topological structure of the latent space formed by the ring. The
stochastic nature of the graph construction necessitated more eigenvectors to fully%
\noindent estimate the
increase in the Kirchhoff index. A similar phenomenon happened with the Barab\'{a}si-Albert
preferential attachment model and the Watts and Strogatz model. In the latter case, $\bfi_2$ was
only able to recover the ring lattice, which corresponds to the regular part of the
graph. Additional eigenvectors were needed to capture the ``disorder'' created by the random
rewiring. As mentioned earlier, the error is a function of the low-rank approximation in
(\ref{partial_sum_thm}) and the greedy algorithm described in section \ref{greedy}, and therefore is
not necessarily monotonically decreasing with $p$.
\section{Analysis of Dynamic Networks with the RP-p Distances 
  \label{experiments}}
We demonstrate in this section how the RP-1 and RP-2 distances can be used to detect anomalies
caused by significant structural changes in dynamic networks. Our analysis is based on a series of
experiments on synthetic and real networks. The results of the experiments clearly show that the
resistance perturbation metric can detect the configurational changes in dynamic graphs that are
triggered by appreciable modifications of the hidden variables controlling the graph dynamics.

The distances $\drpo$ $\drpt$, DeltaCon using $d_\text{rootED}$ (see (\ref{delta_con_def}) and
(\ref{root})), and $d_\text{CAD}$ (see (\ref{CAD})) were computed for all the experiments.  We used
all the edges to compute $d_\text{CAD}$, to wit $F=E$ in (\ref{CAD}).
\subsection{Random  Graphs Models}
The first set of experiments rely on realizations of graphs sampled from ensembles of random
graphs. The experiments were conducted on three different families of random graph models: a random
graph with a latent space, a two-communities block stochastic model, and a Watts and Strogatz
model. All models depend on a single scalar that characterizes the structure of the graph. We first
detail the experimental procedure,
and then describe each graph model.\\

{\noindent \bfseries Experimental procedure.}  All experiments were conducted in the following
manner: a baseline graph $G^{(1)}$ was randomly selected using the baseline value for the parameter
of the corresponding model. We then generated 50 random realizations of a second graph $G^{(2)}$,
using the same value of the parameter.

The parameter that controls the graph was then increased, in 10 increments. For each increment, 50
random realizations of a second graph $G^{(2)}$ were constructed, and all the graph distances were
computed. By modifying the parameter that has an important impact on the structure of the graphs, we
evaluated quantitatively the relationship between the resistance perturbation distance and
(potentially unobserved) changes in the latent parameter that controls the organization of the
graph.

Our experiments show that the resistance perturbation distance is highly correlated with the
evolution of the parameter that governs the structure of the graphs. In contrast, the DeltaCon distance
is very sensitive to the normal fluctuations between the 50 random different realizations of the
same exact graph structure. The DeltaCon distance also exhibits the largest variability between the
different random realizations. The $d_\text{CAD}$ distance, which is biased by changes in the adjacency matrix can
become too sensitive to changes in the graph topology (e.g., in the case of the stochastic block model).

{\noindent \bfseries Unit Circle Latent Space.}  A first graph $G^{(1)}$ was constructed by first
sampling 2,000 points using a uniform distribution on the unit circle in $\R^2$,
\begin{equation*}
  \bx^{(1)}_i = 
  \begin{bmatrix}
    \cos(\theta^{(1)}_i)\\
    \sin(\theta^{(1)}_i) 
  \end{bmatrix}
  \bck,
  \; \text{with} \quad \theta^{(1)}_i \sim \text{U}[0,2\pi],\quad i=1,\ldots, 2,000. 
\end{equation*}
\noindent An unweighted graph $G^{(1)}=(V,E^{(1})$ was then generated by randomly connecting each pair of
vertices $\{i,j\}$ with an edge $[i,j]$ according to a probability prescribed by a Gaussian kernel
in the latent space,
\begin{equation}
  P([i,j] \in E^{(1)})=\frac{20}{\sqrt{\pi}}\exp \left( -400 \| \bx_i - \bx_j \|^2 \right), \quad
  \text{for} \; i \neq j. 
  \label{latent2}
\end{equation}

\begin{figure}[H]
  \centerline{\small \sffamily 
    \hspace*{5pc} unit circle latent space model \hspace*{8pc} two communities stochastic block model
  }\hspace*{1pc}
  \centerline{
    \includegraphics[width=0.5\textwidth]{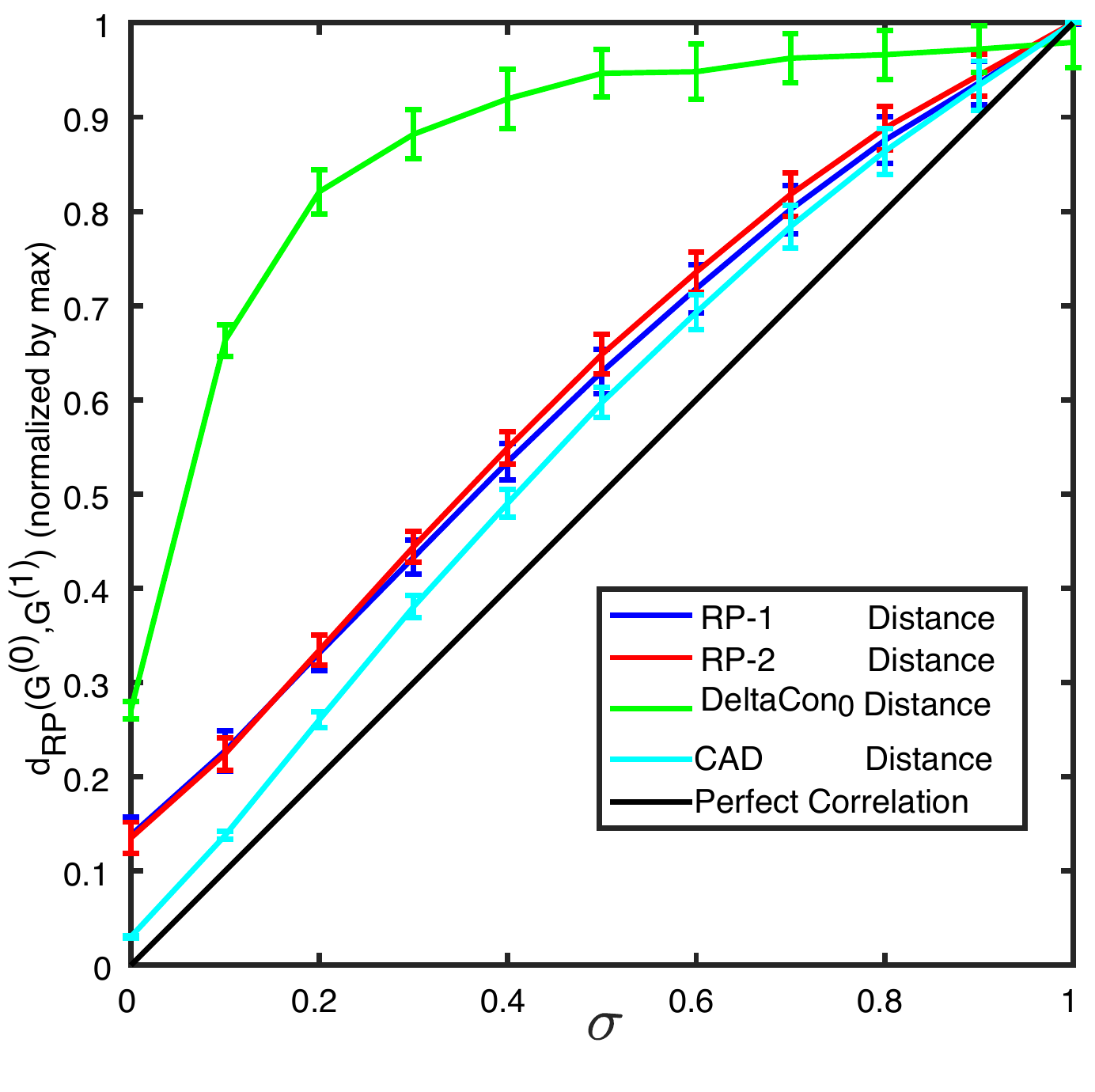}
    \includegraphics[width=0.5\textwidth]{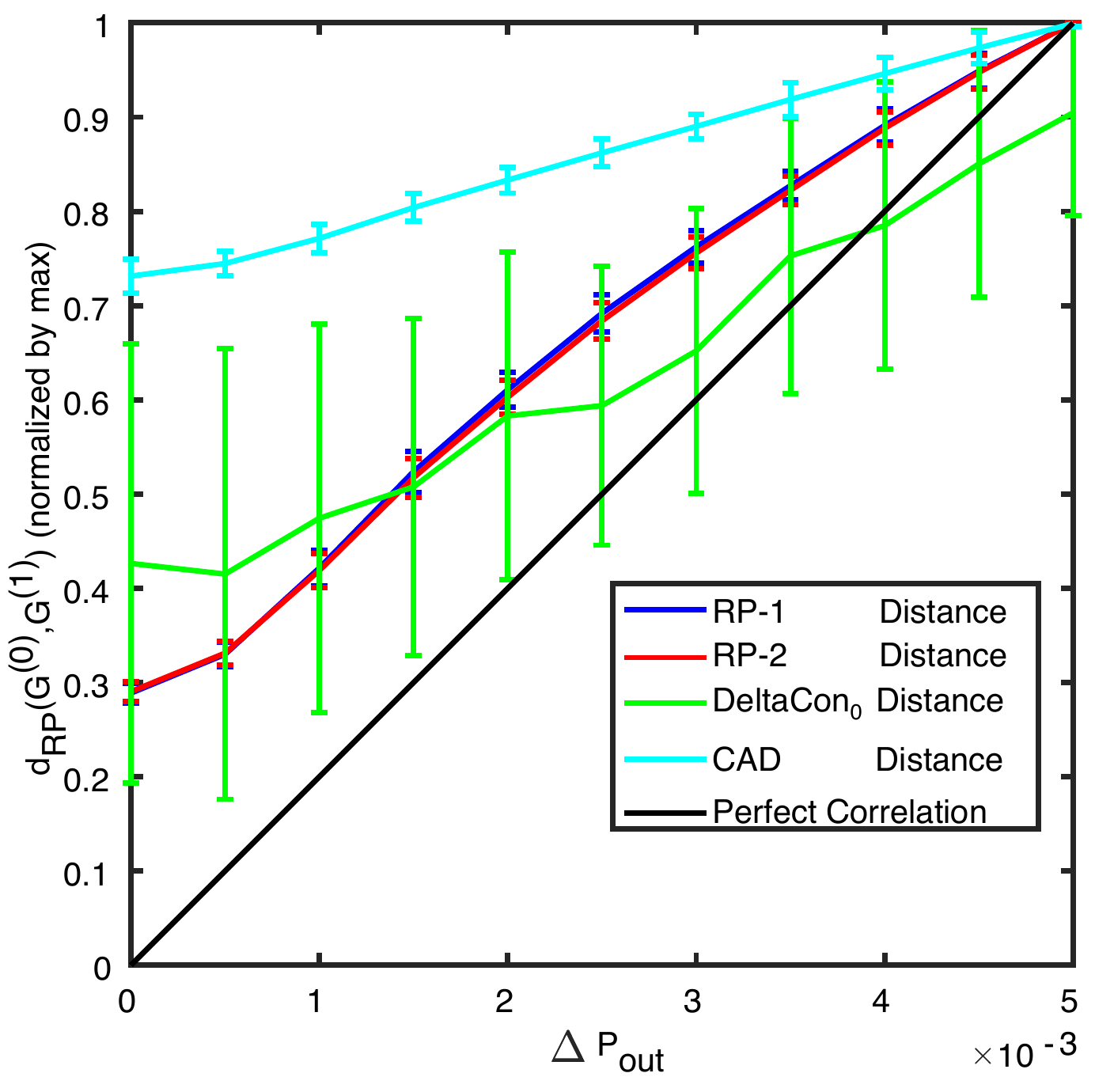}
  }
  \centerline{\hspace*{1pc}\small \sffamily small world (Watts and Strogatz) model}
  \centerline{
    \includegraphics[width=0.5\textwidth]{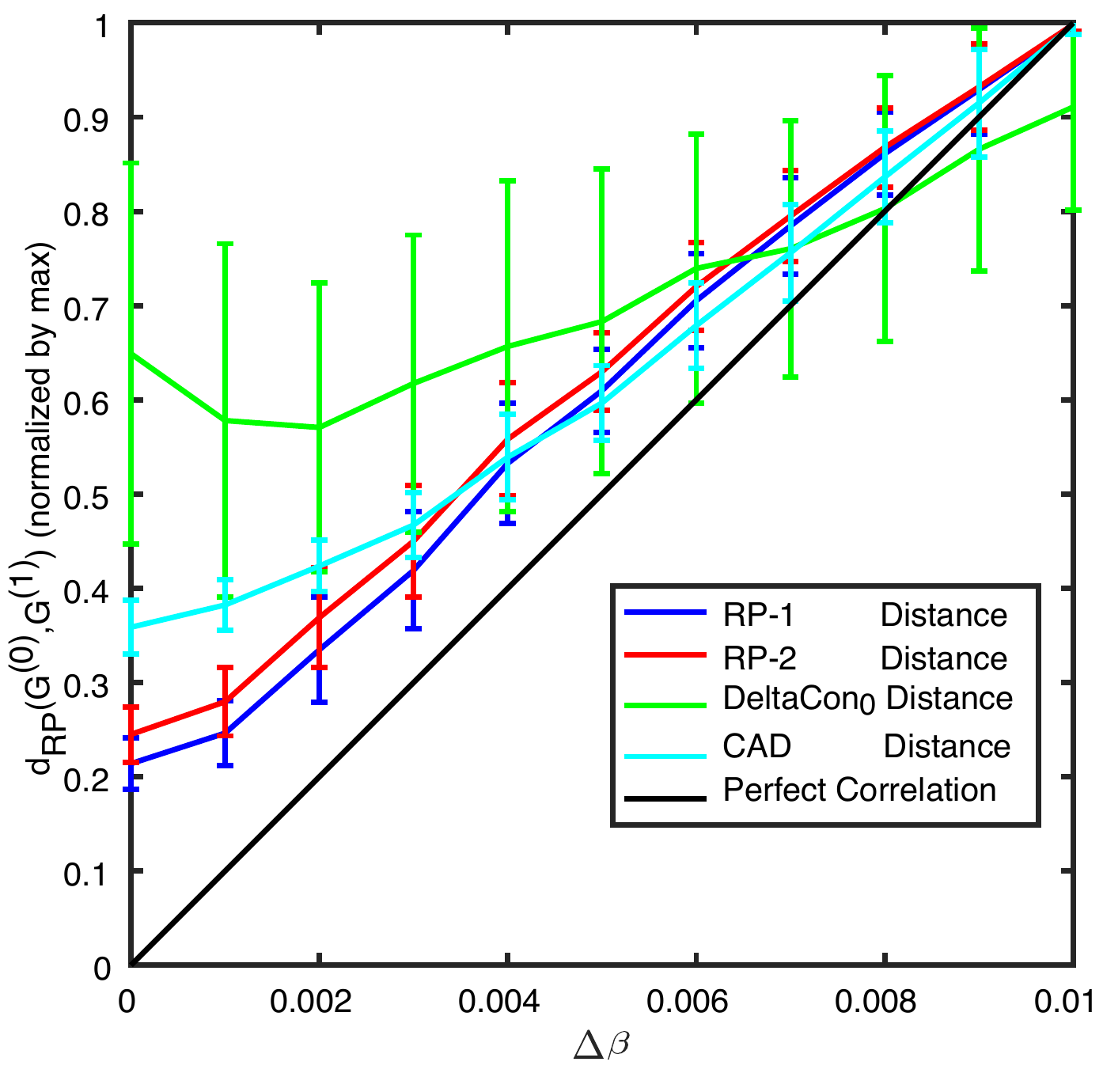}
  }
  \caption{The graph distances $\drpo(G^{(1)},G^{(2)})$, $\drpt(G^{(1)},G^{(2)})$, DeltaCon
      ($\droot{G^{(1)}}{G^{(2)}}$), and $d_\text{CAD}(G^{(1)},G^{(2)})$, as a function of the latent
      parameter that controls the structural difference between $G^{(1)}$ and $G^{(2)}$.  Error bars
      represent the standard deviation computed over 50 random realizations. Top: unit circle latent
      space model (left), and the two communities stochastic block model (right). Bottom: the small
      world (Watts and Strogatz) model. See text for details.
    \label{dynamic_analysis_random_graphs}
  }
\end{figure}
\noindent A second random graph $G^{(2)}=(V,E^{(2)})$ was generated according to the same principle,
but with a second set of latent locations, $\{\bx^{(2)}_i\}$ that was obtained by a perturbation of the
initial set $\{\bx^{(1)}_i\}$,
\begin{equation*}
  \bx^{(2)}_i = 
  \begin{bmatrix}
    \cos(\theta^{(2)}_i)\\
    \sin(\theta^{(2)}_i) 
  \end{bmatrix}\bck, \;
  \text{with} \quad 
  \theta^{(2)}_i = \theta^{(1)}_i + \gamma_i, \;
  \gamma_i \sim {\cal N}(0,\sigma^2),\;
  i=1,\ldots, 2,000. 
\end{equation*}
The random edges $E^{(2)}$ were connected using the same probability distribution given by (\ref{latent2}).
The magnitude of the random Gaussian shifts between the angles of the set $\{\bx^{(1)}_i\}$ and
those of the set $\{\bx^{(2)}_i\}$ is controlled by the standard deviation $\sigma$. For increasing
values of $\sigma\in [0,1]$ we constructed 50 random realizations of $G^{(2)}$, and we computed
$\drpo(G^{(1)},G^{(2)})$ and $\drpt(G^{(1)},G^{(2)})$.

\noindent Figure \ref{dynamic_analysis_random_graphs} top-left displays all the graph distances as a function
of $\sigma$. We first observe that $\drpo(G^{(1)},G^{(2)})$ and $\drpt(G^{(1)},G^{(2)})$ are very
similar.  This is crucial, since we designed a fast algorithm to approximate $\drpt$.  We also note
that both RP distances are highly correlated with the magnitude of the perturbation, $\sigma$.

The increasing difference between $G^{(1)}\bck$ and $G^{(2)}\bck$, created by the increase in
$\sigma$, intensifies the ``disorganization'' of $G^{(2)}$; the latent model is less and less
regularly organized along the unit circle. The DeltaCon distance is able to detect this progression
toward disorder, but quickly reaches its maximum value for unremarkable values of $\sigma$, making
it useless for detecting anomalies. In contrast the $d_\text{CAD}$ distance performed extremely well.

{\noindent \bfseries Two Communities Stochastic Block Model.}  The $n=2,000$ nodes are divided into
two communities of size $n/2$. Every pair of nodes $\{i,j\}$ forms an edge $[i,j]$ with probability
$p_\text{in}$ if they belong to the same community, and with probability $p_\text{out}$ if they
belong to different communities. We fixed $p_\text{in}= 0.9$ for both graphs. We used
$p^{(1)}_\text{out} = 0.005$ for $G^{(1)}$, and we varied $p^{(2)}_\text{out} \in [0.005,0.01]$ for
$G^{(2)}$. Figure \ref{dynamic_analysis_random_graphs} top-right displays the four graph distances
as a function of $\Delta p_\text{out} = p^{(2)}_\text{out} - p^{(1)}_\text{out}$.

In comparison with the latent space model, we note that the changes in the adjacency matrix
  created by the intrinsic randomness of the model confuses the $d_\text{CAD}$ distance very quickly. Indeed,
  the $d_\text{CAD}$ distance immediately reaches 0.73 in the baseline condition when $G^{(1)}$ and $G^{(2)}$
  have the same structure, to wit when they are both realizations of the same random model (same
  $p_\text{in}$ and same $p_\text{out}$).  DeltaCon is equally confused: the standard deviation is
  very large, making it difficult to assess the confidence one should attach to a single measurement
  of the distance.

  Conversely, $\drpo(G^{(1)},G^{(2)})$ are highly correlated with the increase in $p_\text{out}$,
  making the distances suitable to detect changes in community networks. Furthermore, the standard
  deviations for both RP-distances remain very small.
  \\

{\noindent \bfseries Small World Model.}  We generated random realizations of a small world (Watts
and Strogatz) model constructed by randomly re-wiring a regular ring lattice of constant degree 40
using a random rewiring with probability $\beta_2$ that varied from $\beta_2 = \beta_1 = 0.01$ for
$G^{(1)}$, to $\beta_2 = 0.02$. We generated 50 random realizations for each value of
$\beta_2$. Figure \ref{dynamic_analysis_random_graphs}-bottom displays the four graph distances as a
function of $\Delta \beta = \beta_2 - \beta_1$.

  In this model, the initial ring lattice moves toward a state of disorder when $\beta_2$
  increases. In a manner comparable to the latent space model, the increase in disorganization is
  detected by the DeltaCon distance, which is correlated with $\Delta \beta$ over the entire range.
  Both RP-distances as well as the $d_\text{CAD}$ distance are more tightly correlated with the increase of
  $\Delta \beta$, and are therefore more suitable to infer the dynamic underlying changes in the
  graph.

\subsection{Real Dynamic Network}
The second set of experiments involved two real-world dynamic networks, where we can qualitatively
compare the resistance perturbation distance to known events that would likely influence the
behavior of actors in the dynamic networks. These results suggest that the resistance perturbation
metric can identify changes in real dynamic graphs, and could be used to infer changes in the hidden
variables that govern the evolution of such dynamic graphs.  
{\noindent \bf Enron email network.} The Enron email corpus \cite{enroncorpus} is composed of the
email messages between approximately 150 high-level executives (the Enron ``core''); these were
included in the analysis because these individuals were most closely involved in the scandal.
Emails were aggregated on a weekly basis to generate a dynamic series of communication graphs, and
compared to a timeline of events. Undirected edges were assigned between pairs of vertices, with a
weight equal to the number of emails exchanged between the two people during a given week.  In order
to focus on personal communications, emails with greater than three recipients were excluded from
the analysis.  The size of the remaining dynamic graph is: number of vertices
= 151; count of emails = 31534; count of weighted%
\begin{figure}[H]
  \centerline{
    \includegraphics[width=1.0\textwidth]{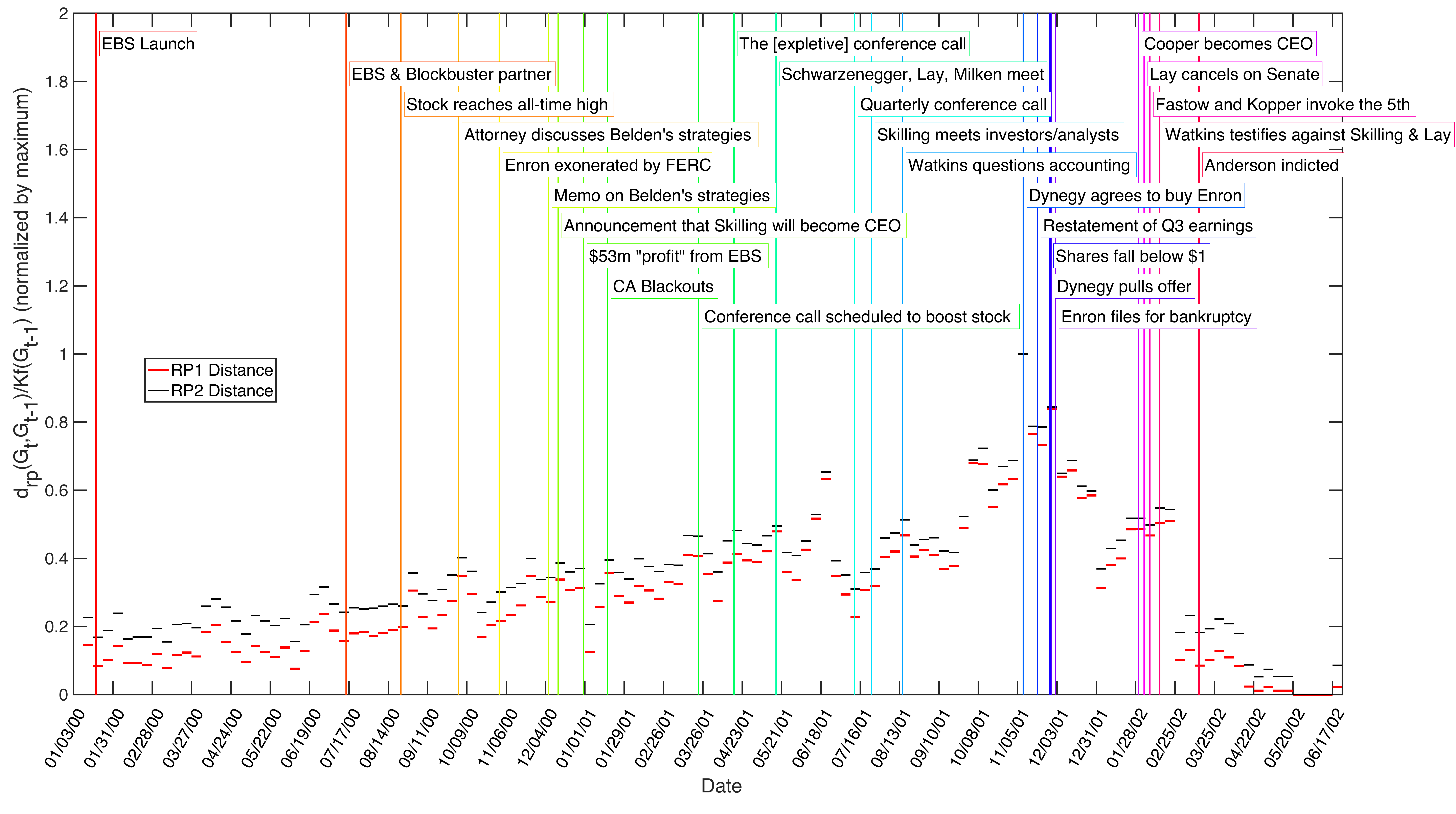}
  }
  \centerline{
    \includegraphics[width=1.0\textwidth]{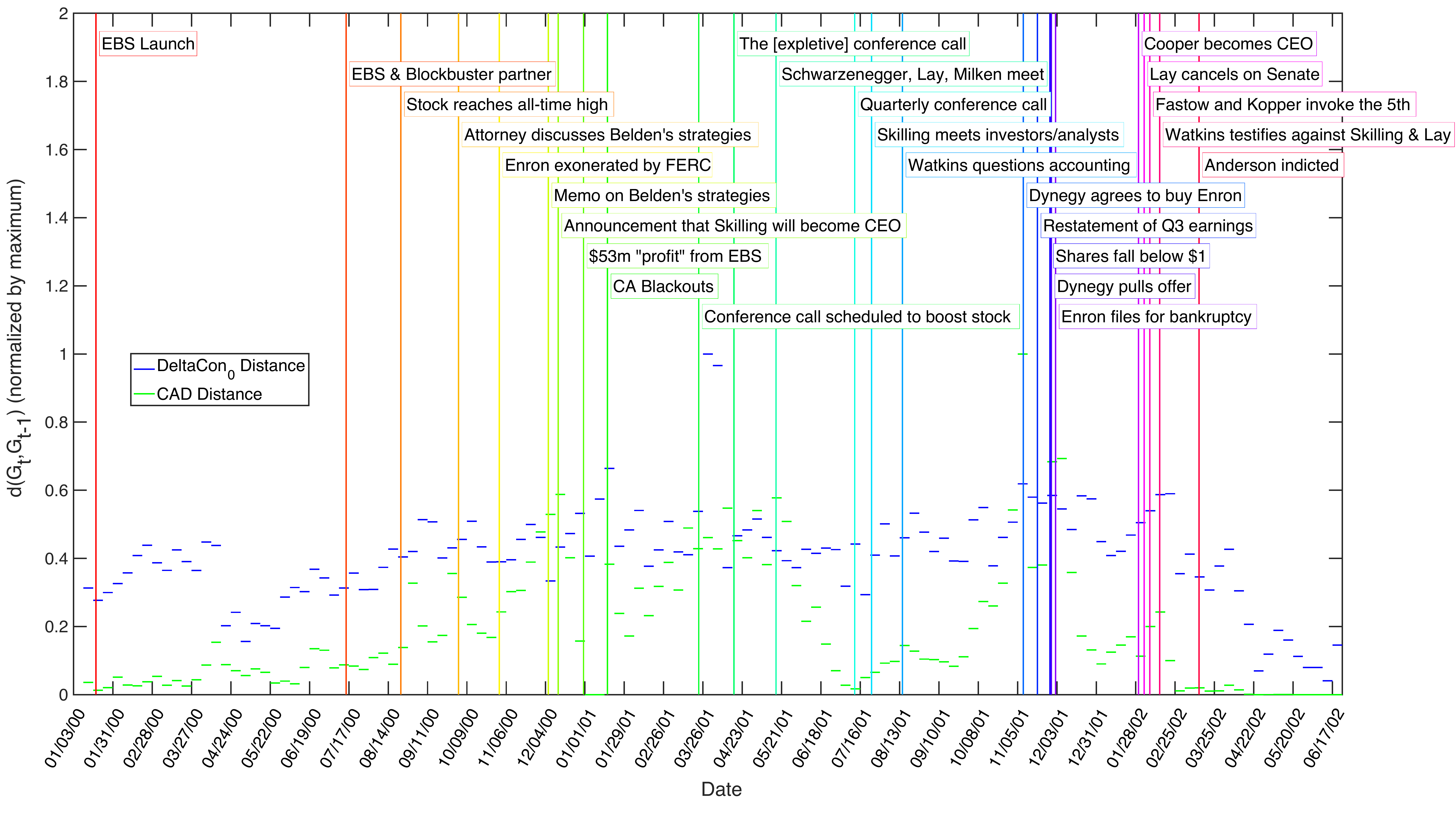}
}
\caption{
  $\drpo$ and $\drpt$ (top) and DeltaCon ($d_\text{rootED}$) and $d_\text{CAD}$ (bottom)
  between consecutive weekly email graphs from the Enron corpus.  Notable events in the timeline of
  the company's collapse are also plotted for reference \cite{peel2014}.}
  \label{enron}
\end{figure}
\noindent edges after weekly aggregation = 7794.
The time period analyzed spans the period leading up to the Enron scandal and
subsequent collapse of the company.

The resistance perturbation metrics, $d_{rp1}$ and $d_{rp2}$, between consecutive weekly email
graphs are plotted in Fig. \ref{enron}-top; $d_{rp1}$ and $d_{rp2}$ have very similar dynamics.
Furthermore, we note that large changes detected by $d_{rp2}$ during the summer and fall of 2001 are
predictive of the events that lead to the ultimate collapse of the company.

An independent analysis of the same dataset \cite{peel14,peel15} confirms that changes in the mean
degree, which are highly correlated to changes in the volume, is a very poor predictor of the
changes detected by the RP-distance.

DeltaCon exhibits a lot of volatility, changing at times when there are no significant events in
  the company, while remaining constant around the time associated with notable events. Changes
  in the $d_\text{CAD}$ distance appear to be tightly coupled with the events described by the
  vertical bars.\\

{\noindent \bf MIT reality mining dataset.} The MIT reality mining dataset \cite{reality} provides
collocation information between a group of students and faculty at MIT during the course of an
academic year. A dynamic undirected graph was built from weekly-aggregated Bluetooth proximity
data.  The weights of the edges in this dynamic graph are proportional to the amount of time each
pair of cellphones registered one-another's presence in close physical proximity.

\begin{figure}[H]
  \centerline{
    \includegraphics[width=\textwidth]{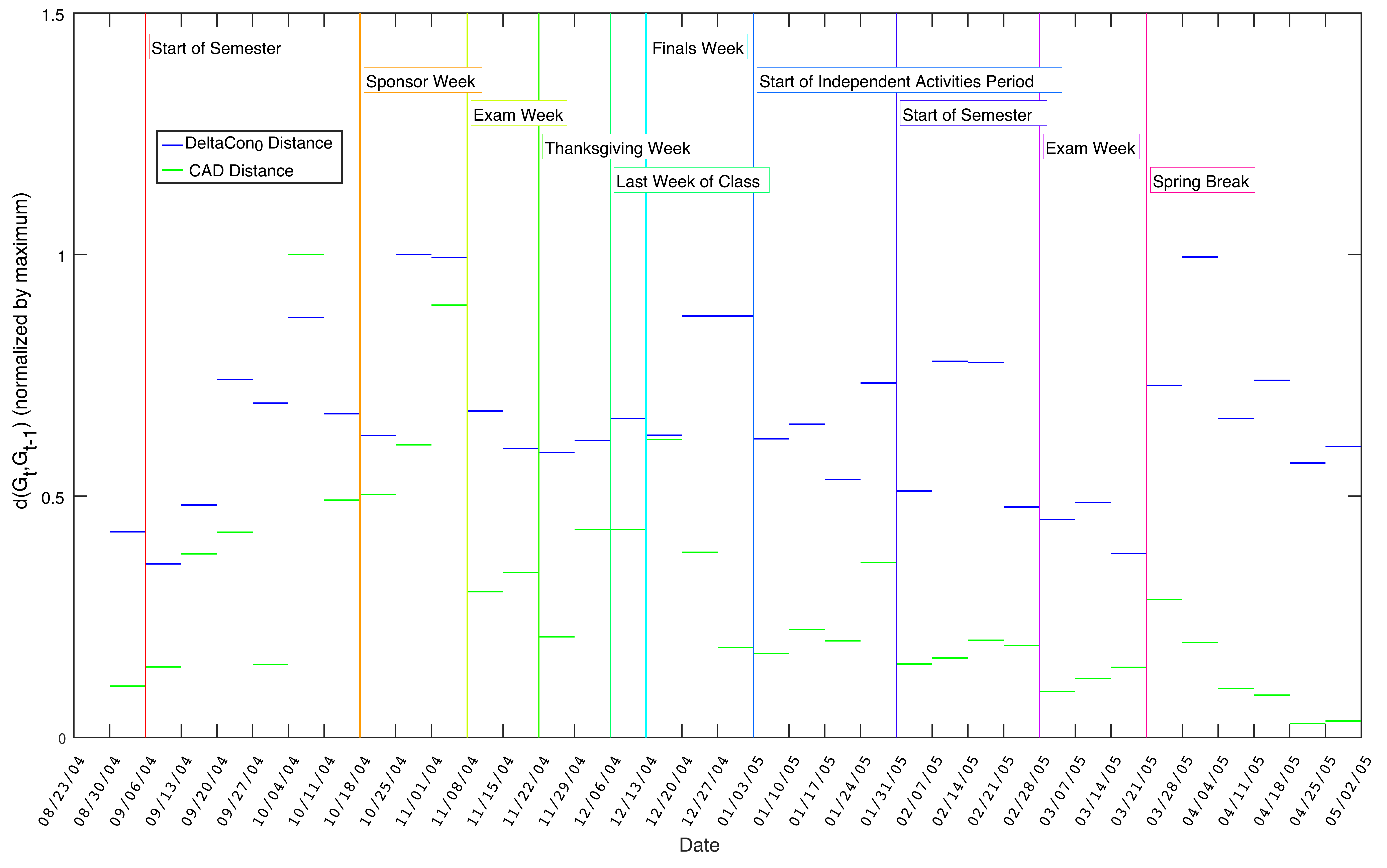}
    }
  \centerline{
    \includegraphics[width=\textwidth]{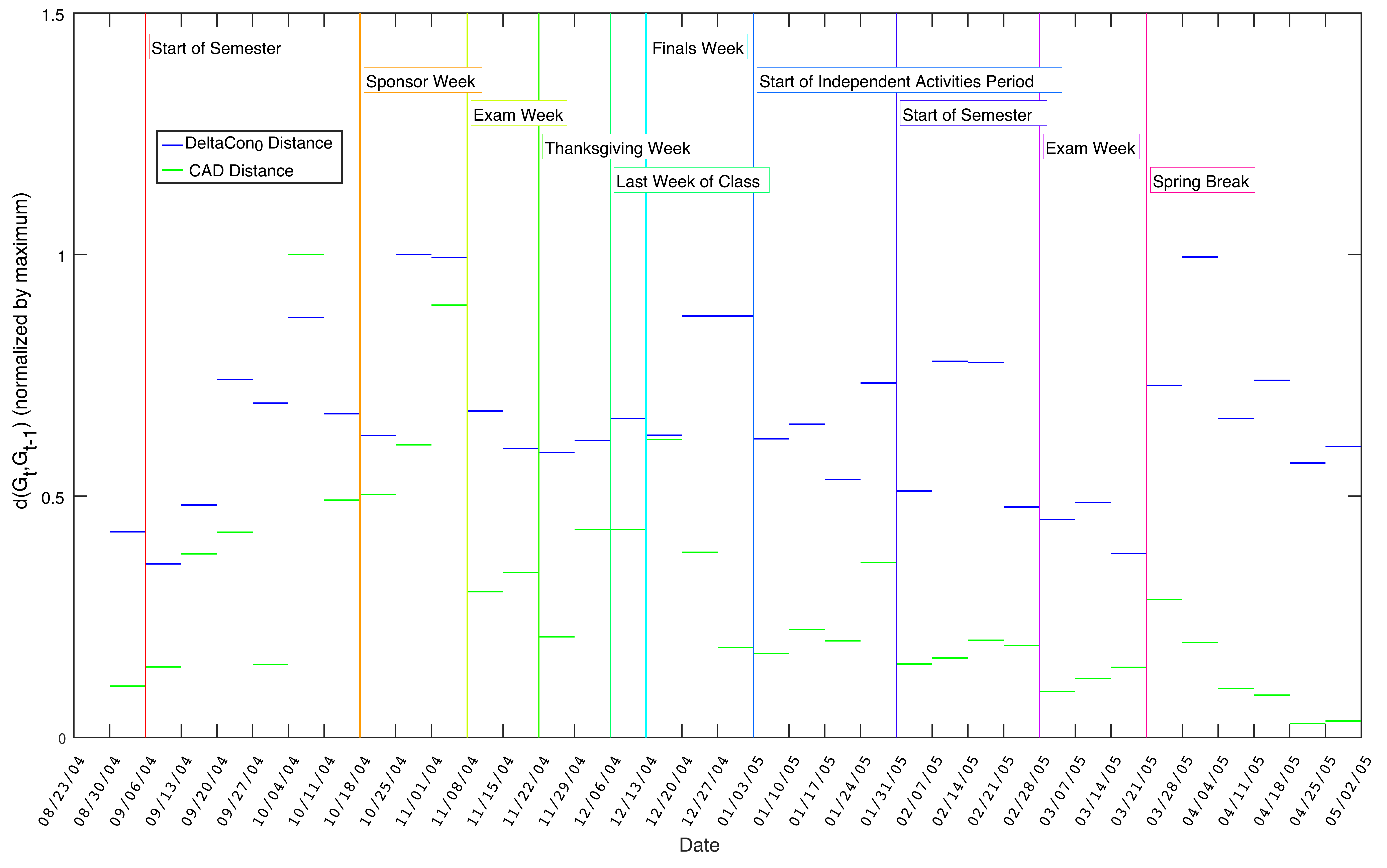}
  }
  \caption{
    $\drpo$ and $\drpt$ (top) and DeltaCon ($d_\text{rootED}$) and $d_\text{CAD}$ (bottom)
    between consecutive weekly Bluetooth proximity graphs from the MIT Reality Mining dataset.
    Important events in the academic calendar are also plotted \cite{peel2014}.}
  \label{mit}
\end{figure}

The $d_{rp1}$ and $d_{rp2}$ metrics between consecutive weekly proximity graphs are plotted in
Fig. \ref{mit}-top.  We again note that $d_{rp1}$ and $d_{rp2}$ appear to be within a constant factor
of one another.  Both metrics can predict events during the course of the academic year. A
substantial change between the first and second week of classes at the beginning of the fall
semester is likely representative of students sorting out their class schedules and friend groups.
The week after finals (the beginning of winter break) and the beginning of the independent
activities period are reflected by significant changes in the network, as measured by the $d_{rp1}$
and $d_{rp2}$ metrics.  The network also changes at the beginning and end of spring break, as
students depart from and return to their campus routine. For comparison, we present a similar analysis using the DeltaCon and $d_\text{CAD}$ distances in
Fig. \ref{mit}-bottom. Both distances appear to detect changes during the academic calendar (e.g.,
sponsor week, finals week, etc.) DeltaCon appears to be more stable then $d_\text{CAD}$.

  Because of the nature of the data, the RP distances can be used to confirm behavioral changes
associated with the geolocation of the different actors (nodes) of the network.  Unlike the Enron
dataset, the RP distance has no predictive value in this case, but can be used to grade the
significance of the changes in behavior: finals are more important than spring break, which is more
important than exam week.  Finals week appears to be more important than the beginning of the
semester.
%
\section{Discussion
  \label{discussion}}
We revisit the goal of the paper and confirm that the resistance perturbation distance RP-p
satisfies the axiom and principles laid out in section \ref{axioms_section}.
\subsection{Adherence to Axioms and Principles}
{\noindent \bfseries Axiom 1.} We have indeed proved in theorem \ref{distance_thm} that
all the RP-p distances were proper distances, and therefore this family of distances satisfies Axiom 1.\\

{\noindent \bfseries Principle 1: Edge Importance.} Remark \ref{target_drpo} in section
\ref{edge_mod_sec} proves that\\ $\drpo(G, G+\dw{i_0j_0}) \rightarrow \infty$ if and only if 
removing the edge $[i_0,j_0]$ disconnects the graph, thereby proving Principle 1.\\

{\noindent \bfseries Principle 2: Weight Awareness} As explained in Remark \ref{target_drpo} in
section \ref{edge_mod_sec}, as $w_{i_0j_0}\rightarrow \infty$, then
$A^{-1}_{i_0j_0} \approx R_{i_0j_0}$, and $1 - w_{i_0j_0} R_{i_0j_0} \approx 0$, leading the
distance $\drpo(G,G+\dw{i_0j_0})$ to go to infinity when the edge is removed, to wit
when $\dw{i_0j_0} = - w_{i_0j_0}$. The second principle of ``weight awareness'' is therefore clearly
satisfied: as the weight of the removed edge grows, the distance
grows to infinity.\\

{\noindent \bfseries Principle 3: Edge-``Submodularity''.} While we do not have a formal proof of
this principle, we can use the comparison of the complete graph (theorem \ref{complete_graph_thm})
with the star graph (theorem \ref{star_graph_thm}) to illustrate the scaling of the distance
$\drpo$.  The complete graph has $O(n^2)$ edges, and $\drpo(K_n,K_n+\dw{i_0j_0}) = O(1/n)$.
The star graph has $O(n)$ edges, and $\drpo(S_n,S_n+\dw{i_0j_0}) = O(n)$.  In this example,
changes made to a sparse graph are more important than equally sized changes made to a denser graph
with the same number of vertices.

  Instead of comparing a single-edge perturbation of graphs that have different topologies, and thus
  different densities, we can evaluate the perturbation of graphs that have the same topology, but
  different densities.

  We illustrate this concept with a stochastic block model composed of size $n=1,000$. The nodes are
  divided into two communities of size $n/2$. Every pair  of nodes $\{i,j\}$ forms an edge $[i,j]$ with probability $p$ if they belong to the same
community, and with probability $q$ if they belong to different communities. We fixed
$p=\log^2(n)/n$, and we increase $q$ from $\log(n)/n^2$ to $7p/n$.

  For each value of $q$ we generate 200 realizations of the model. For each realization $G$, we
  perturb a single edge, $e$. The edge is chosen at random within one of the two communities (within
  community perturbations), or chosen to be one of the cross-community edges (cross community
  perturbation).  We then compute the distances between $G$ and $G\backslash \{e\}$ -- $G$ with the
  edge $e$ removed.
  
\begin{figure}[H]
  \centerline{
    \includegraphics[width=0.8\textwidth]{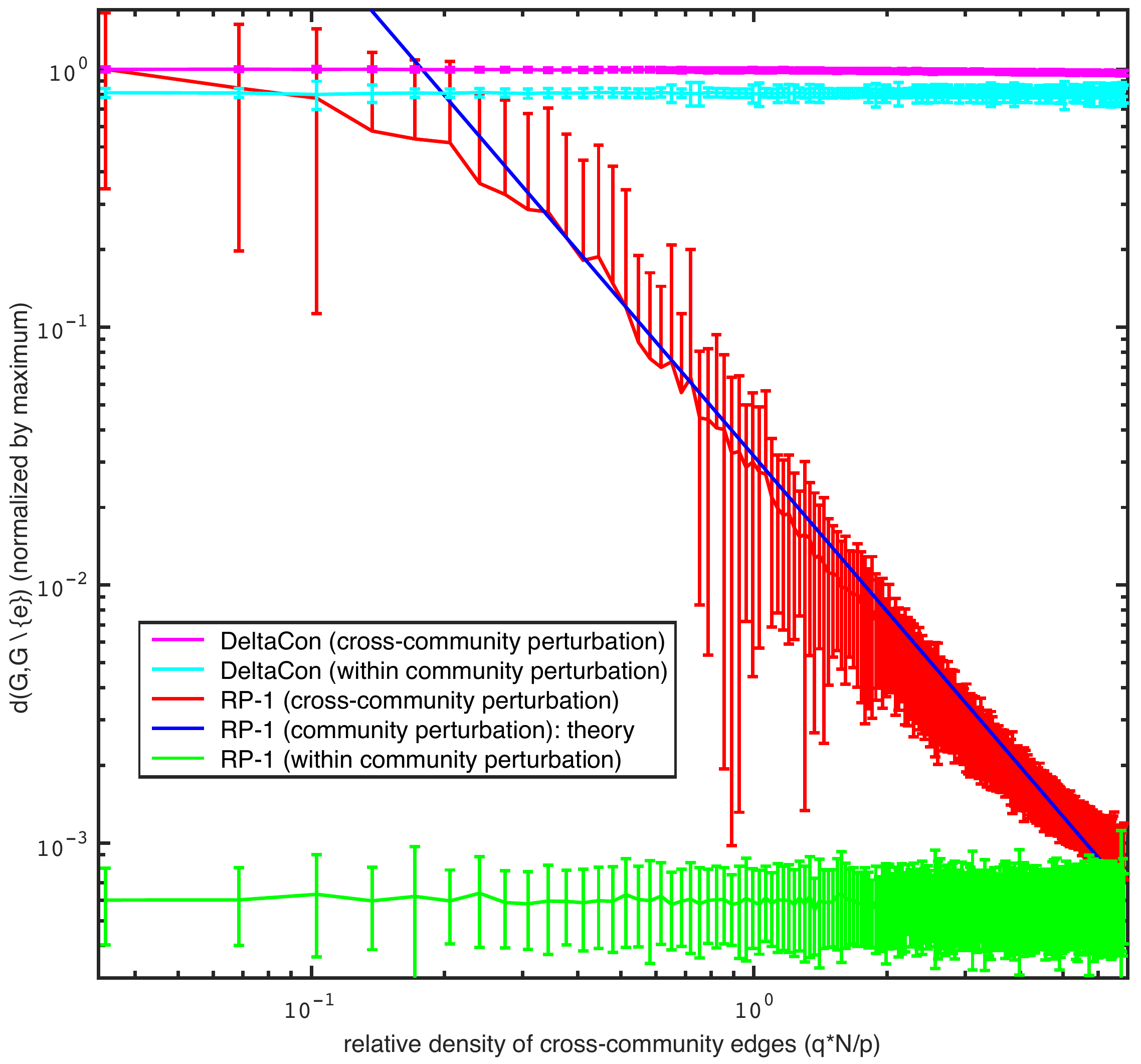}
  }
  \caption{The distances $\dcon{G}{G\backslash \{e\}}$ and
      $\drpo\left(G, G\backslash \{e\}\right)$ as a function of the probability $q$ of connecting
      two balanced communities of size $n/2 = 500$. Each time-series is normalized by its maximum
      value, and the error-bars display the standard deviation computed over 200 realizations. For
      each distance, the perturbed edge $e$ can either belong to one of the two communities (within
      community perturbation), or be one of the cross-community edges (cross community
      perturbation).
    \label{dc-rp-sbm}}
\end{figure}

  Figure \ref{dc-rp-sbm} displays the distances $\dcon{G}{G\backslash \{e\}}$ and
  $\drpo\left(G, G\backslash \{e\}\right)$ as a function of the probability $q$ of connecting the
  two balanced communities. Each of the distance time-series is normalized by its maximum value, and
  the error-bars display the standard deviations computed over 200 realizations. The blue line
  corresponds to the theoretical analysis of $\drpo\left(G, G\backslash \{e\}\right)$ performed in
  \cite{wills17a}, which corresponds to a power-law decay (note the logarithmic scale).

  We note that $d_{\DC_0}$ is sensitive to the type of edges that is being removed: the distance is
  larger for cross-community edges (see Fig. \ref{dc-rp-sbm} magenta). However, $d_{\DC_0}$ is
  independent of the increasing density of the graph.

  Similar to $d_{\DC_0}$, $\drpo$ can easily detect whether the deleted edge $e$ was removed from
  within a community, or was a cross-community edge. In contrast to $d_{\DC_0}$, the distance $\drpo$ is
  very sensitive to the density of edges in the graph. In agreement to principle 3, $\drpo$
  decreases as a function of the graph density.\\

{\noindent \bfseries Principle 4: Focus Awareness.}  The fourth principle, ``focus awareness'',
states that {\em random changes in graphs are less important than targeted changes of the same
  extent}. While the notion of targeted versus random changes would need to be defined more
precisely, we argue that remark \ref{block_drpo} in section \ref{edge_mod_sec} addresses this
principle. Indeed, in the example of a network formed by densely connected communities, which are
weakly connected to one another, $\drpo(G,G+\dw{i_0j_0})$ will be maximal if $i_0$ and $j_0$
are in different communities, for the same $\dw{i_0j_0}$. Because there are much fewer edges
bridging the communities, modifying the edge $[i_0,j_0]$,%
\noindent where $i_0$ and $j_0$ are in different communities, is indeed a targeted change.\\ 

We conclude that the resistance perturbation distance satisfies the axiom and principles (see
section \ref{axioms_section}) that a graph distance should obey. These principles were
inspired by the pioneering work of Koutra {\em et al.} \cite{koutra16}, where  the authors
compared the DeltaCon algorithm to vertex edge overlap \cite{papadimitriou2010}, the graph edit
distance \cite{bunke2007}, the signature similarity \cite{papadimitriou2010}, and three variations
of the $\lambda$-similarity \cite{bunke2007,peabody2002,wilson2008}.  The authors in
\cite{koutra16} show that DeltaCon is the only algorithm that adheres to their set of axioms and
principles.  In fact, our asymptotic analyses of the DeltaCon$_0$ similarity for the complete and star
graphs (\ref{DeltaCon_appendix}) demonstrates that this distance fails to meet Principle 3.
\subsection{Future Work}
\label{future_work_section}
The introduction of the resistance perturbation distance prompts several important research
questions.  A current limitation of the RP distance is its inability to measure distances between
disconnected graphs in a meaningful way, which stems from the fact that the effective resistance
between vertices in disconnected components of a graph is infinite.  Thus, it may prove valuable to
consider extensions of the resistance perturbation distance that accommodate disconnected graphs.
One option may be to define a distance based on some comparison of the conductance matrices.

A volume-normalized version of the distance may also be of interest.  In some applications, the user
might be more interested in the overall structure of the graph, and less interested in the magnitude
of the weights along the edges.  For example, if all edge weights are doubled between one graph and
another, this could be viewed as insignificant in some circumstances.  The precise implications of
such a normalization are a worthy direction for future research.

Many applications of the RP distance should be explored.

 In the context of dynamic graphs (see section \ref{experiments}), the $\drpp$ metric can be
  used to study the dynamic evolution of a graph sequence $\{G^{(n)}\}$, where $n$ denotes the
  time index of the corresponding element $G^{(n)}$ in the graph process. There has been some recent
  interest in the detection of anomalies in dynamic graphs
  \cite{akoglu14,lafond14,ranshous15}. Formally, one can construct a statistic $Z_n$, based on the
  distance $D_n = \drpp(G^{(n)}_n,G_{n+1})$ between $G^{(n)}$ and $G^{(n+1)}$, in order to test the
  hypothesis $H_0$ that the graphs $G^{(n)}$ and $G^{(n+1)}$ are {\em structurally the same} against
  the alternate hypothesis that $G^{(n)}$ and $G^{(n+1)}$ are {\em structurally different.} In this
  context, we accept $H_0$ if $Z_n < z_\varepsilon$ and accept $H_1$ otherwise. The threshold
  $z_\varepsilon$ for the rejection region satisfies
  \begin{equation}
    \proba_{H_0}\left(Z_n \ge z_\varepsilon\right) \leq \varepsilon \quad \text{as}\quad n \rightarrow \infty,
  \end{equation}
  and
  \begin{equation}
    \proba_{H_1}\left(Z_n \ge z_\varepsilon\right) \rightarrow 1 \quad \text{as}\quad n \rightarrow \infty.
  \end{equation}
  The test has therefore asymptotic level $\varepsilon$ and asymptotic power 1.  Our recent work
  \cite{wills17a} develops the construction of the statistic $Z_n$ in the context of a dynamic
  community network.

  Our results in section \ref{experiments} on random graph models, clearly show that one can
  quantify the normal random fluctuations of the metric $d_{rp(p)}$ using ensemble of random graphs,
  which defines a notion of normal baseline ``background'' noise to be expected when a graph does
  not experience significant configurational changes. Furthermore, both $\drpo$ and $\drpt$ can
  detect significant structural changes, such as changes in topology, connectivity, or ``disorder''.
  Formally, one can numerically estimate a $1-\varepsilon$ point wise confidence interval for the
  test statistic with a bootstrapping technique; the details of such a construction extend beyond
  the scope of the present report and are the subject of ongoing investigation \cite{wills17a}.

While the metric $d_{rp(p)}$ can provide insightful information about changes at many different
scales in the graph structure, it does not provide any localization about the anomalies. One could
study the problem of attribution of the anomaly. A multiscale approach, where the metric $d_{rp(p)}$
is computed between corresponding subgraphs of $G^{(2)}$ and $G^{(2)}$, could provide insight into
the localization of the metric changes.   Alternatively, one could try to localize the anomalous
  edges using the approach proposed in \cite{sricharan14}, and described in section
  \ref{existing_metrics_intro}.

Spielman and Srivastava \cite{spielman2008} introduced a method for generating sparse spectrally
similar graphs by sampling edges of the original graph according to the effective resistance between
endpoints of the edges.  This strategy suggests a meaningful connection between effective
resistances and spectral similarity.  Indeed, Batson {\em et al.} \cite{batson} observed that spectrally
similar graphs exhibit similar effective resistances between all pairs of vertices.  Improving our
understanding of potential connections between the spectral similarity and resistance perturbation
distance is an avenue of significant interest for future work.

In this work, we have presented and implemented a fast algorithm to compute $d_{rp2}$. This effort
leads to several questions. First, we observed that the computation time for the fast $d_{rp2}$
approximation algorithm is dominated by the $\widetilde{{\cal O}}(\log n)$ Laplacian linear solver
(each of size $n$).  Our current implementation utilizes the combinatorial multigrid solver of
Koutis {\em et al.} \cite{koutis2011}.  Although we observe linear scaling of the algorithm on several
scalable example problems, the constant hidden in the $\widetilde{{\cal O}}$ is unfortunately
significant.

One could explore competing algorithms for the Laplacian solver.  Lean Algebraic Multigrid (LAMG)
\cite{lamg_code,lamg_report} is a competing method for solving graph Laplacian linear systems in
linear time that may reduce the cost of approximating the $d_{rp2}$ metric.  Given the diversity of
structural features in graphs, an adaptive approach may be necessary to handle different types of
graphs efficiently.

Modern high-performance computing architectures demand the development of highly parallelizable
algorithms.  The structure of the $d_{rp2}$ approximation algorithm lends itself to natural
parallelism.  The most direct opportunity for parallelism involves splitting the ${\cal O}(\log n)$
independent calls to the Laplacian solver onto independent processors/cores.  Additionally,
depending on the choice of the Laplacian solver algorithm, each call to the solver could potentially
be parallelized.  A detailed investigation of such algorithmic improvements is an important avenue
for future work.
\section*{Acknowledgements}
The authors are grateful to the anonymous reviewers for their insightful comments and suggestions
that greatly improved the content and presentation of this manuscript.

NDM was supported in part under the auspices of the U.S. Department of Energy under 
grant numbers (SC) DE-FC02-03ER25574, Lawrence Livermore National Laboratory under 
contract B600360. We are very grateful to Tom Manteuffel and Geoff Sanders for supporting 
this work. NDM was also supported in part by NSF DMS 0941476. FGM was supported in part 
by NSF DMS 1407340. 

We are very grateful to Leto Peel for his help with the Enron dataset.
\section*{References}
\bibliographystyle{model4-names}
\bibliography{references.bib}
\clearpage
\section{Notation}	
\label{notation}
  \begin{center}
    \begin{small}
      \begin{tabular}{llr}
        \toprule
        \sf Symbol & \sf Definition & \sf (equation) \\
        \midrule
        $G$ & graph with vertex set $V$, edge set $E$,
                and edge weights $\bw$
                            & (\ref{adjacency_def_eqn}) \\   
        \midrule 
        $[i,j]$ & edge between nodes $i$ and $j$  
                            & (\ref{adjacency_def_eqn})\\ 
        \midrule
        $\bA$ & $n \times n$ adjacency matrix\\
               & $A_{ij} = w_{ij}$ if $i$ and $j$ are connected,  $0$ otherwise  & (\ref{adjacency_def_eqn}) \\     
        \midrule 
        $\bD$ & $n \times n$ diagonal matrix of vertex degrees, $D_{ii} = \sum_{j=1}^n A_{ij}$
                            & (\ref{combinatorial_lap_def_eqn}) \\     
        \midrule 
        $\bL$ & $n\times n$ combinatorial Laplacian matrix
                            & (\ref{combinatorial_lap_def_eqn}) \\    
        \midrule 
        $\bfi_k,\lambda_k$ & eigenvector and eigenvalue of $\bL$, with $0 = \lambda_1 \leq \ldots\leq
                             \lambda_n$ 
                            &    (\ref{spectral}) \\   
        \midrule 
        $\bL^\dagger$ & pseudoinverse of $\bL$ 
                            & (\ref{Ldagger_eqn}) \\   
        \midrule 
        $\bB$ & $m \times n$   edge incidence matrix 
                            & (\ref{edgeincidence}) \\  
        \midrule 
        $\dA$ & $m \times m$ diagonal edge weight matrix 
                            & (\ref{edgeincidence}) \\   
        \midrule 
        $d_\text{rootED}$ & root Euclidean distance 
                            & (\ref{root})\\   
        \midrule 
        ${\bS}^{(i)}$ & $n \times n$ fast belief propagation matrix 
                            & (\ref{belief}) \\     
        \midrule 
        $\be_i$ & $i$-th canonical basis vector $\be_i \in \R^n$ \\   
        \midrule 
        $\bR$ & $n \times n$ matrix of effective resistances
                            & (\ref{effective_resistance_matrix_eqn}) \\      
        \midrule 
        $\ki(G)$ & Kirchhoff index of $G$
                            & (\ref{kirchhoff_index_eqn}) \\  
        \midrule 
        $\kappa_{ij}$ & commute time between vertices $i$ and $j$
                            & (\ref{commute_time_eqn}) \\   
        \midrule 
        $\left\lVert \cdot \right\rVert_p$ & element-wise p-norm
                            &(\ref{rp_dist_def_eqn}) \\  
        \midrule 
        $d_{\rp(p)}$ & resistance perturbation distance
                            & (\ref{rp_dist_def_eqn})\\  
        \midrule 
        $\tZ$ & ${\cal O}(\log n) \times n$ embedding matrix
                            & (\ref{SS08_embedding_eqn}) \\ 
        \midrule 
        $\bQ$ & ${\cal O}(\log  n) \times m$ random projection matrix
        & Algorithm 1\\
        \midrule  
        $p$ & number of eigenvectors used\\
        & in the low-rank approximation of $\drpo$
                            & (\ref{low_rank_approx_bound_eqn}) \\ 
        \bottomrule
      \end{tabular}
    \end{small}
  \end{center}
\appendix
\section{DeltaCon$_0$ Analysis}	
\label{DeltaCon_appendix}
\subsection{Introduction}
In this section we compute analytically the DeltaCon$_0$ similarity for small perturbations of the complete
and the star graphs. We restrict our attention to simple perturbations, where $G^{(2)}$ is generated from
$G$ by a change in edge weight of size $\dw{i_0j_0}$ between vertices $i_0$ and $j_0$.

Our main tool is  the Sherman--Morrison--Woodburry theorem \cite{golan12}
that provides a closed form expression for the inverse of a low-rank perturbation of a non-singular
matrix. For completeness, we recall the Sherman--Morrison--Woodburry formula.

\noindent If $\bX$ is an $n\times n$ non singular matrix, and $\bU$ and $\bV$ are two $n \times k$
matrices, then $\bT = \bI + \bV^T\bX^{-1}\bU$ is non singular if and only if
\begin{equation}
  \bY = \bX + \bU\bV^T
\end{equation}
is non singular. Furthermore, when $\bY^{-1}$ exists, we have
\begin{equation}
  \left[\bX + \bU\bV^T\right]^{-1} 
  = \bX^{-1} - \bX^{-1}\bU\left[\bI + \bV^T\bX^{-1}\bU\right]^{-1}\bV^T\bX^{-1}.
  \label{SMW}
\end{equation}
In the following, we use this theorem with $k=1$ or $k=2$. When $k=1$, we have the
Sherman--Morrison formula,
\begin{equation}
  \left[\bX + \bu\bv^T\right]^{-1}
  = \bX^{-1} - \frac{1}{1 + \bv^T\bX^{-1}\bu}\bX^{-1}\bu\bv^T\bX^{-1}.
  \label{SM}
\end{equation}
For the purpose of writing concise equations, we extend the big $\cal O$ notation for matrices.
If $\widetilde{\bA}(n)$ and  $\bA(n)$ are two sequences of matrices that depend on $n$, then 
the notation
\begin{equation}
  \widetilde{\bA}(n) = \bA(n) + \O{1/n^q},
\end{equation}
means that 
\begin{equation}
  \forall i,j,\quad \widetilde{a}_{ij}(n) = a_{ij}(n)+ \O{1/n^q}.
\end{equation}
In other words, there exists a sequence of matrices $\bB(n)=\left[b_{ij}^{(n)}\right]$ such that,
\begin{equation}
  \exists \; c_2 > c_1 \ge 0, \quad \widetilde{\bA}(n) = \bA(n) + \bB(n),\quad  \text{and} \quad \forall
  n,\quad c_1 \leq n^q b_{ij}(n) \leq c_2.
\end{equation}
\subsection{Complete Graph}
\noindent In the case of the complete graph $G$ on $n$ vertices we have,
\begin{equation}
  \bA = \bJ - \bI, \quad 
  \bD = (n-1) \bI, \quad \text{and}\quad
  \varepsilon = \frac{1}{n}.
\end{equation}
Then, the  fast belief propagation matrix (\ref{belief}) is given by
\begin{equation}
  \bS = \left[\bI + \frac{n-1}{n^2}\bI - \frac{1}{n} \bJ +
    \frac{1}{n}\bI \right]^{-1} 
  = \left[\frac{n^2+2n-1}{n^2}\bI - \frac{1}{n} \bone\bone^T\right]^{-1}
\end{equation}
The Sherman--Morrison--Woodbury formula (\ref{SMW}) yields
\begin{equation}
  \bS = \frac{n^2}{n^2+2n -1} \left\{\bI - \frac{n}{2n-1}\bJ\right\}.
  \label{Scomplete}
\end{equation}
We now consider the perturbed graph $G^{(2)}$. Without loss of generality, we can assume that the
edge $[1,2]$ was modified, and thus $G^{(2)}$ is obtained by the change
$w_{12} \rightarrow w_{12} + \dw{12}$. The perturbed adjacency matrix $\bA^{(2)}$ is given by
\begin{equation}
  \bA^{(2)}\mspace{-6mu}
  = 
  \begin{bmatrix}
    1 & \mspace{-12mu} 1+ \dw{12} & 1 & \cdots\\
    1 + \dw{12} & 1 & 1 & \cdots\\
    1 & & \mspace{-12mu}\ddots& \\
    \vdots
  \end{bmatrix},
\end{equation}
and degree matrix $\bD^{(2)}$ is given by
\begin{equation}
  \bD^{(2)}\mspace{-6mu}
  = 
  \begin{bmatrix}
    n-1 +\dw{12} & &   &   & 0\\
    & \mspace{-50mu} n-1 + \dw{12} &  & &\\
    & &  \mspace{-70mu} n-1 &  &  \\
    &  & & \mspace{-50mu} \ddots  & \\
    0 &  & & & \mspace{-20mu} n-1
  \end{bmatrix}.
  \label{A2D2complete}
\end{equation}
The inverse of the fast belief propagation matrix of $G^{(2)}$  is given by
\begin{equation}
  \left[\bS^{(2)}\right]^{-1} = \bI + \varepsilon_2^2 \bD^{(2)} - \varepsilon_2 \bA^{(2)},
  \quad\text{with} \quad \varepsilon_2 = \frac{1}{n + \dw{12}},
\end{equation}
which simplifies to
\begin{equation}
\begin{split}
  \left[\bS^{(2)}\right]^{-1} \mspace{-20mu}  & = \frac{n^2 + 2n -(3\dw{12}+1)}{n^2}\bI 
      -\frac{1}{n}\left(1-\frac{\dw{12}}{n}+ \frac{\dw{12}^2}{n^2}\right)\bJ\\
 - &\frac{\dw{12}}{n}\left(1-\frac{\dw{12}}{n}\right)\left(\be_1\be_2^T + \be_2\be_1^T\right)
+ \frac{\dw{12}}{n^2}\left(\be_1\be_1^T + \be_2 \be_2^T\right) + \O{1/n^3}.
\end{split}
\end{equation}
We break $\left[\bS^{(2)}\right]^{-1}$ into two parts. First, we 
apply the Sherman-Morrison formula (\ref{SM}) to get the inverse of a rank-one perturbation of a
diagonal matrix,
\begin{equation}
  \begin{split}
    &\left[\frac{n^2 + 2n -(3\dw{12}+1)}{n^2}\bI 
      -\frac{1}{n}\left(1-\frac{\dw{12}}{n}+ \frac{\dw{12}}{n^2}\right)\bone\bone^T\right]^{-1}\\
     &= \frac {n^2}{n^2 + 2n -(3\dw{12}+1)} \bI + \frac{1}{2 + \dw{12}}
    \left(1 -\frac{3 +\dw{12}}{n(2 +\dw{12})}\right)\bJ 
    + \O{1/n^2}.
  \end{split}
\end{equation}
Then we add the rank-two perturbation,
$- \frac{\dw{12}}{n}\left(1-\frac{\dw{12}}{n}\right)\left(\be_1\be_2^T + \be_2\be_1^T\right)$, and
apply the Sherman--Morrison--Woodbury formula (\ref{SMW}) to arrive at
\begin{equation}
  \begin{split}
    \bS^{(2)} & = \frac {n^2}{n^2 + 2n -(3\dw{12}+1)} 
    \left(\bI + \frac{1}{2 + \dw{12}}\left(1 + \frac{1+ \dw{12}}{n(2 + \dw{12})} \right)\bJ\right) \\
    & + \frac{\dw{12}}{n (2 + \dw{12})^2} \left\lgroup
      2\bJ + (2 + \dw{12})\left(\bone (\be_1 + \be_2)^T + (\be_1 + \be_2)
        \bone ^T\right) \right.\\
    &  \mspace{150mu}\left. +(2+\dw{12})^2\left(\be_1\be_2^T + \be_2\be_1^T\right) \right\rgroup
    + \O{1/n^2}.
  \end{split}
\end{equation}
Which simplifies to 
\begin{equation}
  \begin{split}
    \bS^{(2)} \bck &= \frac {n^2}{n^2 + 2n -(3\dw{12}+1)} \bI  + 
    \frac{1}{2 + \dw{12}}\bck \left[1 +\frac{\dw{12} - 3}{n(2 +\dw{12})}\right]\bck \bJ \\
    &+ \frac{\dw{12}}{n}\bck \left[\be_1\be_2^T + \be_2\be_1^T\right]
     + \frac{\dw{12}}{n (2 + \dw{12})}\bck 
     \left(\bone [\be_1 + \be_2]^T + [\be_1 + \be_2]\bone ^T\right)\bck \\
    &+ \O{1/n^2}.
    \label{S2complete}
  \end{split}
\end{equation}
\noindent From (\ref{Scomplete}) we derive the following first order approximation of $\bS$
\begin{equation}
  \bS = \left(1 - \frac{2}{n}\right)\bI + \left(1 - \frac{3}{2n}\right)\frac{1}{2}\bJ + \O{1/n^2}
\end{equation}
We now proceed to compute the DeltaCon$_0$ similarity between $G$ and $G^{(2)}$. We need to estimate
the size of the terms $\sqrt{S_{ij}} - \sqrt{S^{(2)}_{ij}}$. Because we expect a linear growth of
the distance, we only need an approximation up to order 1.

\noindent We start with the off-diagonal entries. If $i\neq j$,  we have
\begin{equation}
  S_{ij} = \frac{1}{2} + \O{1/n}, \quad \text{and}\quad
  S^{(2)}_{ij}  = \frac{1}{2 + \dw{12}} + \O{1/n},
\end{equation}
from which we get
\begin{equation}
  \sqrt{S_{ij}}  - \sqrt{S^{(2)}_{ij}} = 
  \frac{1}{\sqrt{2}} - \frac{1}{\sqrt{2 + \dw{12}}} + \O{1/n}.
\end{equation}
And therefore,
\begin{equation}
  \left(\sqrt{S_{ij}}  - \sqrt{S^{(2)}_{ij}}\right)^2 =
  \left(\frac{1}{\sqrt{2}} - \frac{1}{\sqrt{2 + \dw{12}}}\right)^2 + \O{1/n}.
\end{equation}
Thus
\begin{equation}
  \sum_{i,j=1; i\neq j}^n \left(\sqrt{S_{ij}}  - \sqrt{S^{(2)}_{ij}}\right)^2 =
  n(n-1) \left(\frac{1}{\sqrt{2}} - \frac{1}{\sqrt{2 + \dw{12}}}\right)^2
+ \O{n}
\end{equation}
From which we conclude that
\begin{equation}
  \sum_{i,j=1; i\neq j}^n \left(\sqrt{S_{ij}}  - \sqrt{S^{(2)}_{ij}}\right)^2=
  \left(\frac{1}{\sqrt{2}} - \frac{1}{\sqrt{2 + \dw{12}}}\right)^2 n^2
  + \O{n}.
\end{equation}
A similar calculation shows that the terms on the diagonal only contribute to a linear term,
\begin{equation}
  \sum_{i=1}^n \left(\sqrt{S_{ii}}  - \sqrt{S^{(2)}_{ii}}\right)^2=
  \O{n},
\end{equation}
since there are only $n$ such terms and they have the same order as the off-diagonal terms.
Combining all the terms, and keeping only the highest order, we conclude that 
\begin{equation}
  \droot{G}{G^{(2)}} = \left\vert\frac{1}{\sqrt{2}} - \frac{1}{\sqrt{2 + \dw{12}}} \right\rvert n + \O{1}.
\end{equation}
Figure \ref{RP_vs_DC}-left confirms experimentally the linear growth of $\droot{G}{G^{(2)}}$, which implies the
decay of the DeltaCon$_0$ similarity, and contradicts Principle 3 from Koutra et
al. \cite{koutra16}, which asserts that {\em ``A specific change is more important in a graph with
  few edges than in a much denser, but equally sized graph.''} Indeed, one would expect that the
similarity between $G$ and $G + \dw{12}$ should increase with $n$, since the relative importance of
the edge perturbation $\dw{12}$ becomes negligible for large $n$.
\subsection{Star Graph}
We proceed with the analysis of $\droot{G}{G^{(2)}}$ in the case of the star graph. The indices of
the leaf nodes are $2,\ldots, n$, and the index of the hub is $1$. We follow the same sequence of
steps as in the complete graph. First, we compute the exact expression of the fast belief
propagation matrix (\ref{belief}), $\bS$, using the Sherman--Morrison--Woodbury formula
(\ref{SMW}). We then perturb a single edge, and we compute the fast belief propagation matrix of the
perturbed graph, $\bS^{(2)}$, using again the Sherman--Morrison--Woodbury formula. Our analysis will
be performed with precision $1/n^2$, since we expect that $\droot{G}{G^{(2)}}$ decays as a
function of $n$.

\noindent For the star graph we have,
\begin{equation}
  \bA  = \be_1 \bb^T + \bb \be_1^T =
  \begin{bmatrix} 
    0 & 1 & \cdots & 1 \\ 
    1 & 0 & \cdots & 0 \\ 
    \vdots & \vdots & \ddots & \vdots \\ 
    1 & 0 & \cdots & 0 
  \end{bmatrix}
  ,
  \quad
  \bD  = \bI + (n-2)\be_1 \be_1^T,
  \quad
  \text{and}\quad  \varepsilon = \frac{1}{n},
\end{equation}
where 
\begin{equation}
\bb = \bone - \be_1 = \begin{bmatrix} 0 & 1 & \cdots 1\end{bmatrix}^T.
\end{equation}
Then, the fast belief propagation matrix (\ref{belief}) is given by
\begin{equation}
  \begin{split}
    \bS &= \left[\bI + \frac{1}{n^2}\left(\bI + (n-2)\be_1\be_1^T\right)
      - \frac{1}{n}\left(\be_1 \bb^T + \bb \be_1^T\right)\right]^{-1} \\
    & = \left[\frac{n^2+1}{n^2}\bI + \frac{n-2}{n^2}\be_1\be_1^T -\frac{1}{n}
      \left(\be_1\bb^T + \bb\be_1^T\right)
    \right]^{-1}
  \end{split}
  \label{Sstar}
\end{equation}
Now,
\begin{equation}
  \bSm = \frac{n^2+1}{n^2}\bI + \frac{n-2}{n^2}\be_1\be_1^T
\end{equation}
is diagonal matrix, and its inverse is the following diagonal matrix
\begin{equation}
\begin{split}
  \bSm^{-1} & = \frac{n^2}{n^2+1}\bI - \frac{n^2(n-2)}{(n^2+1)(n^2+n-1)}\be_1\be_1^T\\
  &= 
  \begin{bmatrix}
    \frac{n^2}{n^2 + n-1} & 0 & \cdots & 0\\
    0 & \frac{n^2}{n^2+1} &  & \vdots\\
    \vdots &  & \ddots & 0\\
    0 & \cdots & 0 & \frac{n^2}{n^2+1}
  \end{bmatrix}.
\end{split}
\end{equation}
We then get $\bS$ using a rank-two perturbation of $\bSm^{-1}$, and we use the
Sherman--Morrison--Woodbury formula to compute the corresponding inverse. We have
\begin{equation}
  \begin{split}
    \bS & = \left[\bSm - \frac{1}{n}\left(\be_1\bb^T + \bb\be_1^T\right)\right]^{-1}  
    = \left[\bSm - \bU\bV^T\right]^{-1}\\
    &  = \bSm^{-1} + \bSm^{-1}\bU\left[\bI -  \bV^T\bSm^{-1}\bU\right]^{-1}\bV^T\bSm^{-1}.
  \end{split}
\end{equation}
where 
\begin{equation}
  \bU
  =\frac{1}{n}
  \begin{bmatrix}
    \be_1 & \bb
  \end{bmatrix}
  =\frac{1}{n}
  \begin{bmatrix}
    1 & 0\\
    0 & 1\\
    \vdots & \vdots\\
    0 & 1
  \end{bmatrix}
  ,
  \quad \text{and} \quad
  \bV
  =
  \begin{bmatrix}
    \bb & \be_1
  \end{bmatrix}
  =
  \begin{bmatrix}
    0 & 1\\
    1 & 0\\
    \vdots & \vdots\\
    1 & 0
  \end{bmatrix}
  .
  \label{UandV}
\end{equation}
We have
\begin{equation}
  \left[\bI -  \bV^T\bSm^{-1}\bU\right]^{-1} = 
  \frac{(n^2+1)(n^2 + n -1)}{n^4+n^2+n-1}
  \begin{bmatrix}
    1 & \frac{n(n-1)}{n^2+1}\\
    \frac{n}{n^2+n-1} & 1
  \end{bmatrix},
\end{equation}
and
\begin{equation}
  \bSm^{-1}\bU=
  \begin{bmatrix} 
    \frac{n^2}{n^2 + n-1}& 0\\
    0 & \frac{n^2}{n^2+1}\\
    \vdots & \vdots\\
    0 & \frac{n^2}{n^2+1}
  \end{bmatrix}
  \quad \text{and} \quad
  \bV^T
  \bSm^{-1}=
  \begin{bmatrix} 
    0 & \frac{n^2}{n^2+1} & \cdots & \frac{n^2}{n^2+1}\\
    \frac{n^2}{n^2 + n-1}& 0 & \cdots & 0
  \end{bmatrix}
\end{equation}
Combining all the terms, and after some elementary calculations, we obtain 
\begin{equation}
  \begin{split}
    \bS  = \bSm^{-1} & + \frac{n^3}{n^4 + n^2 + n -1}
    \left(\be_1\bb^T + \bb\be_1^T\right)\\
    & + \frac{n^4}{n^6+2n^4+n^3+n-1}\bb\bb^T\\
    & + \frac{n^5-n^4}{n^6+n^5+2n^3}\be_1\be_1^T.
  \end{split}
\end{equation}
Therefore, we have the following approximation of order $1/n^2$ of $\bS$,
\begin{equation}
  \bS  = \frac{n^2-1}{n^2} \bI  + \frac{1}{n} \left(\be_1\bb^T + \bb\be_1^T\right)
  + \frac{1}{n^2}\bb\bb^T + \frac{1}{n^2}\be_1\be_1^T + \O{1/n^3},
  \label{belief_star}
\end{equation}
or
\begin{equation}
  \bS  = 
  \begin{bmatrix}
    1 & \frac{1}{n} & \frac{1}{n} &  \cdots & \frac{1}{n}\\
    \frac{1}{n} & 1 & \frac{1}{n^2} & \cdots & \frac{1}{n^2}\\
    \frac{1}{n} & \frac{1}{n^2} &  1 &  & \frac{1}{n^2}\\
    \vdots      & \vdots &  & \ddots & \vdots\\
    \frac{1}{n} & \frac{1}{n^2}  & \cdots & \frac{1}{n^2} &  1 \\
  \end{bmatrix}
  + \O{1/n^3}.
\end{equation}
We now consider the perturbed graph $G^{(2)}$ created by adding an edge between two leaves. The
perturbation created by modifying the weight of an edge connecting the hub (1) to a leave yields the
exact same asymptotic for $\droot{G}{G^{(2)}}$, and for the sake of conciseness is not displayed here.

Without loss of generality, we can assume that the edge $[2,3]$ was modified, and thus $G^{(2)}$ is
obtained by the change $w_{23} \rightarrow w_{23} + \dw{23}$. The perturbed adjacency matrix
$\bA^{(2)}$ and degree matrix $\bD^{(2)}$ are given by
\begin{equation}
  \bA^{(2)}\mspace{-6mu}
  = 
  \bA + \dw{23} \left(\be_2 \be_3^T + \be_3\be_2^T\right),
  \bD^{(2)}\mspace{-6mu}
  = 
  \bD + \dw{23} \left(\be_2 \be_2^T + \be_3\be_3^T\right).
  \label{A2D2star}
\end{equation}
The inverse of the fast belief propagation matrix of $G^{(2)}$  is given by
\begin{equation}
  \left[\bS^{(2)}\right]^{-1} = \bI + \varepsilon_2^2 \bD^{(2)} - \varepsilon_2 \bA^{(2)},
  \quad\text{with} \quad \varepsilon_2 = \frac{1}{n}.
\end{equation}
$\left[\bS^{(2)}\right]^{-1}$ can be expressed as a low-rank perturbation of $\bS^{-1}$,
\begin{equation}
  \left[\bS^{(2)}\right]^{-1} 
  \mspace{-18mu}= \bS^{-1} 
  -\frac{\dw{23}}{n} \left(\be_2 \be_3^T + \be_3\be_2^T\right) 
  + \frac{\dw{23}}{n^2}\left(\be_2 \be_2^T + \be_3\be_3^T\right).
\end{equation}
We break $\bS^{(2)}$ into two parts,
\begin{equation}
  \bS^{(2)}= \bS + \Delta\bS_i+ \Delta\bS_{ii},
\end{equation}
with
\begin{equation}
  \Delta\bS_i = 
  \left[\bS^{-1} - \frac{\dw{23}}{n} \left(\be_2 \be_3^T + \be_3\be_2^T\right)\right]^{-1} 
  - 
  \bS, 
  \label{DSi}
\end{equation}
and
\begin{equation}
  \Delta\bS_{ii}  = \bS^{(2)}
  - \left[\bS^{-1} - \frac{\dw{23}}{n} \left(\be_2 \be_3^T + \be_3\be_2^T\right)\right]^{-1}. 
\end{equation}
Since
$\bS+ \Delta \bS_i = \left[\bS^{-1} -\frac{\dw{23}}{n} \left(\be_2 \be_3^T + \be_3\be_2^T\right)
\right]^{-1}$,
we apply the Sherman--Morrison--Woodbury formula to calculate $\Delta \bS_i$.  The calculation is
very similar to the computation of $\bS$. For the sake of brevity, we only give the important
steps.\\

\noindent Using the same $\bU$ and $\bV$ as defined in (\ref{UandV}), we have
\begin{equation}
  \begin{split}
    \left[\bS^{-1} 
      -\frac{\dw{23}}{n} \left(\be_2 \be_3^T + \be_3\be_2^T\right) 
    \right]^{-1}\mspace{-16mu} & = \left[\bS^{-1} -  \dw{23}\bU\bV^T\right]^{-1}\\
    = \bS + \dw{23}\bS\bU& \left[\bI -  \dw{23}\bV^T\bS\bU\right]^{-1}\bV^T\bS.
  \end{split}
\end{equation}
Injecting the expression for $\bS$ given by (\ref{belief_star}), and after some elementary
calculations, we get 
\begin{equation}
  \begin{split}
    \Delta \bS_i & = \dw{23}\bS\bU\left[\bI -  \dw{23}\bV^T\bS\bU\right]^{-1}\bck\bV^T\bS\\
    & =
    \frac{\dw{23}}{n} 
    \begin{bmatrix}
      0 & 1/n & 1/n & 0 & \cdots\\
      1/n & \dw{23}/n & 1  & 0 & \cdots\\
      1/n & 1  & \dw{23}/n  & 0 & \cdots\\
      0 & 0 & 0 &  & \\
      \vdots & \vdots & \vdots &  &  \\
    \end{bmatrix}
    + \O{1/n^3}.
  \end{split}
\end{equation}
We now carry on with the estimation of $\Delta\bS_{ii}$. We have
\begin{equation}
  \begin{split}
    \bS^{(2)}& = 
    \left[
      \bS^{-1} 
      -\frac{\dw{23}}{n} \left(\be_2 \be_3^T + \be_3\be_2^T\right) 
      + \frac{\dw{23}}{n^2}\left(\be_2 \be_2^T + \be_3\be_3^T\right).
    \right]^{-1}\\
    & =
    \left[
      \bS^{-1} 
      -\frac{\dw{23}}{n} \left(\be_2 \be_3^T + \be_3\be_2^T\right) 
      + \frac{\dw{23}}{n^2} \bU_{23}\bU_{23}^T \right]^{-1},
  \end{split}
\end{equation}
where
\begin{equation}
  \bU_{23}
  =
  \begin{bmatrix}
    \be_2 & \be_3
  \end{bmatrix}
  .
\end{equation}
We recall that
\begin{equation}
\bS^{-1} -\frac{\dw{23}}{n} \left(\be_2 \be_3^T + \be_3\be_2^T\right) = 
\left[\bS + \Delta \bS_i\right]^{-1},
\end{equation}
and thus $\left[\bS^{(2)}\right]^{-1}$ is a rank-two perturbation of $\left[\bS + \Delta \bS_i\right]^{-1}$.
The Sherman--Morrison--Woodbury formula yields
\begin{equation}
  \bS^{(2)}  = \bS + \Delta \bS_i
  - \frac{\dw{23}}{n^2} \mspace{-6mu}\left[\bS +\Delta \bS_i \right] \bck \bU_{23}\bck
    \left[
      \bI + \frac{\dw{23}}{n^2}\bU_{23}^T\left[\bS +\Delta \bS_i \right]
    \right]^{-1} \mspace{-18mu}\bU_{23}^T
    \bck \left[\bS +\Delta \bS_i \right].
\end{equation}
Therefore,
\begin{equation}
  \Delta \bS_{ii} = 
  -\frac{\dw{23}}{n^2} \mspace{-6mu}\left[\bS +\Delta \bS_i \right] \bck \bU_{23}\bck
  \left[
    \bI + \frac{\dw{23}}{n^2}\bU_{23}^T\left[\bS +\Delta \bS_i \right]\bU_{23}
  \right]^{-1} \mspace{-18mu}\bU_{23}^T
  \bck \left[\bS +\Delta \bS_i \right].
  \label{DSii}
\end{equation}
We only want to recover the terms of order $1/n^2$. We can thus neglect $\Delta \bS_i $ in
(\ref{DSii}), since its contribution only creates terms of size $\O{1/n^3}$. As a result,
\begin{equation}
  \Delta \bS_{ii} = 
  -\frac{\dw{23}}{n^2}\bS \bU_{23}
  \left[
    \bI + \frac{\dw{23}}{n^2}\bU_{23}^T\bS\bU_{23}
  \right]^{-1} \mspace{-12mu}\bU_{23}^T \bS + \O{1/n^3}.
\end{equation}
A simple calculation yields,
\begin{equation}
  \Delta \bS_{ii}  = - \frac{\dw{23}}{n^2}\left(\be_2\be_2^T + \be_3\be3^T\right) = - \frac{\dw{23}}{n^2} 
  \begin{bmatrix}
    0 & 0 & 0 & 0 & \cdots\\
    0 & 1 & 0  & 0 & \cdots\\
    0 & 0  & 1  & 0 & \cdots\\
    0 & 0 & 0 &  & \\
    \vdots & \vdots & \vdots &  &  \\
  \end{bmatrix}
  + \O{1/n^3}.
\end{equation}
Finally, we advance to the computation of the DeltaCon$_0$ similarity between $G$ and $G^{(2)}$. We
need to estimate the size of the terms $\sqrt{S_{ij}} - \sqrt{S^{(2)}_{ij}}$. 

\noindent We start with the two non-zero entries on the diagonal of $\Delta S_{ii}$. If $i=2,3$,
\begin{equation}
  S_{ii} = 1 + \O{1/n^3}, \quad \text{and}\quad
  S^{(2)}_{ii}  =1 +  \frac{\dw{23}(\dw{23} -1)}{2n^2} + \O{1/n^3},
\end{equation}
from which we get
\begin{equation}
  \left(\sqrt{S_{ii}}  - \sqrt{S^{(2)}_{ii}}\right)^2 = 
  \frac{\dw{23}^2(\dw{23} -1)^2}{4n^4} + \O{1/n^5}
\end{equation}
And therefore,
\begin{equation}
  \sum_{i=2,3}\left(\sqrt{S_{ii}}  - \sqrt{S^{(2)}_{ii}}\right)^2 =
  \frac{\dw{23}^2(\dw{23} -1)^2}{2n^4} + \O{1/n^5}
\end{equation}
A similar calculation shows that the four terms on the first row and first column contribute to
\begin{equation}
  \sum_{(i,j) \in \{(1,2),(1,3),(2,1),(3,1)\}}\left(\sqrt{S_{ii}}  - \sqrt{S^{(2)}_{ii}}\right)^2 =
  \frac{\dw{23}^2}{n^3} + \O{1/n^4}.
\end{equation}
Finally, we consider the off-diagonal terms for $(i,j) \in \{(2,3), (3,2)\}$,
\begin{equation}
  \sum_{(i,j) \in \{(2,3),(3,2)\}}\left(\sqrt{S_{ii}}  - \sqrt{S^{(2)}_{ii}}\right)^2 =
  \frac{2\dw{23}}{n}\left( 1 - \frac{2}{\sqrt{\dw{23} n}} + \O{1/n}\right).
\end{equation}
Combining all the terms, and keeping only the highest order, we get
\begin{equation}
  \sum_{i,j=1}^n\left(\sqrt{S_{ii}}  - \sqrt{S^{(2)}_{ii}}\right)^2 =
  \frac{2\dw{23}}{n}\left( 1 - \frac{2}{\sqrt{\dw{23} n}} + \O{1/n}\right),
\end{equation}
and thus
\begin{equation}
  \droot{G}{G^{(2)}} = \frac{\sqrt{2\dw{23}}}{\sqrt{n}}- \frac{\sqrt{2}}{n} + \O{1/n^{3/2}}.
\end{equation}
Figure \ref{RP_vs_DC}-left confirms experimentally the decay of $\droot{G}{G^{(2)}}$, which implies the
growth of the DeltaCon$_0$ similarity.
\section{Proofs}	
\label{main_proofs_appendix}
\subsection{Proof of Theorem \ref{edge_mod_thm}}
\label{edge_mod_thm_proof}
\noindent Let $\bL(w + \dw{i_0j_0})$ and $\bL^\dagger(w + \dw{i_0j_0})$ denote the
Laplacian and pseudo-inverse of the Laplacian of the graph after modifying the edge $[i_0~j_0]$,
from $w_{i_0j_0}$ to $w_{i_0j_0} + \dw{i_0j_0}$, respectively. We first observe that we can
apply the Sherman--Morrison theorem to the perturbed pseudo-inverse
$\bL^\dagger(w + \dw{i_0j_0})$. Indeed, using (\ref{Lpseudo}), we have
\begin{equation}
  \bL^\dagger(w + \dw{i_0j_0}) = \left(\bL(w + \dw{i_0j_0})+ \frac{1}{n}\bJ\right)^{-1} - \frac{1}{n}\bJ.
\end{equation}
But $\bL(w + \dw{i_0j_0})$ is a simple rank-one modification of $\bL(w)$,
\begin{equation}
  \bL(w + \dw{i_0j_0}) = \bL(w) -\dw{i_0j_0} \bnabla_{i_0j_0}\bnabla^T_{i_0j_0},
\end{equation}
where $\bnabla_{i_0j_0}$ is $n$-dimensional column vector, with entries given by
\begin{equation}
  \nabla_{i_0j_0}(i) =
  \begin{cases}
    1 & \text{if} \quad i = i_0,\\
    -1 & \text{if} \quad i = j_0,\\
    0 & \text{otherwise.}
  \end{cases}
\end{equation}
From Sherman--Morrison, we have 
\begin{equation}
  \begin{split}
    \left(\bL(w + \dw{i_0j_0})+ \frac{1}{n}\bJ\right)^{-1} 
    &= \left(\bL(w )+ \frac{1}{n}\bJ\right)^{-1} \\
    + \dw{i_0j_0} & \frac{
      \left(\bL(w ) + \frac{1}{n}\bJ\right)^{-1} 
      \bnabla_{i_0j_0}\bnabla_{i_0j_0}^T
      \left(\bL(w ) + \frac{1}{n}\bJ\right)^{-1} 
    }
    {1 + 
      \dw{i_0j_0}
      \bnabla_{i_0j_0}^T
      \left(\bL(w ) + \frac{1}{n}\bJ\right)^{-1} 
      \bnabla_{i_0j_0}
    }.
  \end{split}
\end{equation}
Now, $\bJ \bnabla_{i_0j_0} = \bzero$, and therefore
\begin{equation}
  \begin{split}
    \left(\bL(w ) + \frac{1}{n}\bJ\right)^{-1} 
    \mspace{-24mu}\bnabla_{i_0j_0}\bck\bnabla_{i_0j_0}^T
    \bck
    \left(\bL(w ) + \frac{1}{n}\bJ\right)^{-1} 
    & = \bL^\dagger
    \bnabla_{i_0j_0}\bck\bnabla_{i_0j_0}^T
    \bL^\dagger\\
    & = \bL^\dagger \bnabla_{i_0j_0}
    \left[
      \bL^\dagger
      \bnabla_{i_0j_0}
    \right]^T\bck,
  \end{split}
\end{equation}
since $\bL^\dagger$ is symmetric. The entry $i,j$ of the matrix
$\bL^\dagger \bnabla_{i_0j_0} \left[\bL^\dagger\bnabla_{i_0j_0}\right]^T$ can be found to be 
\begin{equation}
  \bL^\dagger \bnabla_{i_0j_0} \left[ \bL^\dagger \bnabla_{i_0j_0}\right]^T_{ij}
  = \left[L^\dagger_{ii_0} -L^\dagger_{ji_0} +L^\dagger_{jj_0} -L^\dagger_{ij_0} \right]^2.
\end{equation}
Using (\ref{effective_resistances_formula}), we have
\begin{equation}
  L^\dagger_{ii_0} -L^\dagger_{ji_0} +L^\dagger_{jj_0} -L^\dagger_{ij_0} 
  = - \frac{1}{2} \left[R_{i i_0}+R_{j j_0}-R_{i j_0}-R_{j i_0}\right].
  \label{r2l}
\end{equation}
We also have
\begin{equation}
  \bnabla_{i_0j_0}^T \left(\bL(w ) + \frac{1}{n}\bJ\right)^{-1} 
  \mspace{-20mu}\bnabla_{i_0j_0}
  = \bnabla_{i_0j_0}^T \bL^\dagger \bnabla_{i_0j_0} =  R_{i_0j_0}.
\end{equation}
We conclude that the change in effective resistance between vertices $i$ and $j$, $\Delta R_{ij}$,
resulting from a change in edge weight $\dw{i_0j_0}$ between vertices $i_0$ and $j_0$ is given
by
\begin{equation}
  \Delta R_{ij} = - \frac{\dw{i_0j_0} \left(R_{i i_0}+R_{j j_0}-R_{i j_0}-R_{j i_0}
    \right)^2}{4 ( 1 + \dw{i_0j_0} R_{i_0j_0})}. 
  \label{delta_Rij_eqn}
\end{equation}
We now proceed to compute $\drpo(G,G+\dw{i_0j_0})$ by summing the entries
in the numerator of (\ref{delta_Rij_eqn}). In fact, we come back to (\ref{r2l})
and compute
\begin{equation}
  \sum_{i,j=1}^n \left[L^\dagger_{ii_0} -L^\dagger_{ji_0} +L^\dagger_{jj_0}
    -L^\dagger_{ij_0} \right]^2. 
\end{equation}
We use the spectral decomposition of $L^\dagger$ given by (\ref{Ldagger_eqn}) to
express
\begin{equation}
  L^\dagger_{ii_0} -L^\dagger_{ji_0} +L^\dagger_{jj_0} -L^\dagger_{ij_0}  = 
  \sum_{k=2}^n \frac{1}{\lambda_k}
  \left[\bfi_k(i) - \bfi_k(j)\right]\left[\bfi_k(i_0) - \bfi_k(j_0)\right],
  \label{c11}
\end{equation}
so that
\begin{equation}
  \begin{split}
    \sum_{i,j=1}^n[L^\dagger_{ii_0} -L^\dagger_{ji_0} & +L^\dagger_{jj_0} -L^\dagger_{ij_0}]^2   = \\
    & \sum_{i=1}^n \sum_{j=1}^n 
    \left\{
      \sum_{k=2}^n \frac{1}{\lambda_k}
      \left[\bfi_k(i) - \bfi_k(j)\right]\left[\bfi_k(i_0) - \bfi_k(j_0)\right]
    \right\}^2.
  \end{split}
\end{equation}
The above equation can be written as 
\begin{equation}
  \begin{split}
    \sum_{k=2}^n \sum_{l=2}^n 
    \frac{1}{\lambda_k} \frac{1}{\lambda_l}
    \left[\bfi_k(i_0) - \bfi_k(j_0)\right]&
    \left[\bfi_l(i_0) - \bfi_l(j_0)\right]\\
    & \left\{\sum_{i=1}^n \sum_{j=1}^n 
      \left[\bfi_k(i) - \bfi_k(j)\right]
      \left[\bfi_l(i) - \bfi_l(j)\right]
    \right\}.
  \end{split}
\end{equation}
Now, 
\begin{equation}
  \begin{split}
    \sum_{i=1}^n \sum_{j=1}^n 
    \left[\bfi_k(i) - \bfi_k(j)\right] \left[\bfi_l(i) - \bfi_l(j)\right] 
    &= \sum_{i=1}^n \sum_{j=1}^n \bfi_k(i)\bfi_l(i) + \sum_{j=1}^n \sum_{j=1}^n \bfi_k(j)\bfi_l(j) \\
    & - \sum_{i=1}^n \sum_{j=1}^n \bfi_k(i)\bfi_l(j) - \sum_{i=1}^n \sum_{j=1}^n \bfi_k(j)\bfi_l(i) \\
    & = 2n \delta_{kl},
  \end{split}
\end{equation}
where we have used $\langle \bfi_k,1\rangle = 0, k =2,\ldots, n$, and
$\langle \bfi_k,\bfi_l\rangle = \delta_{kl}, k,l =1,\ldots, n$. We conclude that 
\begin{equation}
  \sum_{i,j=1}^n[L^\dagger_{ii_0} -L^\dagger_{ji_0} +L^\dagger_{jj_0} -L^\dagger_{ij_0}]^2  = 
  2 n \sum_{k=2}^n  \frac{1}{\lambda^2_k} 
  \left[\bfi_k(i_0) - \bfi_k(j_0)\right]^2.
  \label{c15}
\end{equation}
Finally, applying (\ref{c11}) with $i=i_0$ and $j=j_0$, we get
\begin{equation}
  R_{i_0j_0} = \sum_{k=2}^n \frac{1}{\lambda_k} \left[\bfi_k(i_0) - \bfi_k(j_0)\right]^2,
\end{equation}
which provides the denominator of (\ref{delta_Rij_eqn}).
Substituting (\ref{c15}) into (\ref{delta_Rij_eqn}) completes the proof of theorem \ref{edge_mod_thm}.
\subsection{Proof of Theorem \ref{complete_graph_thm}}
\label{complete_graph_proof}
\noindent The spectrum of the Laplacian of the complete graph is given by
\begin{equation}
  \lambda_i = 
  \begin{cases}
    0 & \text{if \; $i = 1$} \\
    n & \text{otherwise}.
  \end{cases}
\end{equation}
The first eigenvector is $\bfi_1 = \bone$.  The remaining eigenvectors, $\bfi_2,\ldots,\bfi_n$, form
an orthonormal basis for $\text{span}(\bone)^\perp$. Without loss of generality we assume $i_0=1$
and $j_0=2$, and let
$\bfi_2 = \begin{bmatrix} \frac{1}{\sqrt{2}} & -\frac{1}{\sqrt{2}} & 0 & \cdots &
  0 \end{bmatrix}^T$.
We construct the remaining eigenvectors, $\bfi_3,\ldots,\bfi_n$, by Gram-Schmidt on
$\be_3,\ldots,\be_n$.  By observing that $\be_i^T \bfi_2=0$ for $i=3,\ldots,n$, we note that
$\bfi_i (i_0)-\bfi_i(j_0) = 0$ for $i = 3,\ldots,n$.  Thus, using the result of theorem
\ref{edge_mod_thm},
\begin{equation}
  \begin{split}
    \drpo(K_n,K_n+\dw{i_0j_0}) & =  
    \frac{  
      2 n \left\lvert \dw{i_0j_0} \right\rvert  \sum_{k=2}^n {(\phi_k(i_0) - \phi_k(j_0) )^2}/{\lambda_k^2} 
    }{  
      1 + \dw{i_0j_0} \sum_{k=2}^n {(\phi_k(i_0)-\phi_k(j_0))^2}/{\lambda_k} 
    } \\
    & = \frac{  2 n \left\lvert \dw{i_0j_0} \right\rvert  {(\phi_{2r} -
        \phi_{2t} )^2}/{\lambda_2^2} }{  1 + \dw{i_0j_0}
      {(\phi_{2r}-\phi_{2t})^2}/{\lambda_2} }  \\ 
    & =  \frac{ 2 n \left\lvert \dw{i_0j_0} \right\rvert \left( {2}/{n^2}
      \right)}{1 + \dw{i_0j_0} \left( {2}/{n} \right) }  
    =  \frac{ 4 \left\lvert \dw{i_0j_0} \right\rvert }{n + 2 \dw{i_0j_0}
    },
  \end{split}
\end{equation}
which proves the result. \hfill $\square$
\subsection{Proof of Theorem \ref{star_graph_thm}}
\label{star_graph_proof}
The simplicity of the star graph allows us to employ simple resistance network reduction techniques
to compute the change in effective resistances between each pair of vertices.  For the first case
(leaf to hub) we assume without loss of generality that $i_0=2$.  In this case, we maintain a tree
structure, and as a result $\Delta R_{ij} = 0$ whenever $i \neq 2$ and $j \neq 2$.  In addition,
every direct path between vertex $i_0=2$ and another leaf passes through the hub (vertex 1) which provides
additional simplification:
\begin{equation}
  \begin{split}
    \drpo(S_n,S_n+\dw{1i_0}) & =  \sum_{i,j = 1}^n \left\lvert \Delta R_{ij} \right\rvert  =  2 (n-1) \left\lvert
      \Delta R_{12} \right\rvert \\ 
    & =  2 (n-1) \left\lvert 1 - \frac{1}{1+\dw{12}} \right\rvert  =  \frac{2 (n-1) \left\lvert \dw{1i_0}
      \right\rvert}{1 + \dw{1i_0}}. 
  \end{split}
\end{equation}
In the second case (connecting two leaves), we assume without loss of generality
that $i_0=2$ and $j_0=3$, and we note that $\dw{i_0j_0} \ge 0$. We have
\begin{equation}
  \begin{split}
    \drpo(S_n,S_n+\dw{i_0j_0}) & =  \sum_{i,j = 1}^n \left\lvert \Delta R_{ij} \right\rvert  =  2 \sum_{j \neq
      3} \left\lvert \Delta R_{2j} \right\rvert + 2 \sum_{j \neq 2} \left\lvert \Delta R_{3j} \right\rvert + 2 \left\lvert
      \Delta R_{23} \right\rvert \\ 
    & =  4 \sum_{j \neq 3} \left\lvert \Delta R_{2j} \right\rvert + 2 \left\lvert \Delta R_{23} \right\rvert  =  4 (n-2)
    \left\lvert \Delta R_{21} \right\rvert + 2 \left\lvert \Delta R_{23} \right\rvert. 
  \end{split}
\end{equation}
Simple circuit reduction techniques yield
\begin{equation}
  \left\lvert \Delta R_{21} \right\rvert = \frac{\dw{i_0j_0} }{1 + 2 \dw{i_0j_0}},
  \quad \text{and} \quad
  \left\lvert \Delta R_{23} \right\rvert = \frac{4 \dw{i_0j_0}}{1+2 \dw{i_0j_0}},
\end{equation}
which leads to
\begin{equation}
  \drpo(S_n,S_n+\dw{i_0j_0}) = \frac{ 4 n  \dw{i_0j_0} }{1 + 2 \dw{i_0j_0}},
\end{equation}
as announced. \hfill $\square$
\subsection{Proof of Theorem \ref{path_graph_thm}}
\label{path_graph_proof}
The path RP-1 distance of the path graph can also be determined analytically using simple circuit's
rules. We decompose $\drpo$ as follows,
\begin{equation}
  \begin{split}
    \drpo(P_n,P_n+\dw{i_0j_0}) 
    & = 2 \sum_{i< j} \mspace{-2mu} \Delta R_{ij}  = \sum_{i=1}^n \sum_{j=i+1}^n \mspace{-8mu} \Delta R_{ij}\\
    & = 2 \sum_{i=1}^{i_0-1} \left (\sum_{j=i+1}^{i_0} \mspace{-8mu}\Delta R_{ij}   + 
      \sum_{j=i_0+1}^{j_0-1} \mspace{-8mu}\Delta R_{ij}   + \sum_{j=j_0}^{n}\mspace{-8mu}\Delta R_{ij}\right) \\
    & + 2\sum_{i=i_0}^{j_0-1} \left( \sum_{j=i+1}^{j_0} \mspace{-8mu} \Delta R_{ij} + \mspace{-16mu}
      \sum_{j=j_0+1}^{n} \mspace{-8mu}\Delta R_{ij} \right) 
    + 2 \sum_{i=j_0}^{n} \sum_{j=i+1}^{n}\mspace{-8mu}\Delta R_{ij}.
  \end{split}
\end{equation}
Now, for $1\leq i < j\leq i_0$  or $j_0+1 \leq i < j\leq n$, $\Delta R_{ij} =0$. We are thus left with
four sums,
\begin{equation}
  \begin{split}
    \drpo(P_n,P_n+\dw{i_0j_0}) 
    & = 2 \sum_{i=1}^{i_0-1}      \sum_{j=i_0+1}^{j_0-1} \Delta R_{ij} + 2 \sum_{i=1}^{i_0-1}      \sum_{j=j_0}^{n}\Delta R_{ij}\\
    & + 2\sum_{i=i_0}^{j_0-1}  \sum_{j=i+1}^{j_0}  \Delta R_{ij}  + 2 \sum_{i=i_0}^{j_0-1} \sum_{j=j_0+1}^{n} \Delta R_{ij}.
  \end{split}
\end{equation}
We compute each of the four sums using simple rules for combining resistances. The simplest case
corresponds to $1 \leq i \leq i_0-1$ and $j_0 \leq j \leq n$, where the two nodes are across the
edges that was added.  In this case, $i$ and $j$ only feel a difference that corresponds to the 
resistor $j_0 - i_0$ being in parallel with $r = 1/\dw{i_0j_0}$,
\begin{equation}
  \Delta R_{ij}  = \frac{(j_0 - i)^2}{r + j_0 - i_0},
\end{equation}
which does not depend on $i$ or $j$. Therefore, 
\begin{equation}
  2 \sum_{i=1}^{i_0-1}      \sum_{j=j_0}^{n}\Delta R_{ij} = 
  \frac{(i_0 -i)(n- j_0 +1)(j_0 - i_0)^2}{r + j_0 - i_0}.
  \label{S1}
\end{equation}
The next simple case corresponds to $1 \leq i \leq i_0-1$ and $i_0 \leq j \leq j_0$. In this
case, we have 
\begin{equation}
  \Delta R_{ij}  = \frac{(j - i_0)^2}{r + j_0 - i_0}.
\end{equation}
By symmetry, we can handle the case where $i_0 \leq i \leq j_0-1$ and $j_0 \leq j \leq n$, where we
have 
\begin{equation}
  \Delta R_{ij}  = \frac{(i - j_0)^2}{r + j_0 - i_0}.
\end{equation}
In both cases, we can compute the corresponding sums, and we get,
\begin{align}
  2 \sum_{i=1}^{i_0-1}      \sum_{j=i_0}^{j_0-1}\Delta R_{ij} 
  & =  \frac{(i_0 -1)(j_0 - i_0 -1)(j_0 - i_0)(2 (j_0 - i_0) -1)}{3(r + j_0 - i_0)}, \label{S2}\\
  2 \sum_{i=1}^{i_0-1}      \sum_{j=j_0}^{n}\Delta R_{ij} 
  & =  \frac{(n-j_0) (j_0 - i_0 +1) (j_0 - i_0) (2 (j_0 - i_0) +1)}{3(r + j_0 - i_0)}. \label{S3}
\end{align}
Finally, the last case is slightly more complicated and involves the scenario were both $i$ and $j$
are in between $i_0$ and $j_0$, $i_0 \leq i < j \leq j_0$ . In this case, we have
\begin{equation}
  \Delta R_{ij}  = \frac{(j - i)^2}{r + j_0 - i_0}.
\end{equation}
The corresponding sum becomes
\begin{equation}
  2\sum_{i=i_0}^{j_0-1}  \sum_{j=i+1}^{j_0}  \Delta R_{ij} = 
  \frac{(j_0 - i_0) (j_0 - i_0 +1)((j_0 - i_0)^2 + 3 (j_0 - i_0) + 2)}{6(r + j_0 - i_0)}.
  \label{S4}
\end{equation}
Grouping all the terms, (\ref{S1}), (\ref{S2}), (\ref{S3}), and (\ref{S4}),  together, and after a
some simple algebra, we get 
\begin{equation}
  \drpo(P_n,P_n+\dw{i_0j_0}) = 
  (j_0 - i_0)
  \frac{
    2n \left[1 + (j_0-i_0)(2j_0 - 4i_0 -3)\right] -3 (j_0 - i_0)(i_0 + j_0 -1)^2 
  }
  {6(r + j_0 - i_0)}.
\end{equation}
Substituting $r = 1/\dw{i_0j_0}$ in the above equation yields the advertised result.\hfill
$\square$
\subsection{Proof of Theorem \ref{ring_graph_thm}}
\label{ring_graph_proof}
\noindent In order to compute the resistance perturbation distance in the case of the cycle, we
break the sum into three terms.  The indexing of the vertices along the cycle makes the derivation
slightly more complicated than in the case of the path. To simplify the derivation of the results,
which eventually only depend on $i_0 \ominus i_0 = i_0 - j_0 \pmod{n}$, we first assume that
\begin{equation*}
  1= j_0 <  i_0.
\end{equation*}
In the end, we substitute $i_0 - j_0 \pmod{n}$ for $i_0-1$ in the final formula.  

We proceed in a manner similar to the path and decompose $\drpo(C_n,C_n+\dw{i_0j_0})$ into
three sums,
\begin{equation}
  \begin{split}
    \drpo(C_n,C_n+\dw{i_0j_0}) 
    & =  \sum_{i,j=1}^n \left\lvert \Delta R_{ij} \right\rvert  =  2\sum_{i=1}^{i_0-1}
    \sum_{j=i+1}^r \left\lvert\Delta R_{ij} \right\rvert  \\
    & + 2\sum_{i=1}^{i_0-1}  \sum_{j=i_0+1}^n \left\lvert \Delta R_{ij}
    \right\rvert +   2\sum_{i=i_0}^{n-1} \sum_{j=i+1}^n \left\lvert \Delta R_{ij} \right\rvert .
    \label{ring_RPdist}
  \end{split}
\end{equation}                             
Assuming that $\dw{i_0j_0} > 0$, then Rayleigh's monotonicity  principle implies that 
\begin{equation}
  \begin{split}
    |\Delta R_{ij} (C_n,C_n+\dw{i_0j_0})|  & = | R_{ij} (C_n) - R_{ij} (C_n+\dw{i_0j_0})| \\
    & = R_{ij} (C_n)  - R_{ij} (C_n+\dw{i_0j_0}) 
  \end{split}
\end{equation}
Now,
\begin{equation}
  R_{ij} (C_n) = (j - i)-\frac{(j-i)^2}{n}.
\end{equation}
The first sum corresponds to the case where $1 \leq i \leq i_0-1$ and $i+1 \leq j \leq i_0$. In this
case we have $j_0 =1 < i< j < j_0 \leq n$, and the chords formed by $(1,i_0)$ and $(i,j)$ do not
intersect. Simple circuit rules yield
\begin{equation}
  \Delta R_{ij} (C_n,C_n+\dw{i_0j_0})  = 
  \frac{\left[(i-j)(n -(i_0-1))\right]^2 \dw{i_0j_0}}
  {n^2 + \dw{i_0j_0} n (i_0 -1)(n - (i_0-1))}.
\end{equation}
The second sum correspond to the case where $1 \leq i \leq i_0-1$ and $i_0+1\leq j \leq n$.
In this case, $j_0 =1 < i< j_0 < j \leq n$, and the chords formed by $(1,i_0)$ and $(i,j)$
intersect. The difference in effective resistance is given by 
\begin{equation}
  \Delta R_{ij} (C_n,C_n+\dw{i_0j_0})  = 
  \frac{\left[(i-j)(i_0-1) + (i_0 - i)n\right]^2 \dw{i_0j_0}}
  {n^2 + \dw{i_0j_0} n (i_0 -1)(n - (i_0-1))}.
\end{equation}
Finally, the last sum corresponds to the case where $i_0 \leq n \leq n-$ and $i+1 \leq j \leq n$. In
this case $j_0 =1 < i_0 < i < j < j_0 \leq n$, and the chords formed by $(1,i_0)$ and $(i,j)$ do not
intersect. The difference in effective resistance is given by 
\begin{equation}
  \frac{\left[(i-j)(i_0-1)\right]^2 \dw{i_0j_0}}
  {n^2 + \dw{i_0j_0} n (i_0 -1)(n - (i_0-1))}.
\end{equation}
Computing the three sums in (\ref{ring_RPdist}) yields
\begin{equation}
  \begin{split}
    \drpo(C_n,C_n+\dw{i_0j_0})  & =\\
    n \left\lvert \dw{i_0j_0}\right\rvert
    \left[i_0 - 1\right]
    &\frac{
      (i_0 -1)^3 -2n ((i_0-1)^2 -1) + (i_0-1)(n^2 -2)
    }
    {6\left\{
        n^2 + \dw{i_0j_0} n (i_0 -1)(n - (i_0-1))
      \right\}}.
  \end{split}
\end{equation}
Finally, we can substitute $i_0 \ominus j_0 = i_0 -j_0 \pmod{n}$ for $i_0-1$, and we obtain
(\ref{ring_graph_eqn}),
\begin{equation}
  \begin{split}
    \drpo(C_n,C_n+\dw{i_0j_0})  & = \\
    n \left\lvert \dw{i_0j_0} \right\rvert \left[i_0 \ominus j_0\right]  
    & \frac{
      \left[i_0 \ominus j_0 \right]^3 
      -2n \left[(i_0 \ominus  j_0 )^2 -1\right] 
      + \left[i_0 \ominus j_0 \right](n^2 -2)
    }
    {6\left\{
        n^2 + \dw{i_0j_0} n \left[i_0 \ominus j_0 \right]
        \left[n - (i_0 \ominus j_0)\right]
      \right\}}.
  \end{split}
\end{equation}

\subsection{Proof of Theorem \ref{frobenius_bounds_thm}}
\label{frobenius_bounds_thm_proof}
Combining inequalities for $\tR^{(1)}_{ij}$ and $\tR^{(2)}_{ij}$ and applying the triangle inequality,
{\small
  \begin{equation}
    (1-\varepsilon)R^{(1)}_{ij}-(1+\varepsilon)R^{(2)}_{ij} \leq
    \tR^{(1)}_{ij}-\tR^{(2)}_{ij} \leq
    (1+\varepsilon)R^{(1)}_{ij}-(1-\varepsilon)R^{(2)}_{ij} 
  \end{equation}
  \begin{equation}
    (R^{(1)}_{ij}-R^{(2)}_{ij})-\varepsilon(R^{(1)}_{ij}+R^{(2)}_{ij}) \leq
    \tR^{(1)}_{ij}-\tR^{(2)}_{ij} \leq
    (R^{(1)}_{ij}-R^{(2)}_{ij})+\varepsilon(R^{(1)}_{ij}+R^{(2)}_{ij}) 
  \end{equation}
  \begin{equation}
    \left\lVert (\bR^{(1)}-\bR^{(2)})-\varepsilon(\bR^{(1)}+\bR^{(2)})\right\rVert_F \leq \left\lVert
      \tR^{(1)}-\tR^{(2)} \right\rVert_F \leq
    \left\lVert(\bR^{(1)}-\bR^{(2)})+\varepsilon(\bR^{(1)}+\bR^{(2)})\right\rVert_F 
  \end{equation}
  \begin{equation}
    \left\lVert \bR^{(1)}-\bR^{(2)}\right\rVert_F-\varepsilon\left\lVert\bR^{(1)}+\bR^{(2)}\right\rVert_F \leq \left\lVert
      \tR^{(1)}-\tR^{(2)}\right\rVert_F \leq
    \left\lVert\bR^{(1)}-\bR^{(2)}\right\rVert_F+\varepsilon\left\lVert\bR^{(1)}+\bR^{(2)}\right\rVert_F. 
  \end{equation}}\hfill $\square$
\subsection{Proof of Theorem \ref{fast_frobenius_thm}}
\label{fast_frobenius_thm_proof}
Let $\bd = \diag\left(\tZoT \tZo - \tZtT \tZt\right) \in \R^n$.  Using
the invariance of the trace under cyclic permutations ($\tr(ABC) = \tr(CAB)$), we have
\begin{align*}
  \left\lVert \tR^{(1)} - \tR^{(2)} \right\rVert^2_F 
  &= \left\lVert  \bd \bone^T + \bone \bd^T  - 2 (\tZoT \tZo - \tZtT \tZt) \right\rVert_F^2 \\
  &= \tr \left\{
    \left[
    \bd \bone^T + \bone \bd^T - 2 \left(\tZoT \tZo - \tZtT \tZt\right)
    \right]^2
    \right\} \\ 
  &= \tr\left\{
    \bd \bone^T \bck \bd \bone^T + \bd \bone^T\bck \bone \bd^T 
    +  \bone \bd^T \bck \bd \bone^T +  \bone \bd^T \bck \bone \bd^T 
    - 2 \; \bd  \bone^T \left[\tZoT \tZo -  \tZtT \tZt\right] \right.\\
  & \mspace{32mu}
    - 2 \; \bone \bd^T\left[\tZoT \tZo - \tZtT \tZt \right] 
    - 2\left[\tZoT \tZo - \tZtT \tZt\right]\bd \bone^T \\
  & \mspace{30mu}
    +4 \left[\tZoT \tZo - \tZtT \tZt\right]\bck
    \left[\tZoT \tZo - \tZtT \tZt\right] \\
  &  \mspace{32mu}\left.
    - 2\left[\tZoT \tZo - \tZtT \tZt\right]\bone \bd^T 
    \right\} \\  
  &= 2 \tr \left[ \bd \bone^T \bck \bd \bone^T \right] 
    + 2\tr \left[ \bd\bone^T \bck \bone \bd^T \right] 
    - 8 \tr \left[\bd \bone^T \bck (\tZoT \tZo -  \tZtT \tZt ) \right] \\
  & \mspace{12mu}  
    + 4 \tr \left[
    \left(\tZoT \tZo - \tZtT \tZt\right)\bck
    \left(\tZoT \tZo - \tZtT \tZt\right) \right] \\ 
  & = 2\tr \left[ \bone^T \bck \bd \bone^T \bck \bd \right] 
    + 2\tr\left[\bd^T \bck \bd \bone^T \bone \right] 
    - 8 \tr\left[\bd \bone^T \tZoT\tZo\right] 
    +8\tr\left[\bd \bone^T \tZtT \tZt \right]  \\
  & \mspace{12mu}  
    +4 \tr\left[
    \left(\tZoT \tZo - \tZtT \tZt\right)\bck
    \left(\tZoT \tZo - \tZtT \tZt\right)\right] \\ 
  &= 2 \left(\bone^T \bd \right)^2 
    + 2n\left\lVert \bd \right\rVert_2^2 
    - 8 \left[ \bone^T \tZoT \tZo\bd \right] 
    +8\left[  \bone^T \tZtT \tZt \bd \right] \\ 
  & \mspace{24mu}
    +4 \tr \left[\tZo\tZoT\tZo\tZoT\right] 
    + 4 \tr \left[ \tZt\tZtT\tZt\tZtT\right] \\
  & \mspace{24mu}  
    -8 \tr \left[\tZt\tZoT \tZo \tZtT \right] \\
  &= 2\left(\bone^T \bd \right)^2 + 2n\left\lVert d \right\rVert_2^2 
    - 8 (\bone^T \tZoT) (\tZo \bd) +8  (\bone^T \tZtT) (\tZt \bd)  \\ 
  & \mspace{24mu}  
    +4 \left\lVert \tZo \tZoT \right\rVert^2_F + 4 \left\lVert \tZt \tZtT \right\rVert_F^2 
    -8 \left\lVert \tZt \tZoT  \right\rVert_F^2. 
\end{align*}
Computation of $d$ is ${\cal O}(sn)$, and thus so is the computation of the
$1^{st}$ and $2^{nd}$ terms.  The $3^{rd}$ and $4^{th}$ terms cost
${\cal O}(sn)$, as they involve multiplication of $(s \times n)$-matrices with
length-$n$ vectors.  The $5^{th}$, $6^{th}$, and $7^{th}$ terms cost
${\cal O}(s^2n)$, due to the multiplication of $(s \times n)$ with
$(n \times s)$-matrices.  Recalling that $s={\cal O}(\log n)$ we see that the
total computational complexity of computing the Frobenius norm is reduced to
$\widetilde{{\cal O}}(n)={\cal O}(n \log^2 n)$.
\subsection{Proof of Theorem \ref{partial_sum_thm}}
\label{partial_sum_thm_proof}
We will prove the first of the two inequalities.  The proof of the second is
identical if we replace $\lambda_k$ with $\lambda^2_k$.  We will employ the
observation that $\sum_{k=2}^n (\bfi_k(i)-\bfi_k(j))^2 = 2$.  To show this, we
note that $\bFi = \begin{bmatrix} \bfi_1& \cdots & \bfi_n\end{bmatrix}$ is an
orthogonal matrix, and thus its rows are orthonormal,
\begin{equation}
  \begin{split}
    \sum_{k=2}^n  (\bfi_k(i)-\bfi_k(j))^2 & = \sum_{k=1}^n (\bfi_k(i)-\bfi_k(j))^2 \\
    &= \sum_{k=1}^n \bfi_k(i)^2 + \sum_{k=1}^n \bfi_k(j)^2 -2 \sum_{k=1}^n \bfi_k(i) \bfi_k(j) =2,
  \end{split}
\end{equation}
since each of the first two terms is equal to the squared norm of a row of
$\bFi$, and the second is the inner-product of two rows.  To bound the effective
resistance we break into two partial sums,
\begin{equation}
  R_{ij}=\sum_{k=2}^n \frac{1}{\lambda_k}(\bfi_k(i)-\bfi_k(j))^2 
  = \sum_{k=2}^p \frac{1}{\lambda_k}(\bfi_k(i)-\bfi_k(j))^2
  +\sum_{k=p+1}^n \frac{1}{\lambda_k}(\bfi_k(i)-\bfi_k(j))^2. 
\end{equation}
Then,
\begin{equation}
  \begin{split}
    \sum_{k=2}^p  \frac{1}{\lambda_k}&(\bfi_k(i)-\bfi_k(j))^2 +  \frac{1}{\lambda_n} \sum_{k=p+1}^n (\bfi_k(i)-\bfi_k(j))^2
    \leq R_{ij}\\
    & \leq \sum_{k=2}^p \frac{1}{\lambda_k}(\bfi_k(i)-\bfi_k(j))^2 + \frac{1}{\lambda_p} \sum_{k=p+1}^n (\bfi_k(i)-\bfi_k(j))^2 \\
    \sum_{k=2}^p  \frac{1}{\lambda_k}&(\bfi_k(i)-\bfi_k(j))^2 +  \frac{1}{\lambda_n} \sum_{k=p+1}^n (\bfi_k(i)-\bfi_k(j))^2
    \leq R_{ij}\\
    & \leq \sum_{k=2}^p \frac{1}{\lambda_k}(\bfi_k(i)-\bfi_k(j))^2 + \frac{1}{\lambda_p} \sum_{k=p+1}^n (\bfi_k(i)-\bfi_k(j))^2 \\
    \sum_{k=2}^p  \frac{1}{\lambda_k}&(\bfi_k(i)-\bfi_k(j))^2 +  \frac{1}{\lambda_n}\left\{2-\sum_{k=2}^p (\bfi_k(i)-\bfi_k(j))^2 \right\}
    \leq R_{ij}\\ 
    & \leq \sum_{k=2}^p \frac{1}{\lambda_k}(\bfi_k(i)-\bfi_k(j))^2 +
    \frac{1}{\lambda_p}\left\{2-\sum_{k=2}^p (\bfi_k(i)-\bfi_k (j))^2 \right\}.
  \end{split}
\end{equation}
Combining terms proves the desired result. \hfill $\square$
\subsection{Low-rank edge modification Corollary
  \label{corollary2}}
\begin{corollary}[Low-rank edge modification]
  Assume $G+\dw{i_0j_0}$ is the graph obtained from $G$ by a perturbation $\dw{i_0j_0}$ to the edge connecting $i_0$ and $j_0$.  If $\dw{i_0j_0}>0$ we have
  \begin{equation}
    \drpo(G,G+\dw{i_0j_0}) \geq \frac{  2 n \left\lvert \dw{i_0j_0} \right\rvert \left[ \displaystyle \frac{1}{\lambda_p^2} + \sum_{k=2}^p \left( \frac{1}{\lambda_k^2}-\frac{1}{\lambda_p^2} \right) (\phi_k(i_0) - \phi_k(j_0) )^2 \right] }{  1 + \dw{i_0j_0} \left[ \displaystyle \frac{1}{\lambda_n} + \sum_{k=2}^p \left( \frac{1}{\lambda_k}-\frac{1}{\lambda_n} \right) (\phi_k(i_0) - \phi_k(j_0) )^2 \right] },
  \end{equation}
  and
  \begin{equation}
    \drpo(G,G+\dw{i_0j_0}) \leq \frac{  2 n \left\lvert \dw{i_0j_0} \right\rvert \left[ \displaystyle \frac{1}{\lambda_n^2} + \sum_{k=2}^p \left( \frac{1}{\lambda_k^2}-\frac{1}{\lambda_n^2} \right) (\phi_k(i_0) - \phi_k(j_0) )^2 \right] }{  1 + \dw{i_0j_0} \left[ \displaystyle \frac{1}{\lambda_p} + \sum_{k=2}^p \left( \frac{1}{\lambda_k}-\frac{1}{\lambda_p} \right) (\phi_k(i_0) - \phi_k(j_0) )^2 \right] }.
  \end{equation}
  If $\dw{i_0j_0}<0$ we have
  \begin{equation}
    \drpo(G,G+\dw{i_0j_0}) \geq \frac{  2 n \left\lvert \dw{i_0j_0} \right\rvert \left[ \displaystyle \frac{1}{\lambda_n^2} + \sum_{k=2}^p \left( \frac{1}{\lambda_k^2}-\frac{1}{\lambda_n^2} \right) (\phi_k(i_0) - \phi_k(j_0) )^2 \right] }{  1 + \dw{i_0j_0} \left[ \displaystyle \frac{1}{\lambda_n} + \sum_{k=2}^p \left( \frac{1}{\lambda_k}-\frac{1}{\lambda_n} \right) (\phi_k(i_0) - \phi_k(j_0) )^2 \right] },
  \end{equation}
  and
  \begin{equation}
    \drpo(G,G+\dw{i_0j_0}) \leq \frac{  2 n \left\lvert \dw{i_0j_0}
      \right\rvert \left[ \displaystyle \frac{1}{\lambda_p^2} + \sum_{k=2}^p \left(
          \frac{1}{\lambda_k^2}-\frac{1}{\lambda_p^2} \right) (\phi_k(i_0) -
        \phi_k(j_0) )^2 \right] }{  1 + \dw{i_0j_0} \left[ \displaystyle
        \frac{1}{\lambda_p} + \sum_{k=2}^p \left(
          \frac{1}{\lambda_k}-\frac{1}{\lambda_p} \right) (\phi_k(i_0) - \phi_k(j_0)
        )^2 \right] }. 
  \end{equation}
  \label{approx_opt_cor}
\end{corollary}
\begin{cproof}
  Straightforward application of bounds from theorem \ref{partial_sum_thm} to
  theorem \ref{edge_mod_thm}. \hfill $\square$ 
\end{cproof}
\end{document}